\documentclass{aa}  
\usepackage[switch]{lineno}
\usepackage{xcolor}
\usepackage{graphicx}
\usepackage{txfonts}
\usepackage[T1]{fontenc}
\usepackage{comment}
\usepackage{orcidlink}
\usepackage{hyperref}
\hypersetup{
  colorlinks,
  citecolor=blue,
  linkcolor=blue,
  urlcolor=blue}

\begin{document} 

      \title{The role of triple evolution in the formation of \\ LISA double white dwarfs}

   \titlerunning{The role of triple evolution in the formation of LISA double white dwarfs}

   \author{Abinaya Swaruba Rajamuthukumar\inst{1}\orcidlink{0000-0002-1872-0124}
        \and
        Valeriya Korol\inst{1}\orcidlink{0000-0002-6725-5935}
        \and
        Jakob Stegmann\inst{1}\orcidlink{0000-0003-2340-8140}
        \and
        Holly Preece\inst{2}\orcidlink{0000-0001-7984-7033}
        \and
        R\"udiger Pakmor\inst{1}\orcidlink{0000-0003-3308-2420}
        \and
        Stephen Justham\inst{1}\orcidlink{0000-0001-7969-1569}
        \and
        Silvia Toonen\inst{3}\orcidlink{0000-0002-2998-7940}
        \and
        Selma E. de Mink\inst{1,4}\orcidlink{0000-0001-9336-2825}
        }
\institute{Max-Planck-Institut für Astrophysik, Karl-Schwarzschild-Straße 1, 85748 Garching bei München, Germany\\
              \email{abinaya@mpa-garching.mpg.de}
        \and
            Radboud University,
              Nijmegen, The Netherlands
              \and
              Anton Pannekoek Institute for Astronomy, University of Amsterdam, Science Park 904, 1098, XH, Amsterdam, The Netherlands
              \and
              Ludwig-Maximilians-Universität München, Geschwister-Scholl-Platz 1, 80539 München, Germany
            }
    \authorrunning{A.S.Rajamuthukumar et al}

  \abstract
  {Galactic double white dwarfs will be prominent gravitational wave sources for the Laser Interferometer Space Antenna (LISA).  While previous studies have primarily focused on formation scenarios in which binaries form and evolve in isolation, we present the first detailed study of the role of triple stellar evolution in forming the population of LISA double white dwarfs.  We used the multiple stellar evolution code (\texttt{MSE}) to model the stellar evolution, binary interactions, and the dynamics of triple star systems and then used a Milky Way-like galaxy from the \texttt{TNG50} simulations to construct a representative sample of LISA double white dwarfs. In our simulations, about $7\times10^6$ Galactic double white dwarfs in the LISA frequency bandwidth originate from triple systems, whereas $\sim4\times10^6$ are in isolated binary stars. The properties of double white dwarfs formed in triples closely resemble those formed from isolated binaries, but we also find a small number of systems, $\sim\mathcal{O}(10)$, that reach extreme eccentricities $(>0.9)$, a feature unique to the dynamical formation channels. Our population produces  $\sim\mathcal{O}(10^4)$ individually resolved double white dwarfs (from triple and binary channels) and an unresolved stochastic foreground below the level of the LISA instrumental noise. About $57\,\%$ of the double white dwarfs from triple systems retain a bound third star when entering the LISA frequency bandwidth. However, we expect the tertiary stars to be too distant to have a detectable imprint in the gravitational wave signal of the inner binary.}


   \maketitle
%
\section{Introduction}

Observations indicate that triple star systems are common 
 across various stellar evolutionary stages, including main-sequence stars, evolved giant stars, brown dwarfs, and black holes \citep{2008MNRAS.389..869E,2017ApJS..230...15M,2017A&A...598L...7K,2020NatAs...4..650T,2021A&A...653A..40L,2024Natur.635..316B}. In particular, white dwarfs have been found in triple systems with main-sequence stars (for a review within 20\,pc, see \citealt{2014A&A...562A..14T}), other white dwarfs \citep{10.1046/j.1365-8711.2000.03343.x,2019MNRAS.483..901P}, neutron stars \citep{2014Natur.505..520R}, and brown dwarfs \citep{2022ApJ...927L..31R}. However, the observed number of white dwarfs in triple systems remains significantly lower than theoretical predictions, possibly due to observational biases. Indeed, \citet{2023ApJ...955L..14S,2024arXiv240706257S} propose that many observed local double white dwarfs could have originated from triples. For example, a recent spectroscopic study of wide double white dwarfs by \citet{2024ApJ...969...68H} reveals some systems in which the more massive white dwarf companion has a shorter cooling age compared to the less massive one, which contradicts the initial-final mass relation if both stars were formed simultaneously in a non-interacting binary. One proposed explanation is that the progenitors of the double white dwarf were originally in a triple system, where the massive white dwarf was formed by the merger of two stars, resulting in a shorter cooling age. Furthermore, triples could significantly contribute to the rate of Type Ia supernovae, making a substantial contribution to the Type Ia supernova rate from isolated binary stars \citep{2011PhRvL.107r1101K,2013MNRAS.430.2262H,2023ApJ...950....9R}. 

 Hierarchical triple systems are characterized by a close inner orbit and with a tertiary component in a wider orbit. When the orbits of the inner and outer stars are sufficiently inclined, the gravitational perturbation from the tertiary star can cause large-amplitude von Zeipel-Lidov-Kozai (ZLK) oscillations \citep{Zeipel1910,1962P&SS....9..719L,1962AJ.....67..591K} of the inner binary eccentricity while the semi-major axis remains unchanged. This process can play a key role in the formation of close binaries. The combination of ZLK oscillations with dissipative effects such as tidal friction \citep{10.1046/j.1365-8711.1998.01903.x,Eggleton_2001,2007ApJ...669.1298F} and gravitational wave radiation can lead to a reduction in the inner binary’s orbital semi-major axis. Thus, perturbations from the tertiary star can facilitate close-binary processes such as mass transfer, common-envelope phases, mergers, and collisions in the inner binary \citep{2007ApJ...669.1298F,2012ApJ...760...99P,2013ApJ...766...64S,2013MNRAS.430.2262H,2014ApJ...794..122M,2017ApJ...841...77A,2020A&A...640A..16T,2019ApJ...882...24H,2022MNRAS.516.1406S,2022PhRvD.106b3014S}.

Gravitational waves from compact double white dwarfs (with  frequencies ranging from $10^{-4}$ to $10^{-1}\,\rm Hz$) detectable with the upcoming Laser Interferometer Space Antenna (LISA) mission  offer a unique way to explore these triple systems \citep{2023LRR....26....2A}. By detecting the gravitational wave signals from double white dwarfs, LISA could uncover a population of systems formed through the triple evolution channel, which are inaccessible to electromagnetic observations. This capability has the potential to provide new insights into the formation mechanisms of double white dwarfs, their contribution to Type Ia supernovae \citep{1984ApJS...54..335I,2024A&A...691A..44K}, and the broader implications for the chemical evolution of the Galaxy \citep{1997nceg.book.....P}.

 LISA is expected  to detect $\sim\mathcal{O}(10^6)$ of Galactic double white dwarfs as part of an unresolved confusion gravitational wave background and individually resolve around $10^3$~--~$10^4$ of the ``loudest'' double white dwarfs \citep[e.g.,][]{2017MNRAS.470.1894K,2019MNRAS.490.5888L,2021MNRAS.500.4958W,2023ApJ...945..162T,2023A&A...669A..82L,2024arXiv240520484T}. While previous studies have focused on double white dwarfs formed from the evolution of isolated binary stars, increasing evidence suggests that hierarchical triple systems, in which a close inner binary is orbited by a distant tertiary companion, may also play a significant role in the formation of double white dwarfs \citep{2020A&A...640A..16T,2024ApJ...969...68H,2024arXiv240706257S}.

There is mounting observational evidence that stars often form with bound companions, with a binary fraction of $30\,\%$ and a triple fraction of $10\,\%$ for F- and G-type stars (i.e., with masses $\sim 1\,\mathrm{M_{\odot}}$, \cite{2008MNRAS.389..869E,2010ApJS..190....1R,2014AJ....147...87T,2017ApJS..230...15M,2023ASPC..534..275O}. Moreover, the inner binaries in triple systems tend to be in closer orbits than the binaries found in isolated systems, which increases the probability for some binary interactions \citep{2020A&A...640A..16T}. In this paper, we show that triple systems offer a greater number of evolutionary pathways for forming short-period inner binaries compared to isolated binary systems. Approximately 10\% of white dwarfs are found in binary systems where both components are white dwarfs (i.e., double white dwarfs; \citealt{1999MNRAS.307..122M, 2018MNRAS.476.2584M, 2020A&A...638A.131N}).

Previous studies have explored the potential for detecting tertiary companions in LISA data \citep[e.g.,][]{2008ApJ...677L..55S, 2018PhRvD..98f4012R,2019NatAs...3..858T}. Similar to electromagnetic observations, the motion of the double white dwarf around the center of mass of the triple system modulates the gravitational wave frequency through the Doppler effect. This modulation produces a periodic shift, causing the observed frequency to oscillate around the intrinsic frequency of the inner binary. Recent studies have focused on leveraging this effect to detect substellar mass tertiaries, such as exoplanets and brown dwarfs \citep{2019NatAs...3..858T, 2019A&A...632A.113D,2021AJ....162..247K,2022MNRAS.517..697K}. Thus, previous studies have primarily focused on either the isolated binary population of double white dwarfs or the possibility of detecting a third star.
Our work is the first evolutionary population synthesis study of Galactic double white dwarfs resulting from triple evolution. In addition, our study assesses the impact of the triple evolution channel to the LISA's astrophysical noise background, thereby influencing the $\rho$ of all other gravitational wave sources.

We aim to quantify the contribution of the triples to the population of double white dwarfs detectable by LISA. We combine population synthesis models using the \texttt{MSE} code \citep{2021MNRAS.502.4479H} with cosmological simulations from the \texttt{TNG50} project \citep{2019MNRAS.490.3234N,2019MNRAS.490.3196P} to construct a representative model of the Galactic double white dwarf population. Our study addresses two critical questions: 1) What fraction of double white dwarfs detectable by LISA originates from the triple evolution channel? 2) Can LISA detect the dynamical effects of the third star in these triple systems?

The paper is structured as follows. In Sect.~\ref{sec:methods} we explain our methodology. In Sect.~\ref{sec:Triple evolution channel} we describe the evolutionary pathways of triples that lead to the formation of LISA double white dwarfs. We detail the population properties of LISA double white dwarfs from isolated binaries and triples in  Sect.~\ref{sec:binaryvstriple}, investigating prospects for direct detection of the third star in Sect.~\ref{sec:thirdstar}. Finally, we discuss the results in Sect.~\ref{sec:Discussion} and summarize our findings in Sect.~\ref{sec:Conclusion}.

\section{Methods}\label{sec:methods}

All simulations were performed using the publicly available population synthesis code \texttt{MSE}\footnote{\url{ https://github.com/hpreece/mse}.} \citep{2021MNRAS.502.4479H}. Our set of simulations of hierarchical triples consists of a main run with a choice of default parameters and three model-variant runs. In each simulated data set, we evolve the triples from the start of the zero-age main-sequence until a maximum integration time $t_{\rm max}=14 \,\rm Gyr$. The main triple data set consists of $10^5$ systems, where we adopt a common envelope efficiency parameter $\mathrm{\alpha_{CE} = 1}$, and all three stars are assumed to have formed at solar metallicity $Z=\rm Z_\odot=0.02$. This initial population results in $3 \times 10^{3}$ LISA double white dwarfs, with a $\sim 2\%$ Poisson uncertainty in this number.  In each of the three model-variant runs, we simulate $10^{4}$ systems and either vary the common envelope efficiency parameter as $\mathrm{\alpha_{CE} = 0.1}$ and 10 or change the metallicity of the stars to subsolar $Z=0.1\,\mathrm{Z_{\odot}}$. The effects of the chosen parameters are discussed in Sect.~\ref{sec:Discussion}. In addition to the triple runs, we simulate a population of $10^5$ isolated binaries to compare the impact of tertiary companions. Additionally, we model the inner binaries of all triples without their tertiary stars to assess their influence on the resulting LISA double white dwarf population. We expect that the isolated binary population differs significantly from the inner binary population of triples \citep{2023ApJ...950....9R}. The primary distinction is that the inner binaries have much more compact semi-major axes due to the dynamical stability constraints imposed by the tertiary star.

This section details the physics of the single, binary, and triple evolution incorporated into \texttt{MSE}, and outlines our initial distributions for the stellar populations. Additionally, we provide an overview of the Milky Way-like galaxy selected from the cosmological simulation \texttt{TNG50}. This Milky Way-like galaxy are then used to seed double white dwarfs in the galaxy. The methodology for constructing the Galactic double white dwarf population from triples is also explained here, while further details on building the Galactic double white dwarf population from isolated binaries are provided in the appendix.

\subsection{Multiple stellar evolution code}
\label{sec:mse}

We used the population synthesis approach using the code \texttt{MSE} to model the stellar evolution, binary interactions (tides, mass transfer), dynamical perturbations from higher-order companions in multiple systems, and flybys from ambient stars. \texttt{MSE} is a\texttt{ C/C++} code with a \texttt{Python} interface that can handle any number of stars as long as they start in a hierarchical arrangement. \texttt{MSE} uses a hybrid approach that switches between the secular approximation for dynamically stable orbits (that satisfy the criterion of \citealt{2001MNRAS.321..398M}) and direct $N$-body integration using \texttt{MSTAR} \citep{2020MNRAS.492.4131R} for dynamically unstable orbits. Throughout the evolution post-Newtonian terms are included to 2.5 order in the secular approximation and to 3.5 order for the direct $N$-body integration.

To follow the evolution of single stars, \texttt{MSE} relies on the fitting formulae from \citet{1996MNRAS.281..257T,2000MNRAS.315..543H}, while binary interactions such as tides, wind mass transfer, stable mass transfer episodes,  and common envelope (CE) evolutions are computed using modified prescriptions of \cite{2002MNRAS.329..897H}. We briefly explain the physical handling of these key processes below. For more detailed explanations see \cite{2021MNRAS.502.4479H}.

Stable mass transfer: In \texttt{MSE}, mass transfer stability is determined either by the critical mass ratio criterion or by comparing mass transfer and dynamical timescales. The critical mass ratio depends on the donor’s stellar type \citep{2021MNRAS.502.4479H}. We assume fully conservative mass transfer ($\beta_{\mathrm{MT}} = 1$), meaning no mass is lost from the system. While \texttt{MSE} generally follows \cite{2002MNRAS.329..897H} for binary interactions, it differs in treating mass transfer in eccentric orbits. \cite{2002MNRAS.329..897H} assumes tides are always efficient in circularizing the orbit. However, in triple systems, the eccentricities are excited secularly. We follow the analytical model from \cite{2019ApJ...872..119H} to model mass transfer at periastron in eccentric orbits

Common envelope evolution: Unstable mass transfer/CE evolution in \texttt{MSE} follows the $\alpha_{\mathrm{CE}}$ prescription \citep{1976IAUS...73...75P,1976IAUS...73...35V,1988ApJ...329..764L,1993PASP..105.1373I,2002MNRAS.329..897H}. The code solves for orbital energies before and after the CE phase, parameterizing the envelope ejection efficiency by $\alpha_{\mathrm{CE}}$ and adopting a binding energy factor $\lambda$ for each donor star taken from fits described in \citet{2014A&A...563A..83C}, assuming all the ionization energy stored in each envelope is useful in unbinding that envelope. For the main runs, we adopt $\alpha_{\mathrm{CE}} = 1$. The post-CE semi-major axis is determined from the corresponding orbital energy.  The timescale of mass ejection from a CE in the inner binary is relevant for the evolution of the outer orbit, and we assume that ejected common-envelope material leaves the system in a timescale of $10^3\,\mathrm{yr}$.

Merger or collision: A ``failed'' CE can result in the merger of two stars. This occurs if the post-CE semi-major axis is too small for either star to avoid Roche lobe overflow. Beyond post-CE mergers, \texttt{MSE} also accounts for physical collisions when the sum of the stellar radii exceeds the semi-major axis or when a periapsis collision occurs in an eccentric orbit. The properties of the merger remnant are assigned following \cite{2002MNRAS.329..897H}.

Contact evolution: If both stars simultaneously fill their Roche lobes, \texttt{MSE} assumes a CE phase if both are giant stars. Otherwise, a merger is assumed.

Triple common envelope (TCE) evolution: Mass transfer from a third star onto the inner binary can lead to a TCE. If CE conditions are met for the third star, \texttt{MSE} employs ``circumstellar triple CE evolution,'' allowing the third star to fill its Roche lobe around the inner binary and undergo unstable mass transfer. This follows a similar approach to the prescription proposed by \citet{2020MNRAS.498.2957C}. The final outer semi-major axis is estimated using an $\alpha_{\mathrm{CE}}$ prescription, assuming the inner binary remains intact and does not change. However, CE modeling is uncertain (see \citealt{2013A&ARv..21...59I} for a review), and TCE evolution is even more so, requiring cautious interpretation of results. For hydrodynamical simulations of triple CE outcomes, see, for example, \citet{10.1093/mnras/staa3242}.

Flybys: \texttt{MSE} also includes the gravitational perturbations from stellar flybys in the vicinity of the system using the impulsive approximation. The flyby mass is sampled from a Kroupa initial mass function \citep{2001MNRAS.322..231K}, and encounters are randomly sampled within an encounter sphere of radius $R_{\text{enc}} = 10^5\,au$ with velocities drawn from a Maxwellian velocity distribution with dispersion $\sigma_{\star} = 30$ km s$^{-1}$. The number density of flybys accounts for the low-density environments, assuming $n_{\star} = 0.1 \, \mathrm{pc}^{-3}$. These flybys become significant for the evolution of the system if the semi-major axis of the outer orbit exceeds about $10^3\,\rm au$ \citep{2010MNRAS.401..977J,GrishinPerets,Stegmann:2024rnk}.

\subsection{Initial conditions}
\label{sec: initial distributions}

\begin{figure*}
    \centering
        \includegraphics[trim = 0 0 0 0 ,width=1\textwidth]{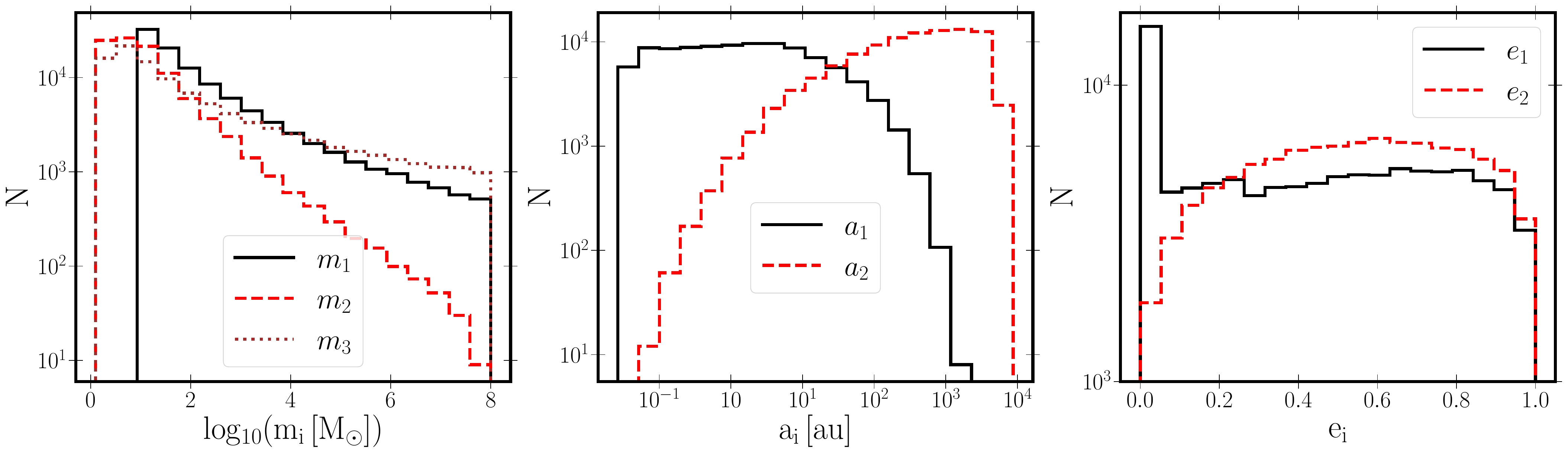}
    \caption{Initial parameter distributions. The left panel shows the mass distributions ($m_{i}$), where $m_1$ and $m_2$ denote the masses of the inner binary components, and $m_3$ represents the mass of the tertiary. The middle panel displays the semi-major axis distributions($a_{i}$) for the inner ($a_1$) and outer ($a_2$) orbits, respectively. The right panel illustrates the eccentricity distributions($e_{i}$) of the inner ($e_1$) and outer ($e_2$) orbits, respectively. (See Sect.~\ref{sec: initial distributions} for more details.)}
    \label{fig:initial_dist}
\end{figure*}

Here, we describe the initial distribution of our synthetic population of stars. We denote the masses of the inner binary components as $m_1$ and $m_2$, where $m_1\ge m_2$, and the mass of the tertiary companion as $m_3$. Semi-major axes are denoted as $a_1$ for the inner binary and $a_2$ for the outer binary. Orbital eccentricities are denoted as $e_1$ and $e_2$, respectively. Figure~\ref{fig:initial_dist} shows the  distribution of the parameters of the initial triple population. 

We draw the primary mass $m_1$ from a Kroupa initial mass function \citep{2001MNRAS.322..231K} between $1$ and $8 \, \mathrm{M_{\odot}}$. Furthermore, we follow functions from \citealt{2017ApJS..230...15M} to sample the orbital period ($0.2 \leq\log({T}_1/\text{days}) \leq 8$) and secondary mass ( $0.08 \, \mathrm{M_{\odot}}\leq m_2\leq m_1$) of the inner binary, and calculate the semi-major axis $a_1$ from Kepler's law. Similarly, the orbital period of the outer binary also follows \citealt{2017ApJS..230...15M}, where we assume that the inner binary is represented as a single star with a mass of $m_1 + m_2$. We allowed the tertiary mass, $m_3$, to be more massive than the total mass of the inner binary in certain cases, and an extrapolated mass ratio distribution from \citealt{2017ApJS..230...15M} is used to sample $m_3$. In addition, we sample the eccentricities of the inner ($e_1$) and outer ($e_2$) orbits from \citealt{2017ApJS..230...15M} and randomly sample the spatial orientations of the inner and outer orbital frames from isotropic distributions.

We rejected any star whose radius exceeds the Roche-lobe radius on the zero-age main-sequence \citep{1983ApJ...268..368E}
and reject any system which would be dynamically unstable \citep{2022MNRAS.516.4146V} at the start of the simulation. Any Roche lobe overflow or dynamical instabilities are modeled using prescriptions in \texttt{MSE} during the evolution (see Sect.~\ref{sec:mse} for more details).

\subsection{Construction of Galactic double white dwarf population} \label{sec:gal_pop}

We used a Milky Way-like galaxy from the large-scale cosmological magneto-hydrodynamical simulations \texttt{TNG50} \citep{2019MNRAS.490.3234N,2019MNRAS.490.3196P}. With a comoving volume of $(50\,\mathrm{Mpc})^3$ the \texttt{TNG50} simulation box contains about 100 Milky Way-like galaxies with a total mass of $10^{14} \, \mathrm{M_{\odot}}$. To select a suitable Milky Way-like galaxy, we randomly chose one of the six halos with a total mass $1$~--~$2\times 10^{12} \, \mathrm{M_{\odot}}$ and whose central galaxy has a stellar mass $5$~--~$7 \times 10^{10} \, \mathrm{M_{\odot}}$. Additionally, we examined the galaxy’s stellar projection to confirm disk dominance. The mass of the selected galaxy (Galaxy ID = 476266) is $\sim 5 \times 10^{10} \, \mathrm{M_{\odot}}$, which is consistent with our Milky Way galaxy \citep[e.g.][]{2016ARA&A..54..529B}.  We extracted the present-day properties such as age, stellar mass, and 3D position of the star particles. We placed our observer at a randomly assigned Sun-like position in the disk, $8.2\,\rm kpc$ from the Galactic center. We then measured the distances to all LISA-detectable double white dwarfs from each star particle to this location. We combined these Galactic properties with the simulated triples to construct a representative Galactic double white dwarf population as follows.

From a population of $10^5$ simulated triple systems generated with \texttt{MSE}, we select only those that evolve into double white dwarf binaries emitting gravitational waves within the LISA sensitivity band ($10^{-4},\mathrm{Hz}$–$0.1,\mathrm{Hz}$) at some point in their evolution. We use this subset to reconstruct the Galactic LISA DWD population by assigning systems to star particles in a Milky Way–like galaxy from the \texttt{TNG50} simulation. Each star particle represents a population of stars spanning a range of masses, with an average mass of $\sim 10^5 \, \mathrm{M_{\odot}}$. By populating each particle with our simulated systems, we imprint our assumed initial conditions (Sect. \ref{sec: initial distributions}) and multiplicity fractions onto the \texttt{TNG50} galaxy.
The number of DWDs assigned to each star particle is obtained by scaling the total number of systems in the \texttt{MSE} sample, $N_{\textrm{DWD, MSE}}$, by the particle's stellar mass, $M_\star$, at redshift $z=0$:
 \begin{equation}\label{starp double white dwarf}
 N_{\textrm{DWD},\star} = \frac{N_{\textrm{DWD, MSE}}}{M_{\textrm{tot, MSE}}} \times M_\star .
 \end{equation}
For each star particle, we then randomly sample $N_{\textrm{DWD},\star}$ binaries from our subset. To ensure consistency in formation history, we only retain systems whose formation times precede the particle's age. For such systems, we account for gravitational wave radiation reaction between their formation and the particle's age. Evolving every binary fully within \texttt{MSE} would be computationally prohibitive, so we adopt a hybrid approach.
For circular binaries, the tertiary companion has no appreciable effect once the system enters the LISA band. These binaries are therefore evolved analytically forward in time, using the gravitational wave emission formula of \citet{1964PhRv..136.1224P}, until they reach the age of the host particle. In contrast, eccentric binaries (much rarer but significantly influenced by the tertiary) are included only if their simulated formation times lie within $\pm 100{,}000$ years of the particle's age (a choice set mainly by computational constraints; see Sect.~\ref{sec:ecc}).

The total stellar mass of the simulated population is given by
\begin{equation}
\begin{aligned} \label{eq: Mtot}
M_{\text{tot, MSE}} &= \frac{N_{t,\,\mathrm{in\,range}}}{f_{t,\,\mathrm{in\,range}} \cdot f_t} \cdot 
\left[ f_t \cdot \bar{m_t} + f_b \cdot \bar{m_b} + (1 - f_t - f_b) \cdot \bar{m_s} \right], \\
&\quad\quad\quad\quad\quad\quad\quad\quad\quad\quad\quad\quad\quad\quad\quad \text{for } f_t \neq 0,
\end{aligned}
\end{equation}
where $N_{t,\,\mathrm{in\,range}} = 10^5$ is the number of simulated triple systems with {\tt MSE}, $f_{t,\,\mathrm{in\,range}}$ is the fraction of triples in the mass range ($1\,\mathrm{M_{\odot}} - \,8\,\mathrm{M_{\odot}}$) relative to a wider mass range of stars ($0.08\,\mathrm{M_{\odot}} - \,100\,\mathrm{M_{\odot}}$). We estimate $f_{t,\,\mathrm{in\,range}}$ numerically and find it to be $f_{t,\,\mathrm{in\,range}} = 0.16$.  Triple fraction $f_t = 0.2$, binary fraction $f_b =0.3$, and single star fraction $1 - f_t - f_b = 0.5$ represent the fractions of triple, binary, and single systems in a full stellar population \citep[][]{2017ApJS..230...15M}. We assumed a more optimistic triple fraction than currently estimated from observations to account for the incompleteness. However, our results can be rescaled for practically any assumed triple fraction. The parameters $\bar{m_t} =3.5\, \mathrm{M_\odot}$, $\bar{m_b} = 0.9 \, \mathrm{M_\odot}$, and $\bar{m_s} = 0.5 \, \mathrm{M_\odot}$ denote the numerically computed average masses of triple, binary, and single systems, respectively.
The first term of Eq.~\eqref{eq: Mtot} represents the total contribution to the stellar mass from triple systems, scaled by their fraction in the population and their average mass. The second term accounts for the contribution from binary systems and the third term represents the mass contribution from single stars. The multiplicative term $N_{t,\,\mathrm{in\,range}}/(f_{t,\,\mathrm{in\,range}} \cdot f_t)$ re-scales the number of simulated triples in the simulated mass range ($1$~--~$ 8 \, \mathrm{M_\odot}$) to the total number of stellar systems in the full mass range ($0.08$~--~$100 \, \mathrm{M_\odot}$), accounting for the fraction of triples in the full population ($f_t$) and their relative contribution to the restricted range ($f_{t,\,\mathrm{in\,range}}$).

Using Eq.~\eqref{starp double white dwarf}, we seed the number $N_{\text{DWD,}\star}$ of LISA-detectable double white dwarfs corresponding to each star particle and randomly select them from our simulated sample (for $N_{\text{DWD},\star}>N_{\text{DWD, MSE}}$ a star particle contains some sampled double white dwarfs more than once). Additionally, we assume that all double white dwarfs are located at a star particle's center of mass.

\section{Key processes that shape the triple evolutionary pathways} \label{sec:Triple evolution channel}
In this section we present a brief overview of the five key processes characteristic of triple evolution that shape the evolutionary pathways leading to the formation of LISA double white dwarfs: 
\begin{enumerate}
    \item Induced mass transfer: The perturbation from the tertiary star triggers mass transfer in the inner binary, altering its timing or evolutionary phase compared to an isolated binary. This process ultimately leads to the formation of a short-period LISA double white dwarf.
    \item Outer binary channel: The inner binary stars merge and form a new rejuvenated star which then evolves with the tertiary star to become a LISA double white dwarf.
    \item Ejected tertiary: The tertiary star aids the formation of the inner double white dwarf but becomes unbound before it enters the LISA frequency bandwidth.
    \item Triple common envelope: When the tertiary star overflows its Roche lobe, mass transfer onto the inner binary initiates a triple common envelope phase, which further drives the formation of a LISA double white dwarf.
    \item Inner binary channel: The tertiary companion remains bound to the inner binary throughout its evolution but is too distant to significantly affect the formation of a double white dwarf; this channel is effectively that of the isolated binary channel. 
\end{enumerate}

\begin{figure*}
    \centering
     \includegraphics[trim = 0 310 0 0 ,width=1\textwidth]{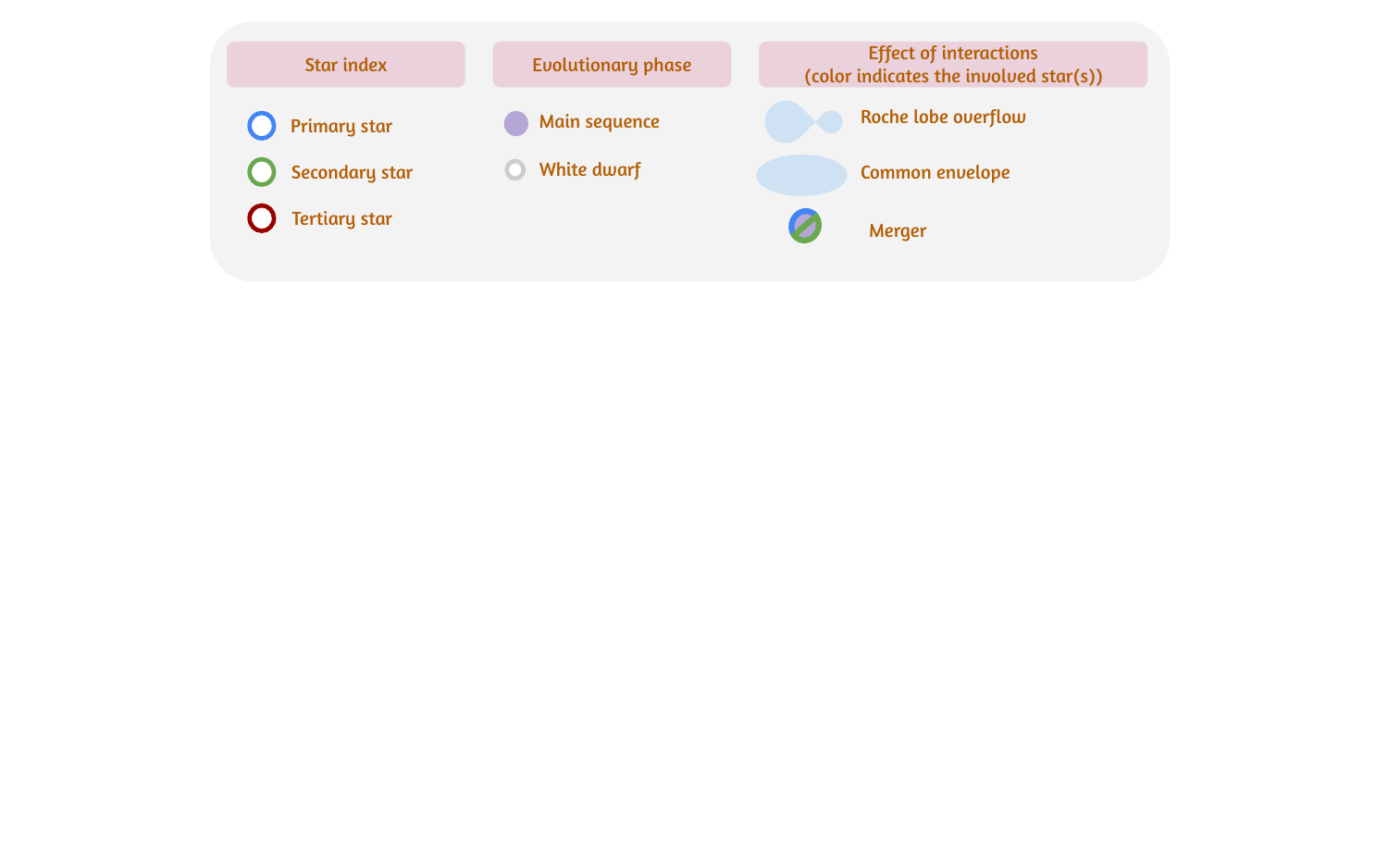}
        \includegraphics[trim = 0 0 0 0 ,width=1\textwidth]{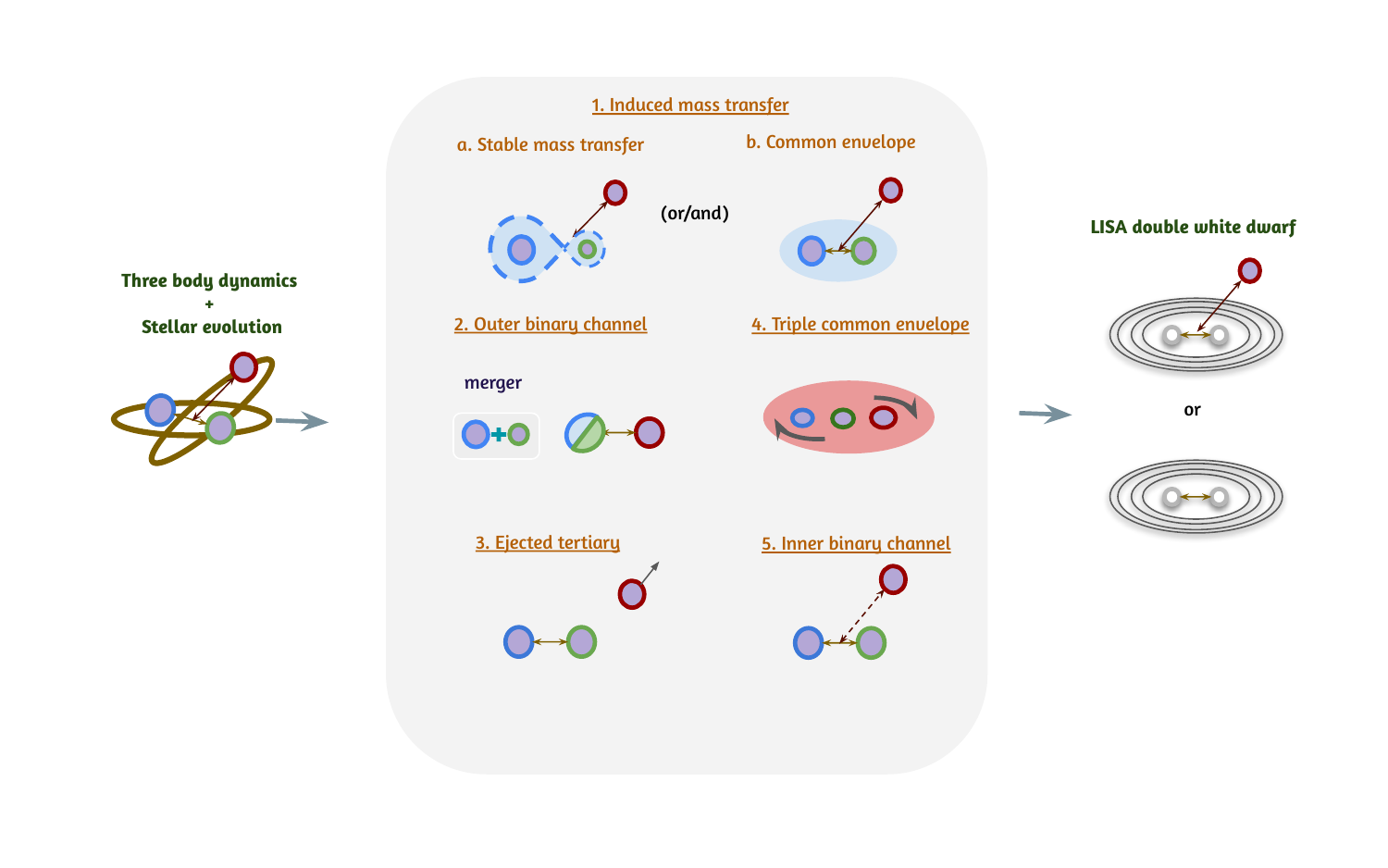}
    \caption{Diagram of possible key processes that drive the evolutionary phases of a triple evolution leading to the formation of a double white dwarf in the LISA frequency bandwidth. The diagram showcases key stages, including mass transfer, common envelope phases, ZLK oscillations that enhance eccentricity, and eventual binary evolution. The tertiary star plays a critical role in shaping the inner binary's dynamics, either by inducing orbital changes or facilitating interactions that lead to the formation of the LISA double white dwarf. The circles represent the index of the star, with blue, green, and red indicating the primary, secondary, and tertiary stars, respectively. The filling inside each circle represents the star's evolutionary phase: purple for the main-sequence and white for a white dwarf. A dashed arrow denotes a distant tertiary star that is too far to significantly influence the inner binary. A multi-colored circle represents a post-merger star, with the two colors signifying the components that have merged.}
    \label{fig:schematic_diagram}
\end{figure*}

\begin{figure}
   
    \includegraphics[width=\columnwidth]{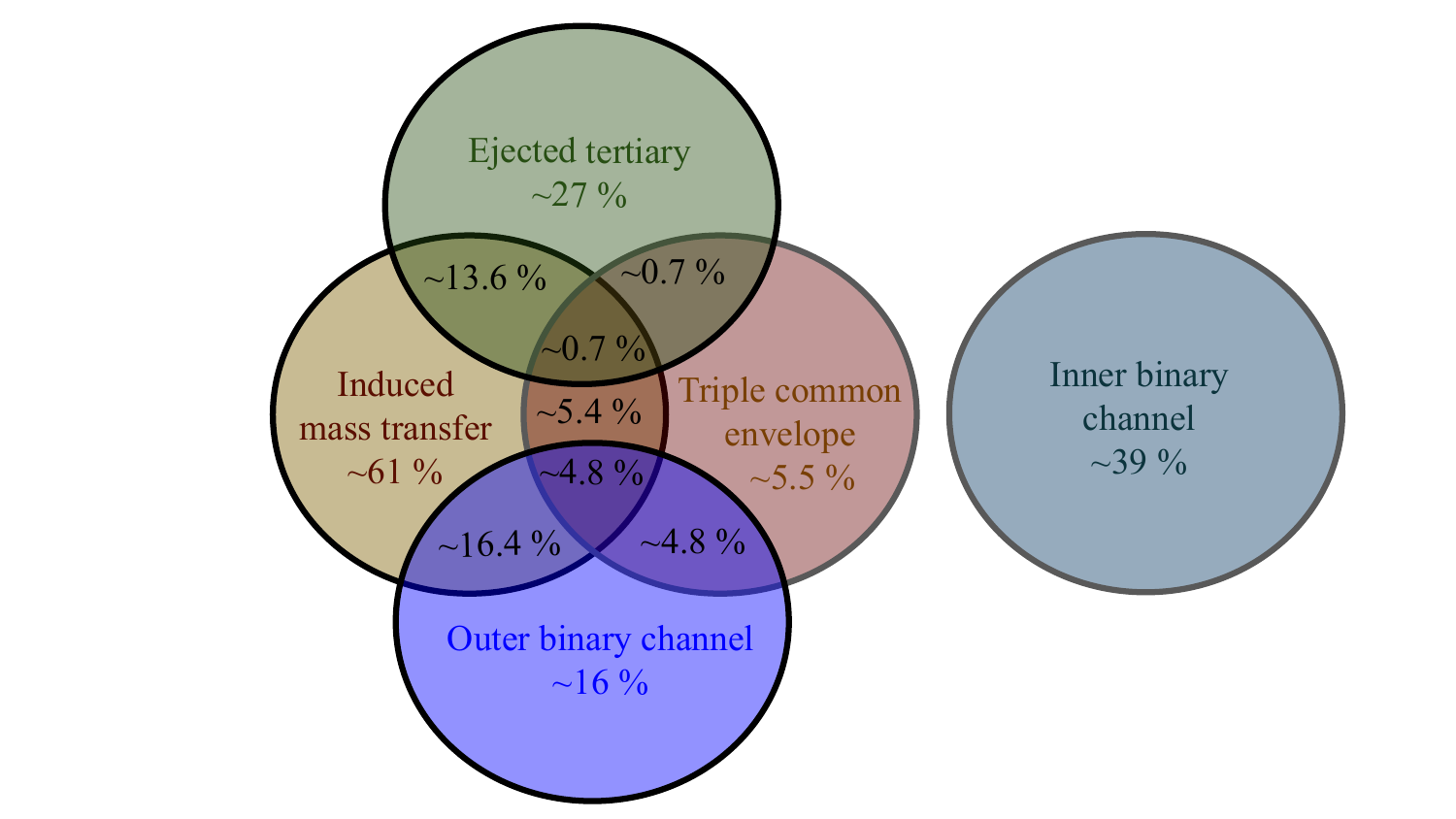}
    \caption{Venn diagram illustrating the overlap of the different evolutionary processes. Among the five processes, the inner binary channel is the only one that does not require assistance from the tertiary star to produce a LISA double white dwarf. In contrast, the other four processes rely on the tertiary star to bring the system into the LISA frequency bandwidth. These processes are not mutually exclusive and exhibit significant overlap.}
        \label{fig:venn_diagram}
\end{figure}

We show a schematic diagram for primary processes that drive evolutionary pathways leading to LISA double white dwarfs from triples in Fig.~\ref{fig:schematic_diagram}. 
In the following subsections, we discuss these processes in greater depth with detailed examples. All uncertainties presented below are estimated by scaling up the fractional Poisson error from the intrinsic population evolved with \text{MSE}. Figure~\ref{fig:venn_diagram} shows that the processes are not mutually exclusive. For instance, about 4.8\% of Galactic LISA double white dwarfs from triples undergo a TCE phase and induced mass transfer, leading to a merger in the inner binary. Hence, the relative percentages quoted below do not add up to $100\,\%$.

\subsection{Induced mass transfer}
\label{sec: Enhanced mass transfer}

\begin{figure*}
    \centering
    \includegraphics[trim = 0 310 0 0 ,width=1\textwidth]{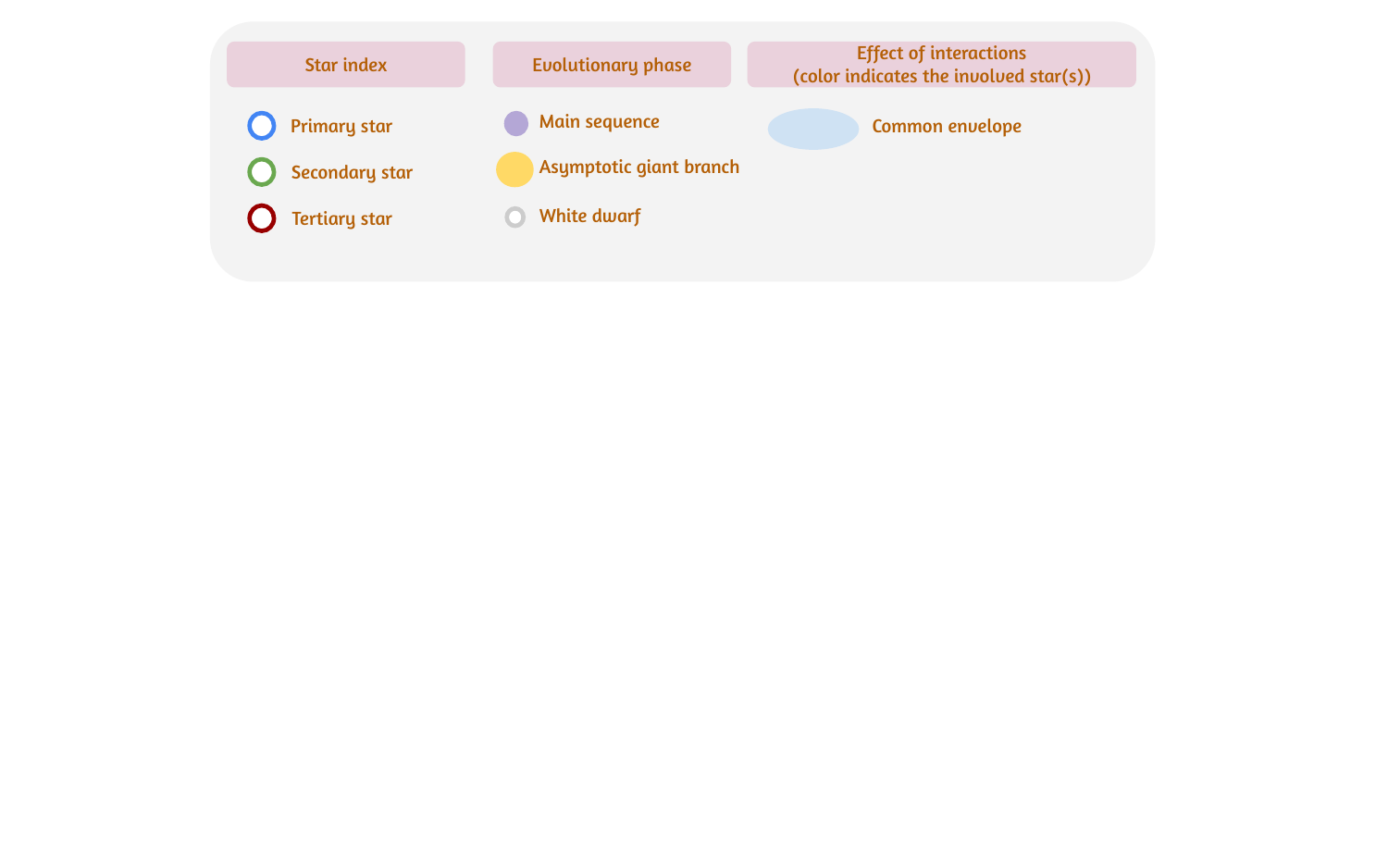}
        \includegraphics[trim = 0 0 0 0 ,width=1\textwidth]{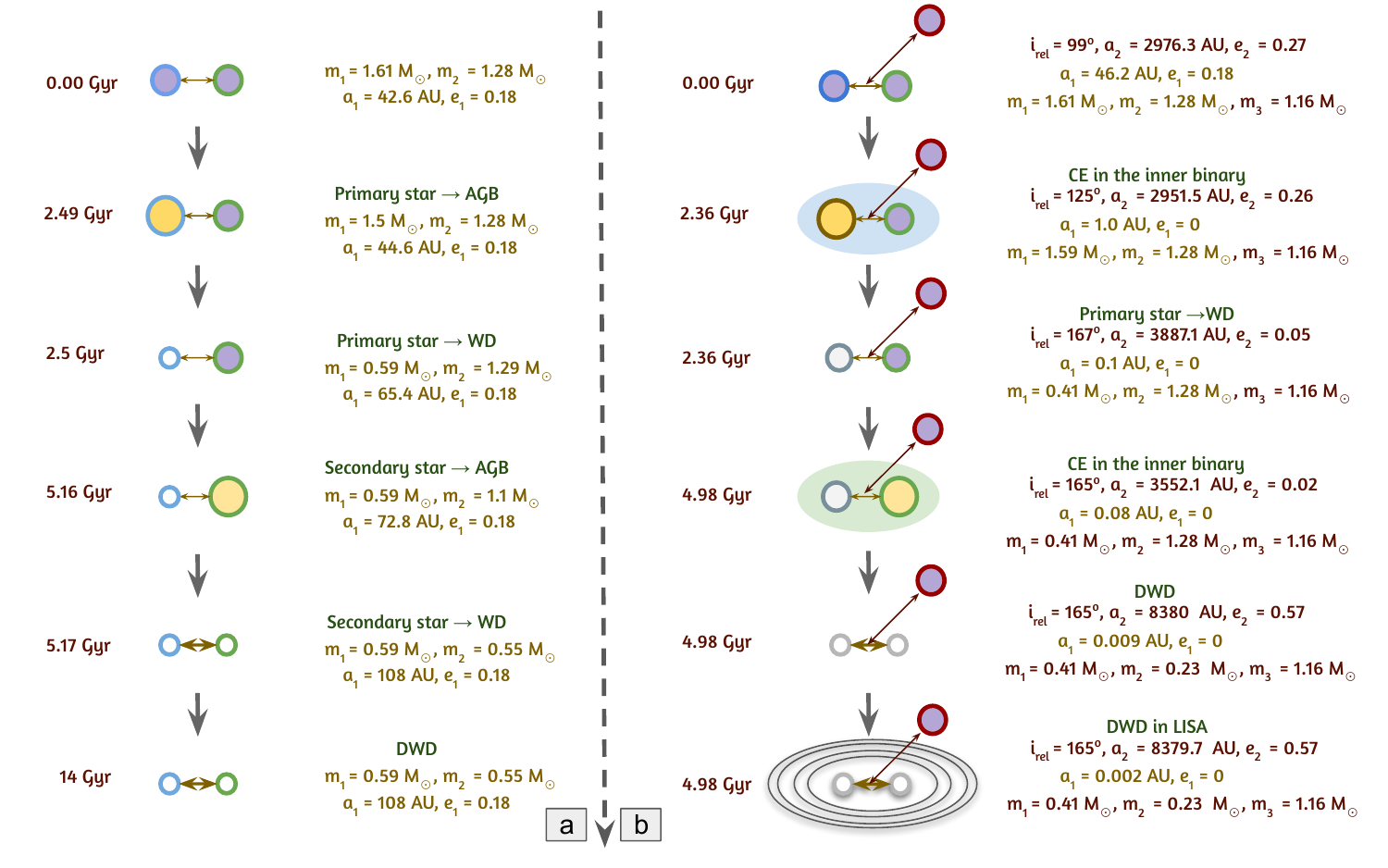}
    \caption{Comparison of the evolution of a triple system with and without a tertiary star. Panels (a) and (b) illustrate the evolution of the inner binary with and without the third star respectively. In the system with the tertiary star, mass transfer is induced by perturbations from the third star, allowing the system to eventually enter the LISA frequency bandwidth. When evolved without a tertiary star the binary components remain too far apart to interact. Such a system does not enter the LISA frequency bandwidth. The legends are similar to those in Fig.~\ref{fig:schematic_diagram}. In addition, the yellow filling represents a star in the Asymptotic giant branch phase.}
    \label{fig:FaciMT2}
\end{figure*}

\begin{figure*}
    \centering
        \includegraphics[trim = 0 0 0 0 ,width=1\textwidth]{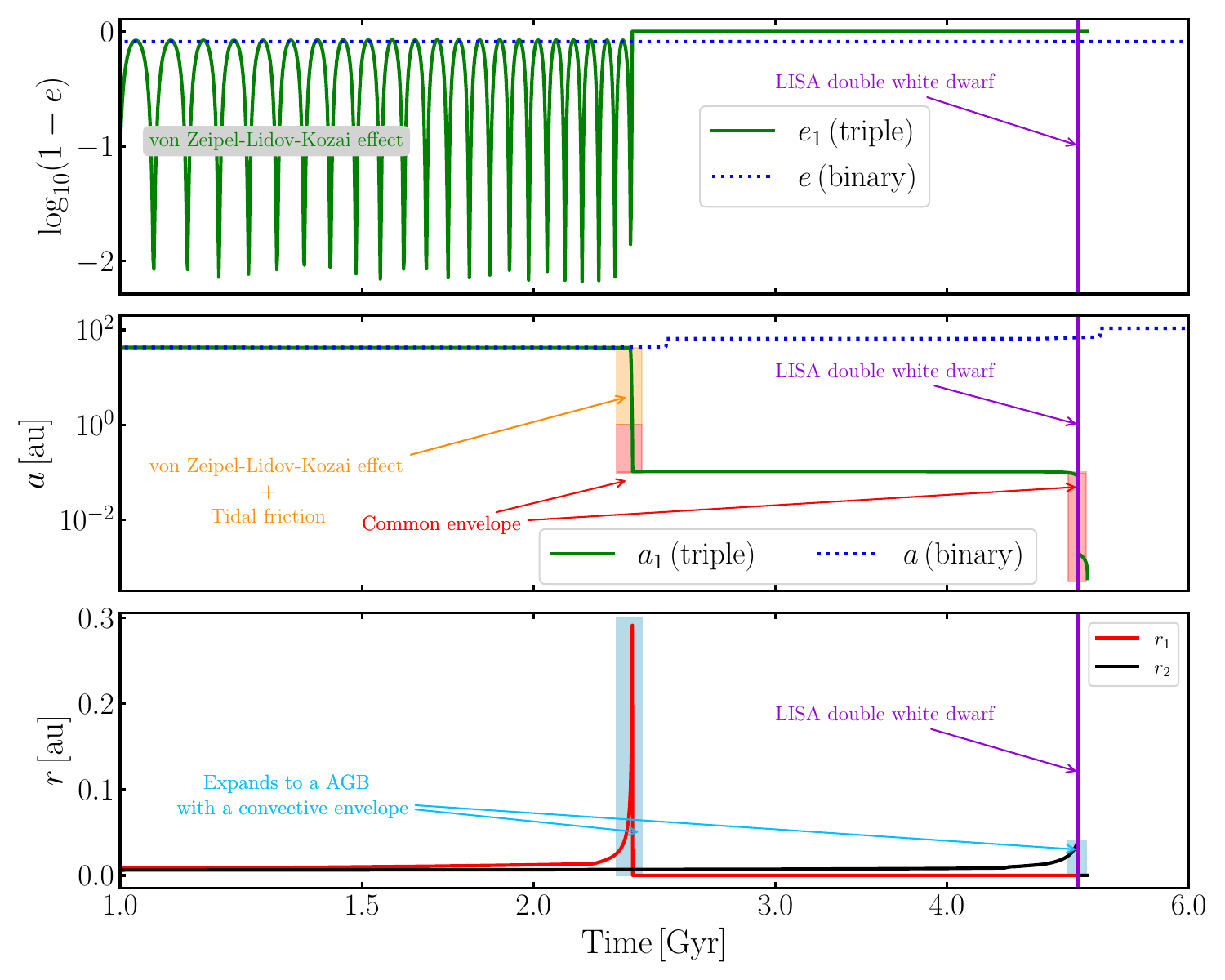}
    \caption{Comparative evolution of the properties of the inner binary of the triple with and without a third star. The three panels, from top to bottom, display the zoomed-in evolution of key parameters of the two stars: eccentricity, semi-major axis, and radius. In the case of the triple system, the inner binary experiences unstable mass transfer, causing it to shrink further and eventually enter the LISA band. Meanwhile, the inner binary when evolved without a tertiary star undergoes mass loss due to winds, resulting in an increase in its semi-major axis and a widening of the orbit. Here $a_1\,(\mathrm{triple})$ and $e_1\,(\mathrm{triple})$ represent the semi-major axis and eccentricity of the inner binary evolved with a tertiary star, while ${a}\,({\rm binary})$ and ${e}\,({\rm binary})$ show the semi-major axis and eccentricity of the same inner binary evolved without the tertiary star. Additionally, $r_1$ and $r_2$ represent the radii of the primary and secondary stars in the inner binary, respectively.}
    \label{fig:evolution_FaciMT}
\end{figure*}

When comparing the triple runs to the inner binary runs (where the inner binary from the same triple population evolves without the tertiary star; see Sect.~\ref{sec:methods}). We find that approximately $9^{+2}_{-3}\,\%$ of the Galactic LISA double white dwarfs undergo a mass transfer episode solely induced by the influence of a third star. These inner binaries would not have interacted over a Hubble time without the presence of the tertiary star. About $52^{+6}_{-7}\,\%$ of Galactic LISA double white dwarfs initiate mass transfer at a different time due to the influence of a tertiary star.  Thus, a total of $61^{+8}_{-10}\,\%$ of systems experience induced mass transfer (see Fig.~\ref{fig:venn_diagram}).
The mass transfer is induced due to the combined effect of dynamical stability constraints and dynamical interactions resulting from the third star. It is not trivial to quantify the precise influence of the third star. This mass transfer is often found to have occurred at various evolutionary stages well before the formation of double white dwarfs. In contrast, the corresponding isolated binaries would either not undergo mass transfer or interact at a different time. This timing difference for the onset of mass transfer is crucial for driving the merger in the inner binary or forming a short-period inner binary that can enter the LISA frequency bandwidth within Hubble time (see the example below for more details). Also, we note that only a negligible fraction ($\sim10^{-6}$) of systems undergo stable mass transfer without eventually becoming unstable.

As examples we show the evolution of a triple system and the inner binary without the tertiary star in Fig.~\ref{fig:FaciMT2}. The triple system starts with $m_1 = 1.61\,\mathrm{\mathrm{M_{\odot}}}$, $m_2 = 1.28\,\mathrm{\mathrm{M_{\odot}}}$, and $m_3 = 1.16\,\mathrm{\mathrm{M_{\odot}}}$ as the tertiary star. The inner binary is initially eccentric, with an eccentricity $e_1 = 0.18$ and a semi-major axis $a_1 = 42.6\,\mathrm{au}$, while the tertiary orbit is much wider, with a semi-major axis of $a_2 \approx 2980\,\mathrm{au}$ and an eccentricity of $e_2 = 0.26$. The initial mutual inclination between the inner and outer orbits is 99$^\circ$. Without the tertiary star the system evolves as an isolated binary. The large initial semi-major axis allows the binary components to evolve into white dwarfs without a mass transfer episode. During the course of evolution of the binary mass loss due to winds further widens the system, as shown in Fig.~\ref{fig:evolution_FaciMT}. The system continues to lose orbital energy via gravitational wave radiation but remains too wide ($a\approx32.1\,\rm{au}$) to enter the LISA frequency bandwidth within a Hubble time. However, with the third star present, the system undergoes significant orbital shrinkage of the inner binary during its evolution and ultimately enters the LISA frequency band. At around $2.2\,\rm{Gyr}$, the primary star of the inner binary becomes a red giant with a convective envelope. The ZLK effect, combined with tidal friction \citep{2007ApJ...669.1298F}, shrinks the inner binary’s orbit from $a_{1}\approx42.6\,\mathrm{au}$ to $1\,\mathrm{au}$. This shortening allows the primary star to fill its Roche lobe and initiate mass transfer to its companion. The mass transfer becomes unstable due to the high mass ratio and the system enters a CE phase, which further reduces the inner binary’s orbit to $0.1\,\mathrm{au}$. This results in a $0.41\,\mathrm{\mathrm{M_{\odot}}}$ white dwarf and a $1.28\,\mathrm{\mathrm{M_{\odot}}}$ main-sequence star. Later, at around $5\,\rm{Gyr}$, the main-sequence companion evolves to an AGB star, eventually leading to a second CE phase. This produces a short-period double white dwarf with component masses of $0.41\,\mathrm{\mathrm{M_{\odot}}}$ and $0.23\,\mathrm{\mathrm{M_{\odot}}}$, which enters the LISA frequency bandwidth after a few million years. Thus, in this example, the binary would not be to become a LISA source without the tertiary. 

\subsection{Outer binary channel}
\label{sec:outer binary channel}
About $16^{+3}_{-3}\,\%$ of Galactic LISA double white dwarfs originate from triples where there was a merger of the inner binary stars. We identified four scenarios which lead to mergers. First, if the inner and outer orbital planes are highly inclined with respect to each other, the perturbations from the tertiary companion shrinks the orbit of the inner binary through the combination of the ZKL effect and tides. The tightened triple system has an earlier CE episode than in an isolated binary. The CE evolution results in a merger, forming a rejuvenated star bound to the former tertiary star as a binary system. This binary can evolve into the LISA frequency bandwidth. Second, in comparison to the first, the triples start out with a short-period inner binary and a distant tertiary. Here, the binary comes into contact and merges without the aid of the tertiary star. The merged star further evolves with the tertiary star to enter the LISA frequency bandwidth. Third, a TCE can cause a merger of the inner binary stars (see Sect.~\ref{sec:Triple common envelope} for more details).  Fourth, the orbit of the inner binary can widen due to mass transfer and winds, making the entire system dynamically unstable which eventually leads a chaotic evolution of the orbits and the merger of the inner binary stars. 

\begin{figure*}
    \centering
    \includegraphics[trim = 0 310 0 0 ,width=1\textwidth]{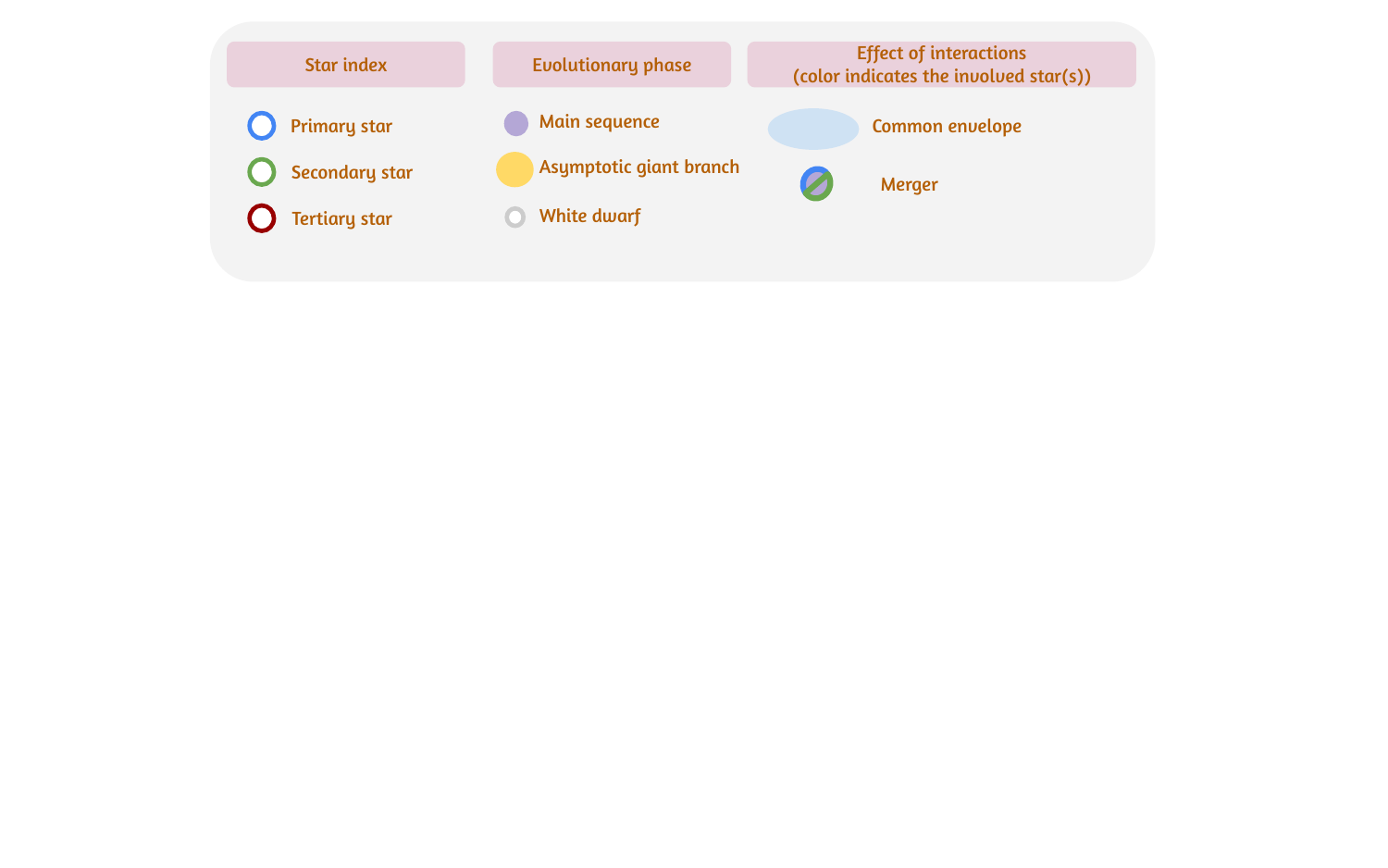}
        \includegraphics[trim = 0 0 0 0 ,width=1\textwidth]{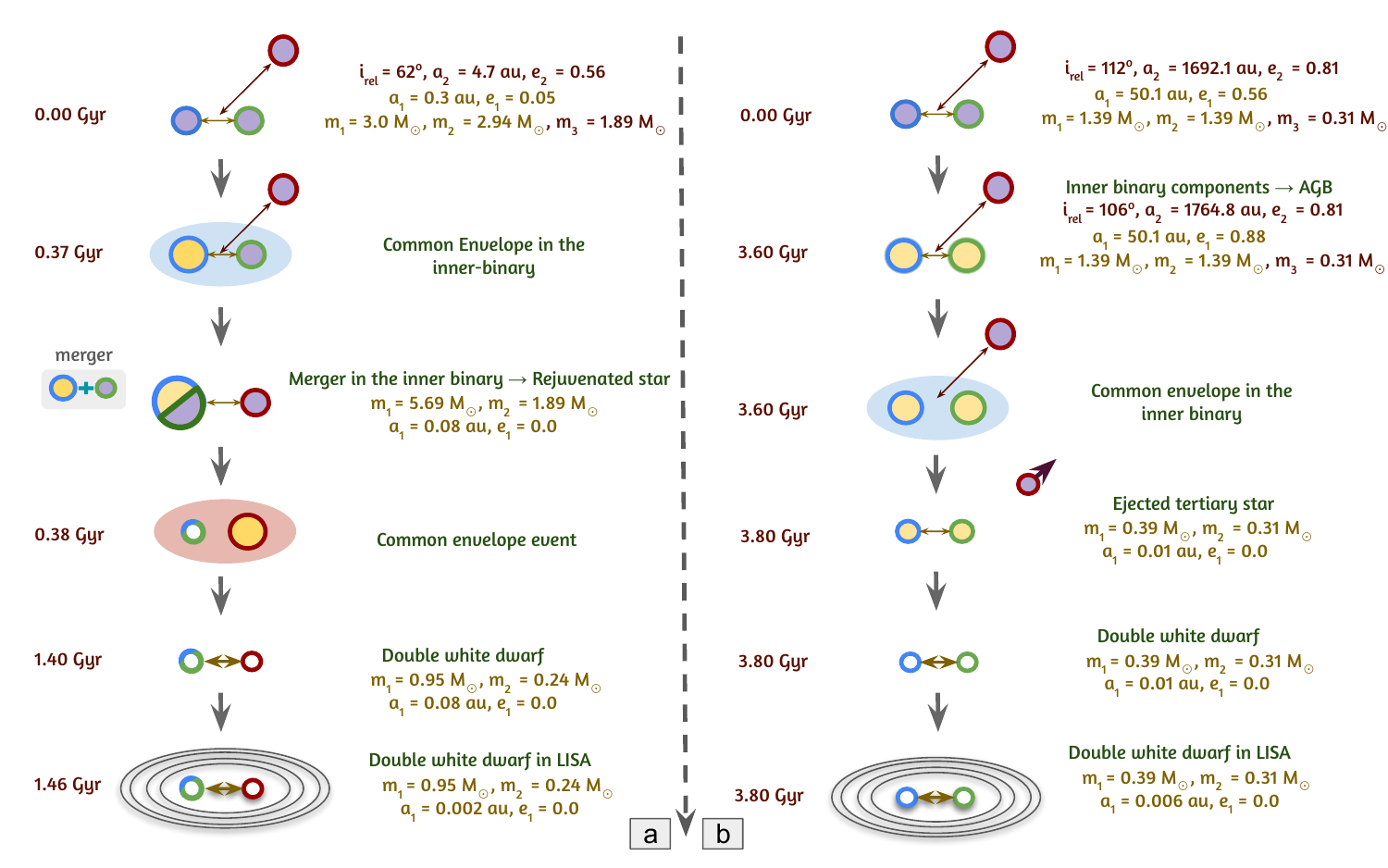}
    \caption{Schematic diagrams of systems that enter the LISA frequency bandwidth after following the outer binary channel and those in which the triple eject the tertiary during the course of evolution. Panel (a) depicts an example of systems following the outer binary channel. In this scenario, the inner binary merges to form a rejuvenated star, which later enters the LISA frequency bandwidth along with the tertiary star. Panel (b) illustrates an example system that initially includes a bound third star, which facilitates a common envelope phase in the inner binary but is later ejected. The inner binary subsequently enters the LISA frequency bandwidth.The legends are similar to those in Fig.~\ref{fig:schematic_diagram}. In addition, the yellow circle represents a star in the asymptotic giant branch phase.}
    \label{fig:outerbin_ejected}
\end{figure*}

Panel (a)  of Fig.~\ref{fig:outerbin_ejected} shows an example where the inner binary merges to form a new star. The system starts with $m_1 = 3\,\mathrm{\mathrm{M_{\odot}}}$ and $m_2 = 2.94\,\mathrm{\mathrm{M_{\odot}}}$, and $m_3 = 1.89\,\mathrm{\mathrm{M_{\odot}}}$ as the tertiary star. The inner binary is initially circular with a semi-major axis $\mathrm{a_1} = 0.3\,\mathrm{au}$. The tertiary orbit is tight, with an outer semi-major axis $\mathrm{a_2} = 4.7\,\mathrm{au}$ and is eccentric with $\mathrm{e_2} = 0.56$. The mutual inclination between the inner and outer orbits is $62^\circ$. The most massive of all three stars is in the inner binary ($m_1 = 3\,\mathrm{\mathrm{M_{\odot}}}$) which, at about $370\,\rm Myr$, initiates an unstable mass transfer episode onto the secondary resulting in the merger of the two stars. This merger leads to the formation of a new star with a mass $m_r = 5.69\,\mathrm{\mathrm{M_{\odot}}}$. The remaining post-merger binary composed of the rejuvenated star and the tertiary companion subsequently undergoes and survives two more CE episodes: one when the rejuvenated star enters the AGB at about $390\,\rm Myr$ and another when the tertiary companion becomes an red giant star at about $1.4\,\rm Gyr$. The second CE leads to the formation of a circular double white dwarf with a semi-major axis of $9 \times 10^{-3}\,\rm{au}$ later entering the LISA bandwidth after $\sim 1\,\rm{Gyr}$. An isolated binary with the same properties as the inner binary of such a triple would not enter the LISA band.    

\subsection{Ejected tertiary}
\label{sec:Unbound tertiary}

About $27^{+4}_{-2}\,\%$ of Galactic double white dwarfs from triples lost the tertiary star. The unbinding of the tertiary companion primarily occurs due to the following reasons. First, during a CE phase the inner binary's orbit shrinks and loses angular momentum, while the prompt mass loss during the CE unbinds the outer orbit. Second, if the inner binary widens due to mass transfer or stellar winds the system becomes less hierarchical and dynamically unstable. Dynamically unstable orbits lead to chaotic evolution which may eject an object from the system. We provide a specific example of this process below.

Panel (b) of Fig.~\ref{fig:outerbin_ejected} shows an example of a system where a tertiary gets ejected after a CE in the inner binary. The system starts with $m_1 = 1.39\,\mathrm{\mathrm{M_{\odot}}}$ and $m_2 = 1.39\,\mathrm{\mathrm{M_{\odot}}}$ as the inner binary components and $m_3 = 0.31\,\mathrm{\mathrm{M_{\odot}}}$ as the tertiary star. The inner binary is initially eccentric with an inner eccentricity $\mathrm{e_1} = 0.56$ and an inner semi-major axis $\mathrm{a_1} = 50.1\,\mathrm{au}$. The tertiary orbit is wide, with an outer semi-major axis $\mathrm{a_2} \approx 1690\,\mathrm{au}$ and is eccentric with $\mathrm{e_2} = 0.81$. The mutual inclination between the inner and outer orbits is 112$^\circ$. The inner binary masses are greater than the tertiary star so they evolve on a shorter timescale. The CE occurs in the inner binary during the AGB phase after about 3.6 Gyr. The CE reduces the inner binary semi-major axis to $10^{-2}\,\rm{au}$ and unbinds the outer tertiary. The remaining binary components evolve into a double white dwarf system. This double white dwarf binary emits gravitational waves and later enters the LISA frequency bandwidth after $1\,\rm{Myr}$. In this example, even though the binary does not have a bound tertiary star by the time it enters the LISA frequency bandwidth, the tertiary plays a role in shrinking the binary before the CE episode. 

\begin{figure*}
    \centering
    \includegraphics[trim = 0 310 0 0 ,width=1\textwidth]{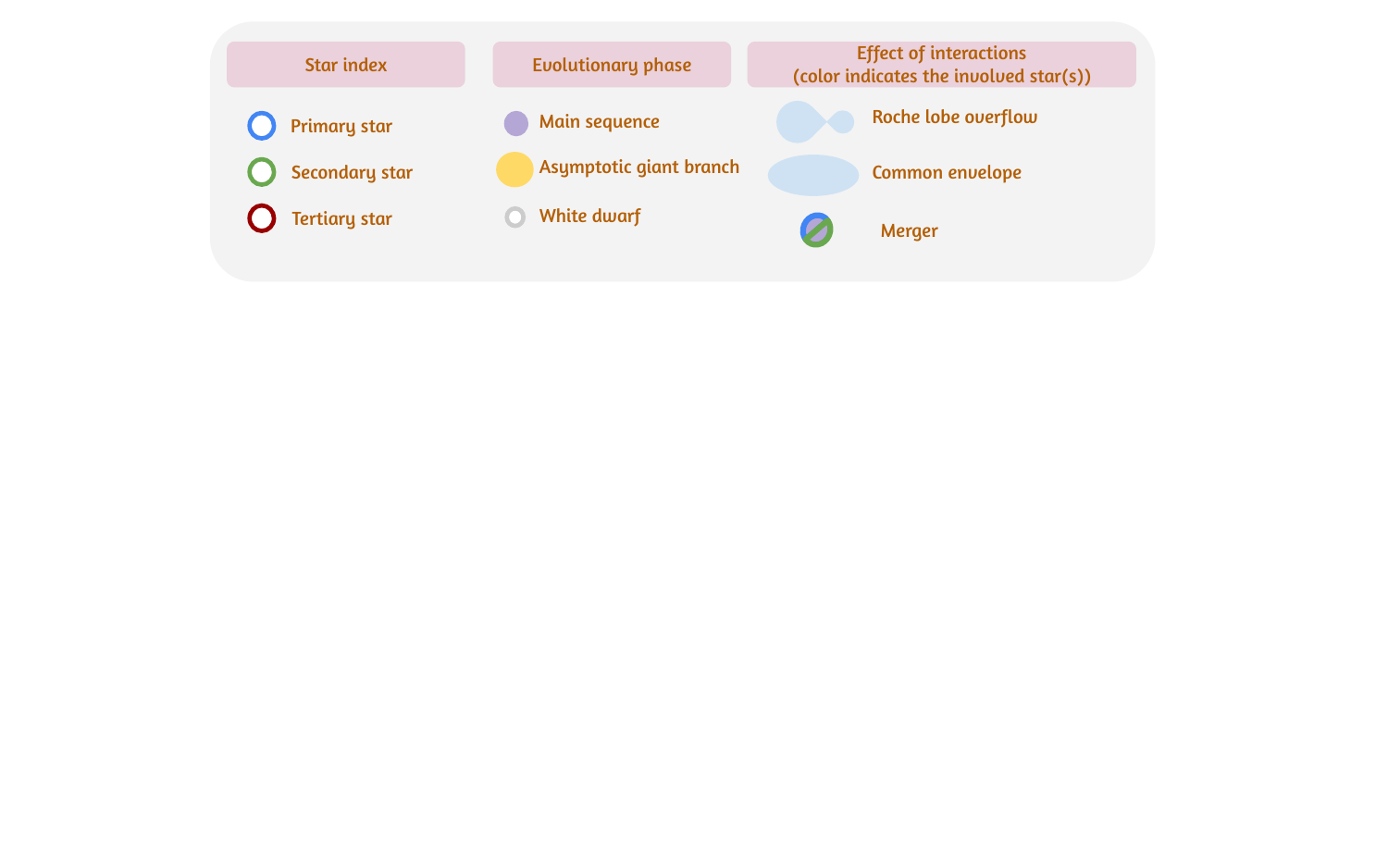}
        \includegraphics[trim = 0 0 0 0 ,width=1\textwidth]{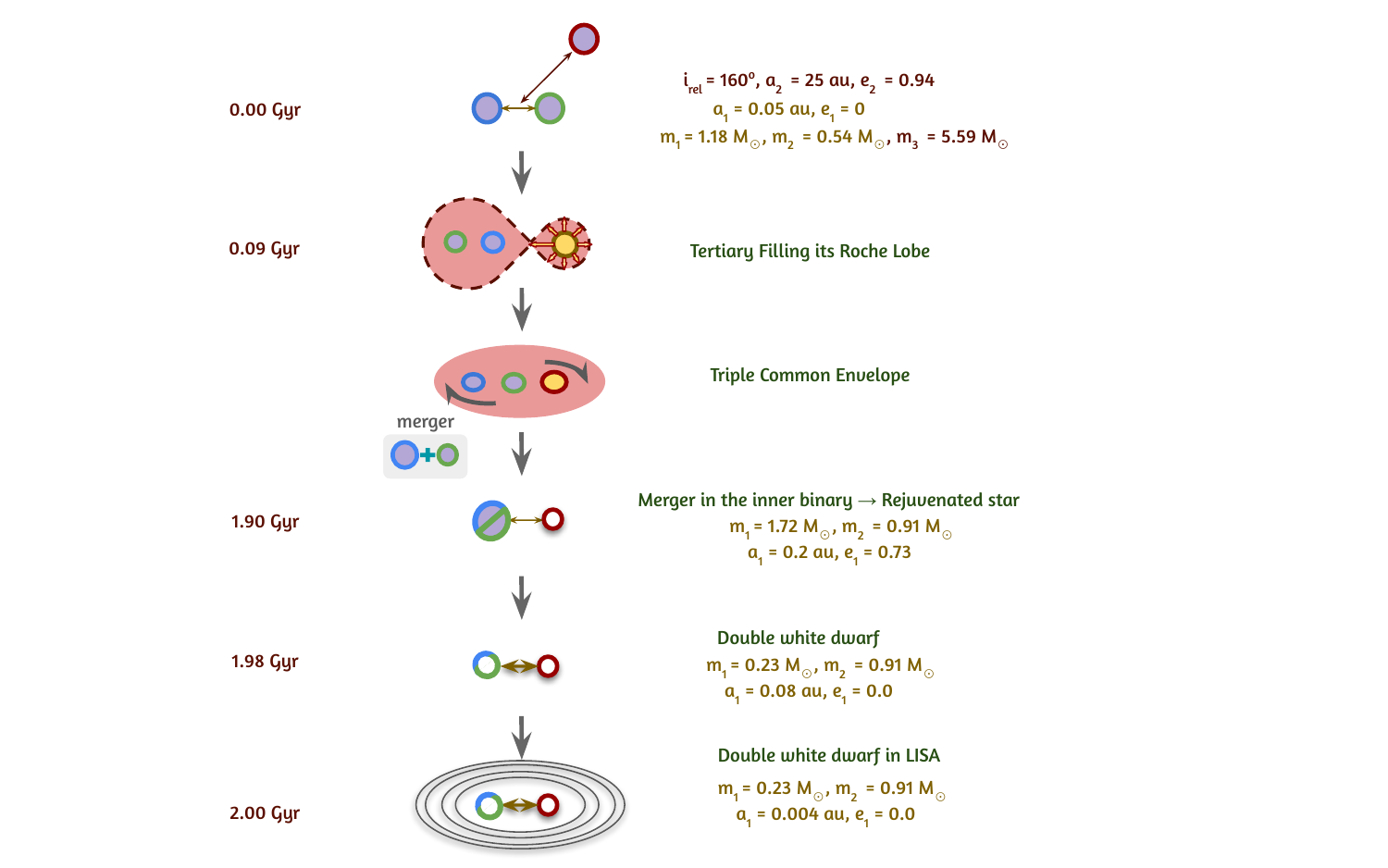}
    \caption{Schematic diagram for an example system that undergoes a TCE. In this scenario, the massive star transfers mass onto the inner binary, leading to its merger and the formation of a rejuvenated star. This rejuvenated star later enters the LISA frequency bandwidth along with the tertiary star. The legends are similar to those in Fig.~\ref{fig:schematic_diagram}. In addition, the yellow circle represents a star in the asymptotic giant branch phase.}
    \label{fig:TCE}
\end{figure*}

\subsection{Triple common envelope}
\label{sec:Triple common envelope}

About $5.5^{+2}_{-1}\,\%$ of Galactic double white dwarfs from triples undergo a phase of TCE  evolution before entering the LISA frequency band. We identified two possible evolutionary outcomes of a TCE event in our simulations.
First, the merger of the inner binary: the inner binary may merge to form a rejuvenated star. The new star forms a binary with the third star which later enters LISA frequency bandwidth. Second, ejection of one component: One of the stars may become unbound from the system, leaving behind a binary formed by the remaining two stars. This binary may enter the LISA band. We explain one of the examples below (see \citealt{2022ApJS..259...25H} for a more detailed study of TCE). 

Figure~\ref{fig:TCE} shows an example of a TCE where the inner binary merges to form a new binary. The system starts with $m_1=1.18\,\mathrm{\mathrm{M_{\odot}}}$,  $m_2=0.54\,\mathrm{\mathrm{M_{\odot}}}$, and a massive tertiary  $m_3=5.59\,\mathrm{M_\odot}$. The inner binary is initially circular, with a semi-major axis $\mathrm{a_1} = 0.05\,\mathrm{au}$. The tertiary orbit is compact, with an outer semi-major axis $\mathrm{a_2} = 25\,\mathrm{au}$ and highly eccentric with $\mathrm{e_2} = 0.94$. The mutual inclination between the inner and outer orbit is 160$^{\circ}$. Since the tertiary star is massive compared to the inner binary masses, it is the first to reach the AGB phase, in about 92 Myr. The inner binary components are still on the main-sequence.

Expansion of the envelope during the AGB phase initiates Roche lobe overflow. The AGB tertiary starts transferring mass onto the inner binary. The mass transfer becomes dynamically unstable owing to a higher mass ratio. The envelope of the AGB tertiary star engulfs both components of the inner binary. The TCE evolution leads to interactions between the three stars. The inner binary merges to form a rejuvenated star which is still bound to the stripped tertiary star. The resulting binary has a large eccentricity of 0.73 and a semi-major axis of $0.2\,\mathrm{au}$.

After approximately $2\,\rm{Gyr}$, the rejuvenated star evolves into a red giant which transfers mass onto the white dwarf, initiating another CE phase. The end of this CE phase leaves a circular and compact double white dwarf in a circular orbit with a semi-major axis of $\sim 8 \times 10^{-2}\,\mathrm{au}$. After a few million years the binary enters the LISA frequency bandwidth due to gravitational wave emission. Here, TCE plays a significant role as it leads to merging the inner binary. The binary with the post-merger rejuvenated star later enters the LISA frequency bandwidth. Hence, the tertiary star plays a vital role in the evolution of the triple to LISA frequency bandwidth.  

\renewcommand{\arraystretch}{1.75}
\setlength{\tabcolsep}{1.5pt}

\begin{table}
\caption{Estimated counts.}
\centering
\makebox[\linewidth][l]{
\begin{tabular}{c c c}
\hline
Category & $N(f<10^{-4}\,\textrm{Hz})\times 10^6$ & $N(\rho > 7)\times10^3$\\ \hline
Triple            &  $7.20^{+0.52}_{-0.50}$  &   $ 10.9^{+0.67}_{-0.50}$                  \\ 
Binary            & $3.80^{+0.17}_{-0.59}$ &   $ 6.5^{+0.22}_{-1.19} $                     \\ 
Triple + Binary & $11.00^{+0.69}_{-1.09}$ & $17.4^{+0.89}_{-1.69}$ \\
Only binaries ($f_t= 0$)    & $9.40^{+0.90}_{-0.67}$ &  $ 14.4^{+0.76}_{-0.45} $           \\ \hline
\end{tabular}
}
\label{tab:counts}
\end{table}
\tablefoot{Table presents the estimated counts of total and individually resolvable (signal-to-noise ratio $\rho > 7$) Galactic LISA double white dwarfs from different formation channels: Triple (originating from triple systems), Binary (originating from isolated binaries), Triple + Binary (the combined population from both channels), and Only binaries (isolated binaries assuming no contribution from triples, i.e., a triple fraction $f_t = 0$). The quoted uncertainty estimates correspond to a 68.3\% confidence interval and are estimated from bootstrap resampling and capture the statistical uncertainty introduced by the stochastic seeding of the Galaxy (see Appendix~\ref{appendix:B} for details).}

\subsection{Inner binary channel}

\label{sec:Binary channel}

We evolve the inner binaries of triple systems in isolation (i.e., without the influence of the third star) to quantify the impact of tertiary companions. About $39^{+8}_{-5}\,\%$ of the Galactic double white dwarfs from triples in our simulations enter LISA frequency bandwidth without any notable contribution from the third star. However, the initial configurations of these binaries (particularly their shorter orbital periods) are themselves a consequence of the dynamical stability constraints imposed by the tertiary companion. Here, the inner binary undergoes phases of CE before forming a short-period double white dwarf. This double white dwarf later enters the LISA frequency bandwidth either with a wide tertiary or as a binary with an ejected tertiary (for details on ejected tertiary, see Sect.~\ref{sec:Unbound tertiary}).

\section{LISA double white dwarfs: Isolated binary versus triple evolution} \label{sec:binaryvstriple}

\begin{figure}
        \includegraphics[width=\columnwidth]{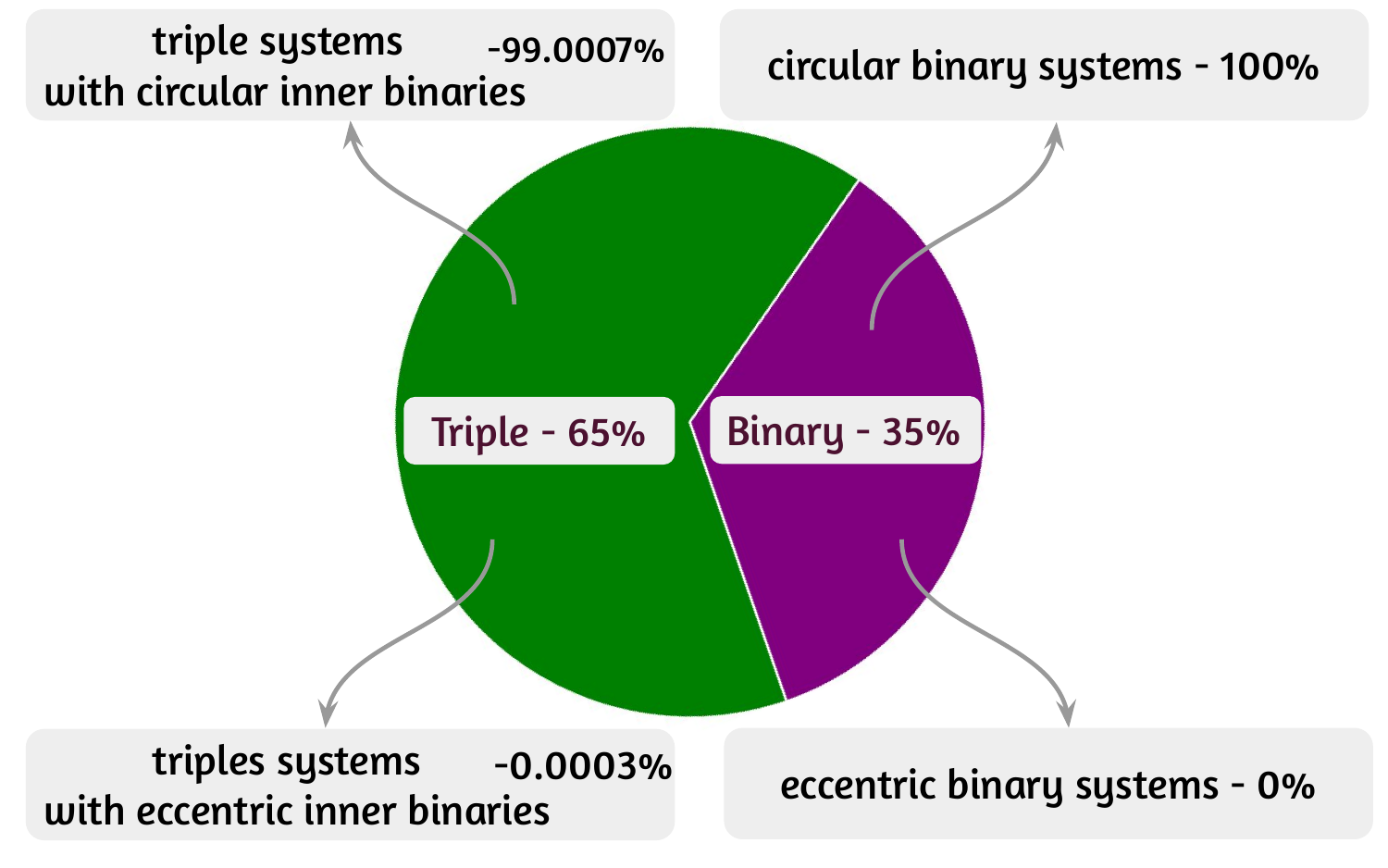}
    \caption{Pie chart showing the fraction of eccentric and circular orbits among double white dwarfs from triple systems and isolated binary systems. While isolated binaries do not produce double white dwarfs with eccentric orbits, approximately $3 \times 10^{-6}$ of LISA-detectable double white dwarfs from triple systems exhibit eccentric orbits.}
    \label{fig:piechart}
\end{figure}

Building on the method in Sect.~\ref{sec:gal_pop}, where we construct the Galactic population by combining \texttt{MSE} with a Milky Way-like galaxy from \texttt{TNG50}, we obtain a total of $ \sim 7.2 \times 10^6 $ Galactic double white dwarfs originating from triples. To compare this result with the isolated binary channel, we adjust our simulation as described in Appendix~\ref{appendix:A}. We emphasize that the initial distribution of the isolated binary population is constructed to represent truly isolated binaries and is different from the initial properties sampled to assemble the inner binaries in triple systems (See Sect.~\ref{sec: initial distributions}). In the isolated binary case, we find that about $ \sim 3.8 \times 10^6 $ double white dwarfs are produced through isolated binary evolution, yielding a total of $ \sim 1.1 \times 10^7 $ Galactic double white dwarfs currently emitting gravitational waves within the LISA bandwidth. Thus, approximately $65\,\%$ ($ \sim 7.2 \times 10^6 $) of all Galactic double white dwarf binaries originated from triple systems as illustrated in Fig.~\ref{fig:piechart}. We note that the isolated binary evolution channel only yields circular binaries. In contrast, the triple channel generates a small fraction ($3\times 10^{-6}$) of eccentric binaries, which we discuss below. 

          Notably, we find that among all the double white dwarfs initially in triples only about half ($\sim$57\%) retain a bound tertiary. In the remaining systems, the third star either became unbound or there was a merger of inner binary stars, reducing the system from a triple to a binary (See Sects.~\ref{sec:Unbound tertiary} and \ref{sec:outer binary channel}). Importantly, all double white dwarfs that retain their tertiary companion in our simulations have the tertiary in a wide orbit. We discuss the potential for detecting the presence of the tertiary companion based on LISA data in Sect.~\ref{sec:thirdstar}.
We recall that we also generated a comparative simulation where the Galactic population only follows the isolated binary channel (i.e., entirely excluding triple channels). In this simulation we obtain a total of $\sim 9.4 \times 10^6$ double white dwarfs emitting at LISA frequencies. All results are summarized in Table \ref{tab:counts}.

In the following subsections we present our estimates for the number of individually resolvable sources and estimate the unresolved stochastic foreground. We discuss the source properties and white dwarf (core composition) types from triple systems then compare them with those from the isolated binary channel.

\subsection{Detectability with LISA}

\subsubsection{Circular systems}\label{sec:circular systems}
Double white dwarfs in the LISA band are typically millions of years away from merging. They are continuous, quasi-monochromatic gravitational wave sources for LISA. Describing these signals requires a set of eight parameters, typically chosen as $\{ {\mathcal{A}_{\rm gw}, f_{\rm gw}, \dot{f}_{\rm gw}, \lambda, \beta, \iota, \psi, \phi_0 } \}$ \citep{LISAWaveform}. Here, $\mathcal{A}_{\rm gw}$ represents the gravitational wave amplitude, $f_{\rm gw}$ and $\dot{f}_{\rm gw}$ denote the gravitational wave frequency and its time derivative (or chirp), $(\lambda, \beta)$ correspond to the ecliptic longitude and latitude, respectively, $\iota$ is the (inner) binary inclination angle with respect to the line-of-sight, $\psi$ is the gravitational wave polarization angle, and $\phi_0$ represents the binary's initial phase. Our population synthesis models provide binary parameters such as component masses, orbital periods and eccentricities, sky positions, and distances. We use these to derive gravitational wave parameters as follows.  As discussed above, the overwhelming majority of double white dwarfs in our simulations are circularized. 

The gravitational wave frequency of a circular binary is twice its orbital frequency ($f_{\rm orb}$),
\begin{equation} \label{eq:fGW}
    f_{\rm gw}=2f_{\rm orb},
\end{equation}
while the amplitude is given by
\begin{equation} \label{eq:ampGW}
    \mathcal{A}_{\rm gw} = \frac{2 (G \mathcal{M}_{\rm c})^{5/3} }{c^4 d} (\pi f_{\rm gw})^{2/3},
\end{equation}
where $G$ and $c$ are the gravitational constant and speed of light respectively.
The amplitude is set by the source's distance, $d$, and chirp mass:
\begin{equation}
    \mathcal{M}_{\rm c} = \frac{(m_1 m_2)^{3/5}}{(m_1 + m_2)^{1/5}}, 
\end{equation}
The chirp mass also sets the rate at which the frequency changes due to the gravitational radiation reaction:
\begin{equation} \label{eq:fdotGW}
    \dot{f}_{\rm gw} = \frac{96}{5}\frac{(G\mathcal{M})^{5/3}}{\pi c^5} (\pi f_{\rm gw})^{11/3}.
\end{equation}
Equations \eqref{eq:fGW}, \eqref{eq:ampGW}, and \eqref{eq:fdotGW} define the first three parameters of the set. The ecliptic coordinates $(\lambda, \beta)$ are inherited from the \texttt{TNG50} Milky Way-like galaxy where the binary was seeded, while the remaining three parameters are assigned randomly. Specifically, $\iota$ is sampled from a uniform distribution in $\cos \iota$, and $\psi$ and $\phi_0$ are sampled from flat distributions. 

As the next step, we estimate the confusion noise produced by these gravitational wave sources in our mock Milky Way using the pipeline\footnote{\url{ https://gitlab.in2p3.fr/Nikos/gwg}.} described in \citet[][see also \citealt{tim06,cro07,nis12}]{kar21}. The pipeline approximates the so-called ``global fit'' analysis, which is the currently adopted approach to handling LISA's complex data analysis \citep[][see also \citealt{Littenberg_2023,katz2024efficientgpuacceleratedmultisourceglobal,deng2025modularglobalfitpipelinelisa}]{LISARedBook}. 

The pipeline employs a signal-to-noise ($\rho$) evaluation to iteratively estimate the strain amplitude spectral density resulting from the combined signals of the unresolved (i.e., low $\rho$) part of the input population. As a result, we also obtain a catalog of individually resolved (i.e., high $\rho$) binaries. For our analysis, we adopt LISA's instrumental noise requirements as defined in the technical note by \citet{LISAdoc}. 
We assume a mission duration of 4 yr and use an $\rho$ threshold of 7 to distinguish between individually resolvable LISA sources and unresolved ones, a choice commonly adopted in detailed simulations \citep[e.g.,][]{cro07,Cornish_2007,2023MNRAS.522.5358F}

We estimate a confusion foreground from our mock population containing double white dwarf binaries from both isolated binary and triple channels to be at the level of the LISA instrument noise. We also obtain $\sim 1.7\times10^4$ sources above the $\rho$ threshold, of which $\sim 6.5\times10^3$ come from the isolated binaries and $\sim 1.1\times10^4$ come from the triples (see Sect.~\ref{sec:previous_works} for comparison to previous works). Figure~\ref{fig:strain vs frequency plot} shows the characteristic strain as a function of gravitational wave frequency for Galactic double white dwarf sources originating from both triples and isolated binaries. The characteristic strain is defined as the effective strain amplitude, incorporating the number of cycles a system completes during the observation time. This quantity is convenient for comparing source signals with the detector sensitivity, since the height of a system above the sensitivity curve represents its signal-to-noise ratio. 

Our estimate of the confusion noise is lower than that presented in the LISA mission proposal \citep{LISAproposal} and the more recent LISA Definition Study Report \citep{LISARedBook}, which used a different population synthesis study \citep{2004MNRAS.349..181N,2017MNRAS.470.1894K}. This discrepancy arises from multiple aspects of population modeling, including variations in binary evolution prescriptions, Milky Way modeling, and, importantly, the inclusion of the triple formation channel in our simulations. However, since the same systematics are applied to both our triple and isolated binary populations, their differential properties remain largely unaffected. 

These factors collectively influence the total number of binaries in the LISA band and their properties. Identifying a single source of the difference is challenging, as these aspects are interrelated and nontrivially correlated. The LISA Consortium's Astrophysics Working Group is currently investigating the differences and uncertainties in predicting the confusion foreground as part of the Ultra-Compact Binaries catalog comparison project \citep[][as well as Breivik et al. and Bobrick et al. in prep.]{2023arXiv231103431V}. We refer the reader to those forthcoming results and provide a comparison using our test simulation, in which all LISA binaries were generated solely via the isolated binary channel.

\subsubsection{Eccentric systems}\label{sec:ecc}

Here, we focus on eccentric systems originating from the triple formation channel. These systems were excluded from the analysis above as they are estimated to be very few (and therefore do not contribute to the overall Galactic confusion signal) and because their gravitational wave signals differ from those of circular systems.

We find that approximately $3 \times 10^{-6}$ of all Galactic double white dwarfs exhibit eccentric orbits. These systems result primarily from two reasons: (1) the ZLK effect excites the eccentricity of the inner binary orbit, or (2) the system becomes dynamically unstable and eventually achieves a stable configuration with high eccentricity.

In our simulations, all eccentric systems show high eccentricities ($e > 0.9$) and wide semi-major axis ($10^1 - 10^6\,\rm au$). For such orbital configurations, gravitational wave emission predominantly occurs near pericenter passage, lasting up to a few hours and producing burst-like gravitational wave signals. Since the orbital periods of these systems in our simulations are significantly longer than the mission duration (> 46 yr), the double white dwarf burst signals will not repeat within LISA's observation window of 4 yr.

\begin{figure*}
          \includegraphics[trim = 0 0 0 0 ,width=1\textwidth]{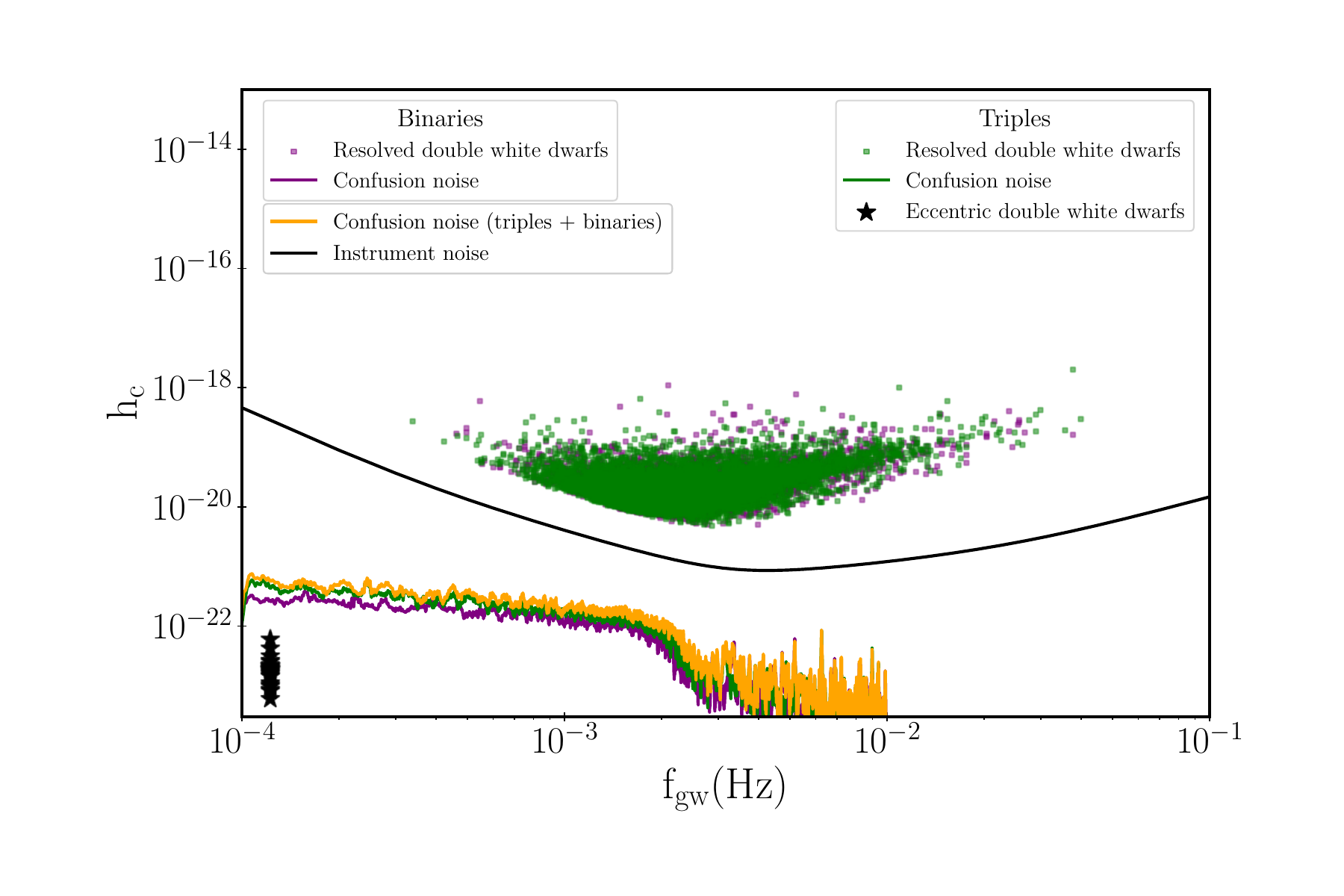}
    \caption{Characteristic strain $h_c = {\cal A}_{\rm gw}\sqrt{f_{\rm gw} T_{\rm obs}}$ of resolved double white dwarf binaries in our mock simulation for a mission duration of $T_{\rm obs}=4\,yr$. Triples are shown with green points, and the isolated binary channel is shown with purple points. This is compared to the LISA instrumental noise (black solid line). Triples with highly eccentric orbits ($e > 0.9$) and other low signal-to-noise sources are marked in black and dark gray markers, respectively. The confusion background from the Galactic double white dwarf population is represented by green, purple, and orange lines for triples, binaries, and their combined contribution, respectively. All the eccentric systems (black markers) tend to have a narrow periapsis time ($10^4 \,\mathrm{s}$), and hence they occupy a narrow frequency range. The astrophysical noise from these backgrounds remains significantly lower than the instrumental noise.
    }
    \label{fig:strain vs frequency plot}
\end{figure*}

The probability to observe a gravitational wave burst from an individual system at periapsis is $\sim\mathrm{T_{obs}/T_{orb}}$, where $\mathrm{T_{obs}}$ is the observational time ( $\mathrm{\sim 4 \, yr}$) and $\mathrm{T_{orb}}$ is the orbital period of the binary. Assuming Poisson binomial distribution, we estimate that LISA will detect at most one eccentric double white dwarf during its operational duration.
Following \citet{2024ApJ...965..148X}, we estimated the frequency and strain amplitude of such a burst signal as
\begin{eqnarray}
f_{\mathrm{burst}}  \sim \frac{2}{T_{\rm orb}(r_{\rm p})},
\end{eqnarray}
\begin{equation}
    \mathrm{h}_{\rm burst} = \frac{2 (G \mathcal{M})^{5/3} }{c^4 d} \left[\frac{2\pi}{T_{\rm orb}(r_{\rm p})} \right]^{2/3},
\end{equation}
where $r_{\rm p}$ is the periapsis distance. We found the (dimensionless) strain amplitudes of the gravitational wave emitted during the periapsis are between $\sim 10^{-24}$ and $10^{-22}$, i.e., below the noise curve (black stars in Fig.~\ref{fig:strain vs frequency plot}).  Therefore, such a signal would not be detectable by LISA.

\subsection{Population properties}
\label{sec:population properties}

\begin{figure*}
    \centering
        \includegraphics[trim = 0 0 0 0,width=1\textwidth]{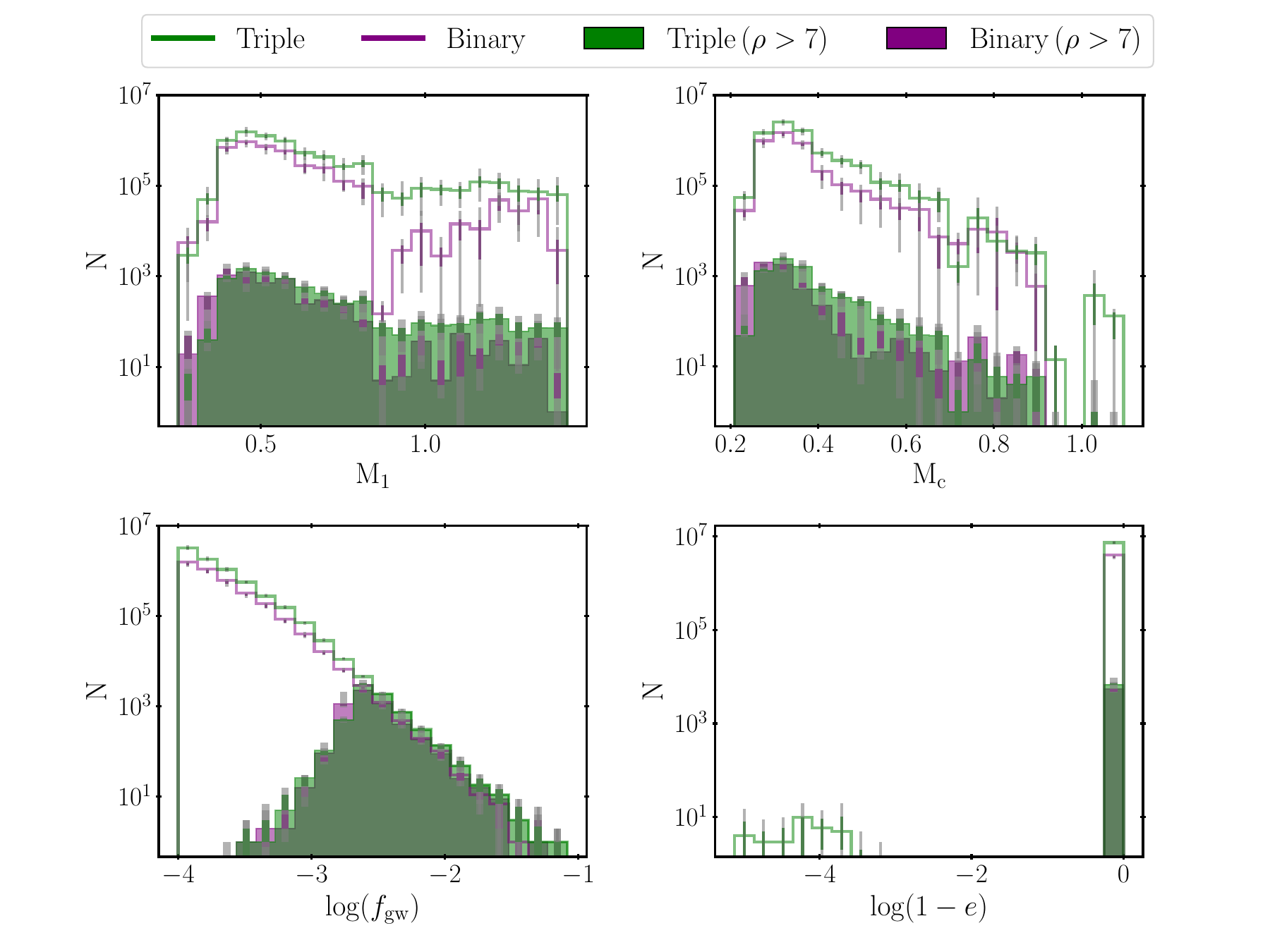}
    \caption{Population properties of LISA-detectable double white dwarfs from triple systems compared to isolated binaries. Triple-origin systems tend to produce more massive white dwarfs than isolated binaries. Outline (step) histograms represent the total LISA-detectable population, while filled histograms show the subset of systems individually resolvable with S/N > 7. For more on the distributions shown, refer to Sect.~\ref{sec:population properties}. Green indicates triple-origin systems, and purple indicates isolated binaries. Thin and thick vertical lines show the 68.3\% confidence intervals for the full and individually resolvable populations, respectively. The vertical gray lines indicate the full range of values spanned in each bin. Confidence intervals are estimated via bootstrap resampling and reflect the sampling uncertainty arising from the stochastic seeding of the Galaxy (see Appendix~\ref{appendix:B} for details).}
    \label{fig:pop_properties}
\end{figure*}

In this section we describe the similarities and differences in the population properties of double white dwarfs formed from triple systems and binary systems. Figure~\ref{fig:pop_properties} shows the distribution of population properties such as chirp mass, primary mass, eccentricity, and gravitational wave frequency. The green color corresponds to the properties of double white dwarfs from triple systems while the purple color shows the properties of double white dwarfs from binary star systems. The shaded green and purple regions represent the properties of resolvable double white dwarfs from triple and binary star systems, respectively. The resolved double white dwarf population traces the features of the total double white dwarf population. It is also interesting to note that all double white dwarfs in our mock Milky Way with $f_{\rm gw} > 2 \times 10^{-3}$ are fully resolvable by LISA. Individually resolving any double white dwarf with $f_{\rm gw} < 2 \times 10^{-3}$ is more difficult for LISA due to the higher degree of overlap in this frequency range.

Triple systems produce about $\sim 1.5$ times more LISA  double white dwarf systems with primary masses greater than  $0.9\,\mathrm{M_{\odot}}$ and $\sim 3.9$ times more super-Chandrasekhar mass double white dwarfs (binary white dwarfs in which total mass of both the white dwarfs combined exceeds $1.44\,\mathrm{M_{\odot}}$) than the isolated binary channel. In addition, triple systems produce $\sim 1.6$ times more extremely low-mass white dwarfs ($m < 0.25\,\mathrm{M_{\odot}}$) that enter the LISA frequency bandwidth than isolated binaries. Isolated binaries do not produce binaries with a chirp mass greater than  $0.9\,\mathrm{M_{\odot}}$. In contrast, triple systems produce $\sim 10^{3}$ binaries whose chirp mass is greater than 0.9 $\mathrm{M_{\odot}}$. There are two possible ways to form massive white dwarfs: 1) The system originates from a massive main-sequence progenitor that evolves to a massive white dwarf; 2) Two less massive progenitors merge to form a massive main-sequence progenitor that subsequently forms a massive white dwarf. In our simulations, initial massive main-sequence progenitors are rare. Hence, it is difficult to form massive white dwarfs in isolated binaries. However, in triple systems, the two less massive progenitors in the inner binary can merge to form a massive progenitor. The merged star co-evolves with the former tertiary component to form a double white dwarf with a massive component. 

There is no significant difference in the frequency distribution of double white dwarf binaries formed from isolated binaries and triple systems. This is because most double white dwarfs emerge as short-period binaries after a common envelope phase, eventually emitting gravitational waves to enter the LISA band.

\subsection{Different types of double white dwarfs}
\label{sec:double white dwarf types}

From the \texttt{MSE} single-star evolution model, helium-core white dwarfs originate from binary interactions where the progenitor loses its envelope before helium ignition, with typical masses of $\lesssim 0.45 \,$M$_\odot$. Carbon-Oxygen core white dwarfs form from intermediate-mass stars that exhaust helium in their cores and expel their outer layers, resulting in masses between $\sim 0.45 - 1.1 \,$M$_\odot$. Oxygen-Neon core white dwarfs arise from more massive stars that undergo carbon burning before shedding their envelopes, with typical masses of $\gtrsim 1.1 \,$M$_\odot$. 

\begin{figure}
        \includegraphics[width=\columnwidth]{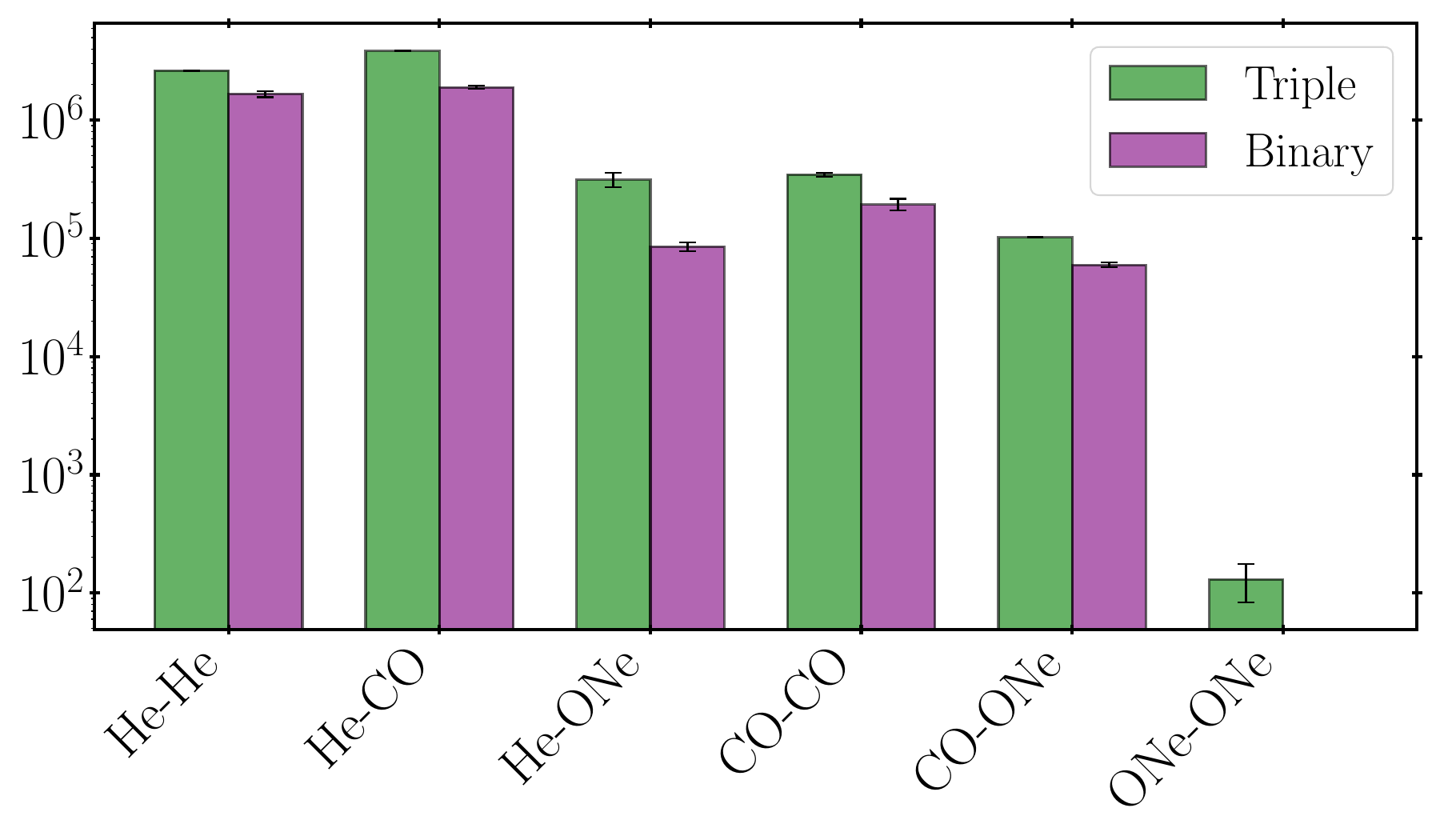}
    \caption{Types of LISA-detectable double white dwarfs formed from triple systems compared to isolated binaries. In our models, ONe-ONe double white dwarfs are produced exclusively in triple systems. Error bars represent a 68.3\% confidence interval and are derived from bootstrap resampling and capture the statistical uncertainty introduced by the stochastic seeding of the Galaxy. (See Appendix~\ref{appendix:B} for details.)}
    \label{fig:Dwhite dwarf types}
\end{figure}

Figure~\ref{fig:Dwhite dwarf types} shows the relative numbers of different double white dwarf (core composition) types from binary and triple systems, respectively. Our simulations produce all types of double white dwarfs in the LISA frequency bandwidth from both triple and binary star systems, including He-He, He-CO, He-ONe, CO-CO, and CO-ONe systems. He-CO systems dominate the population of double white dwarfs originating from both triples and isolated binaries. Our simulations produce ONe-ONe systems only from the triple populations, but the sampling uncertainties in this mass range are too high to draw meaningful conclusions.

\subsection{Detectability of the third star}
\label{sec:thirdstar}
 
We emphasise again that approximately $57^{+8}_{-5}\,\%$ of all double white dwarfs detectable by LISA will have a bound tertiary companion. The bound tertiary star can be found in various evolutionary stages, including as a main-sequence star, a giant star, or a white dwarf. In the remaining $43\,\%$ of systems, the third star either became unbound or there was a merger of two stars, reducing the system from a triple to a binary (See Sect.~\ref{sec:Unbound tertiary}). If the tertiary is retained, it can impart accelerations to the center of mass of the binary, leading to observable Doppler shifts in the gravitational wave signals \citep[e.g.,][]{2008ApJ...677L..55S, 2018PhRvD..98f4012R,2019NatAs...3..858T}.

In the context of LISA Galactic binaries in hierarchical triple systems, \cite{2018PhRvD..98f4012R} identifies three regimes, essentially governed by the ratio of the outer orbital period to the observation time: 1) When the outer period is much larger than the observation time the hierarchical orbit imparts an overall unobservable Doppler shift. 2) When the outer period is up to a factor ten larger than the observation time the influence of the companion can be detected. 3) When the outer period is shorter than or comparable to the observation time, the eccentricity and period of the hierarchical orbit can be inferred.
Specifically, a tertiary companion leaves a detectable imprint in the gravitational wave signal if the outer binary period satisfies
\begin{equation}\label{eq:T2_factor}
\begin{aligned}
T_2 \lesssim T_{\mathrm{lim}} = 43.2 \, \text{yr} 
\left( \frac{\rho}{10} \cdot \frac{m_3}{1.0 M_\odot} \cdot \frac{f}{5 \, \text{mHz}} \right)^{3/4}
\left( \frac{m_1+m_2}{2{\rm M}_\odot} \right)^{-1/2} \\
\times \left( \frac{T_{\text{obs}}}{4 \, \text{yr}} \right)^{3/8}
\left( \frac{1 + \frac{1}{2} e_2^2}{(1 - e_2^2)^{5/2}} \right)^{3/8},
\end{aligned}
\end{equation}
where $\rho$ is the signal-to-noise ratio of the binary. 

\begin{figure}
        \includegraphics[width=\columnwidth]{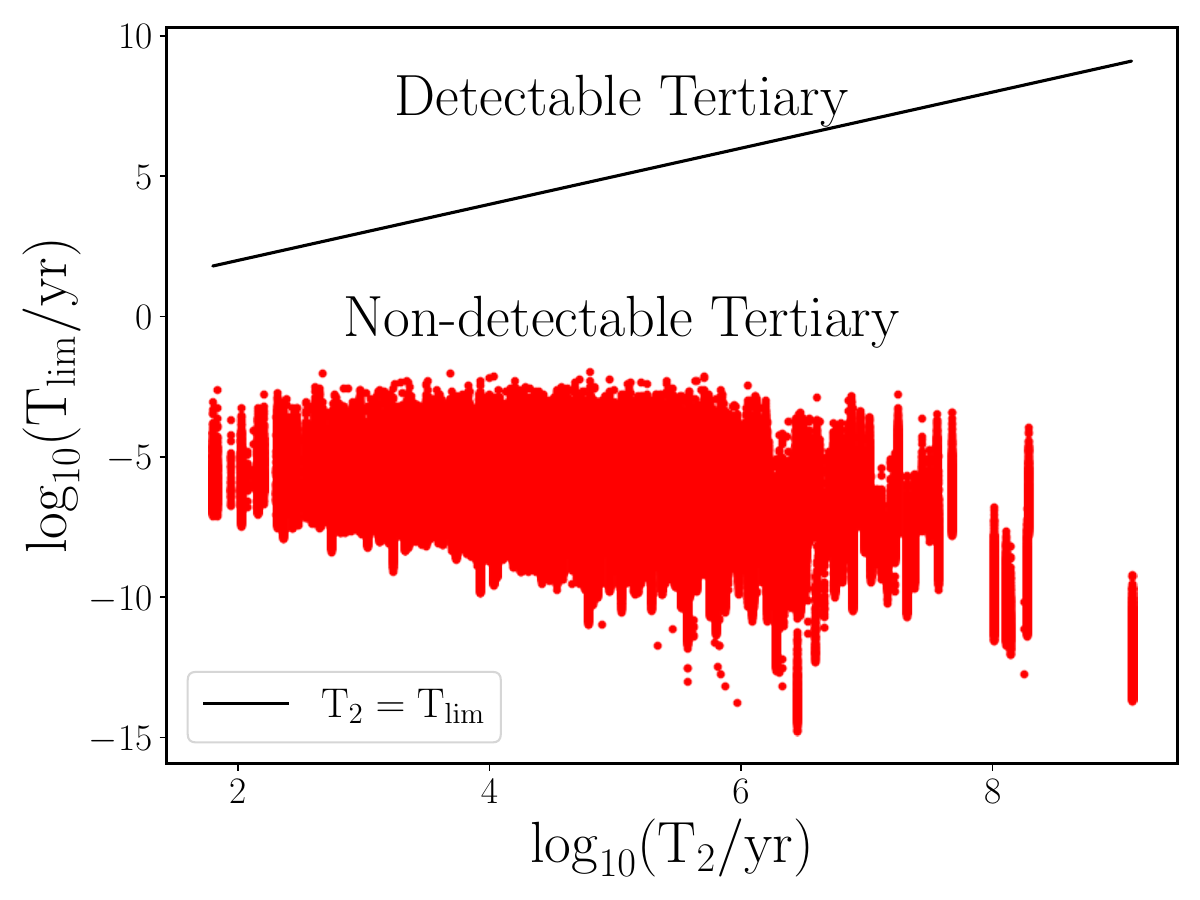}
    \caption{Comparison between the outer orbital period of a triple system and the factor on the right-hand side of Eq.~\eqref{eq:T2_factor}. Red points represent $T_{\mathrm{lim}}$ vs. the outer orbital period $T_{2}$ for all the Galactic LISA double white dwarfs with a bound third star. The black line represents the points where $T_2 = T_{\mathrm{lim}}$. No tertiary star satisfies Eq.~\eqref{eq:T2_factor} for detectability within the LISA frequency bandwidth.}
    \label{fig:a2_thirdstar}
\end{figure}

In our simulations, in all systems that have retained the tertiary companion, the outer orbits are too wide to have any detectable effect within the resulting LISA gravitational wave signal.
Figure~\ref{fig:a2_thirdstar} compares the outer orbital period $\mathrm{T_2}$ with the factor on the right hand side of the Eq.~\eqref{eq:T2_factor}. The black line shows the limit where $\mathrm{T_2}$ equals the factor on the right hand side. We find that none of our surviving triple systems lie within, or close to, the detectable limit. In addition, we find no systems where $T_{\text{orb}} \leq T_{\text{obs}}$ or $T_{\text{orb}} \approx 10 \times T_{\text{obs}}$. The minimum outer semi-major axis across all systems is found to be approximately $21\,\text{au}$, which corresponds to an orbital period of around $63\,\rm yr$.

\section{Discussion}
\label{sec:Discussion}

We discuss our results in the context of previous works, present the uncertainties associated with our models, and describe the constraints imposed by electromagnetic observations.

\subsection{Comparison to previous works}\label{sec:previous_works}

We calculated the number and population properties of LISA double white dwarfs that originated from triple systems (See Sect.~\ref{sec:methods}). We also simulated LISA double white dwarfs from an isolated binary population (See Appendix~\ref{appendix:A}) to compare with triple populations.
Our isolated binary simulation predicts $\approx10^{4}$ individually resolvable double white dwarfs in the LISA frequency bandwidth, which agrees with previous works, including \cite{2004MNRAS.349..181N, 2010ApJ...717.1006R, 2010A&A...521A..85Y, 2017MNRAS.470.1894K, 2019MNRAS.490.5888L, 2023A&A...669A..82L, 2023ApJ...945..162T,2024arXiv240520484T}. In total, our models predict $\sim 1.1 \times 10^{7}$ double white dwarf sources that emit gravitational waves in the LISA frequency bandwidth but have too low $\rho$ to be individually detected by LISA.

In particular, \cite{2017MNRAS.470.1894K} used the binary population model of \cite{too12} based on the \texttt{SeBa} binary population synthesis code \citep{1996A&A...309..179P} and an analytic Galactic potential and star formation history to estimate the number of Galactic double white dwarfs. They predict $\sim2.6 \times10^{7}$ LISA double white dwarfs as foreground noise and $\sim 2.5 \times 10^4$ as individually resolvable LISA double dwarf sources. \cite{2022MNRAS.511.5936K} performed a data-driven analysis using existing observational double white dwarf data and also estimated a LISA double white dwarf population of $\sim 2.6\times10^7$ as foreground noise and $\sim 6.0 \times 10^4$ as individually resolvable LISA double dwarf sources. Using the \texttt{BSE} code (i.e., similar to our isolated binary evolution channel but with different underlying assumptions) combined with the \texttt{Fire} cosmological simulation, \cite{2019MNRAS.490.5888L} constructed the Galactic LISA double white dwarf population and estimated $\sim 6.2\times 10^{7}$ double white dwarf sources as foreground noise and $\sim 1.2 \times 10^{4}$ as individually resolvable LISA double dwarf sources. Further, \cite{2023A&A...669A..82L} also used the \texttt{BSE} code but with a mass transfer stability criterion by adopting critical mass ratios from the adiabatic mass loss model by \citet{Ge_2010,Ge_2015,Ge_2020}, and estimated a foreground LISA double white dwarf population size of $\sim 5.0\times10^7$ and about $4.0\times10^4$ individually resolvable double white dwarfs. Our results are broadly consistent with previous studies in terms of the number of resolved binaries, but we estimate comparatively fewer double white dwarf sources emitting gravitational waves in the LISA frequency band (see Sect.~\ref{sec:circular systems} for details). This discrepancy may arise from multiple factors, including the use of different population synthesis codes, varying assumptions underlying binary evolution, and different approaches to modeling the Galaxy. Most importantly, previous studies model isolated binary evolution only, ignoring triples.

When comparing our isolated binary evolution results (with triple fraction set to zero; see Table~\ref{tab:counts}) to previous studies, several factors can contribute to differences in LISA predictions. 
For instance, assuming a different total stellar mass for the Milky Way would linearly scale our results (See Eq.~\ref{eq: Mtot}). As shown by \citet{2023MNRAS.521.1088K}, the underlying star formation history also plays a role \citep[see also][]{2010A&A...521A..85Y}.
However, the most likely primary source of differences is the modeling of stellar and binary evolution. This was recently demonstrated by \citet{van_Zeist_2024}, who investigated the gravitational wave population of the Large and Small Magellanic Clouds as a case study using the \texttt{BPASS} \citep{2017PASA...34...58E} and \texttt{SeBa} codes. They specifically attributed variations in the predicted number of LISA double white dwarfs to differences in the treatment of CE evolution and the stability of mass transfer. Indeed, even studies using the same population synthesis code report significant differences in LISA predictions when these processes are modeled differently \citep[e.g.,][]{2017MNRAS.470.1894K,2023A&A...669A..82L}. 
As mentioned before, quantifying the impact of all these factors on LISA predictions is a large collaborative effort within the LISA Consortium’s Astrophysics Working Group. We refer the reader to those forthcoming results.

\subsection{Uncertainties in our modeling}

\begin{figure}
        \includegraphics[width=\columnwidth]{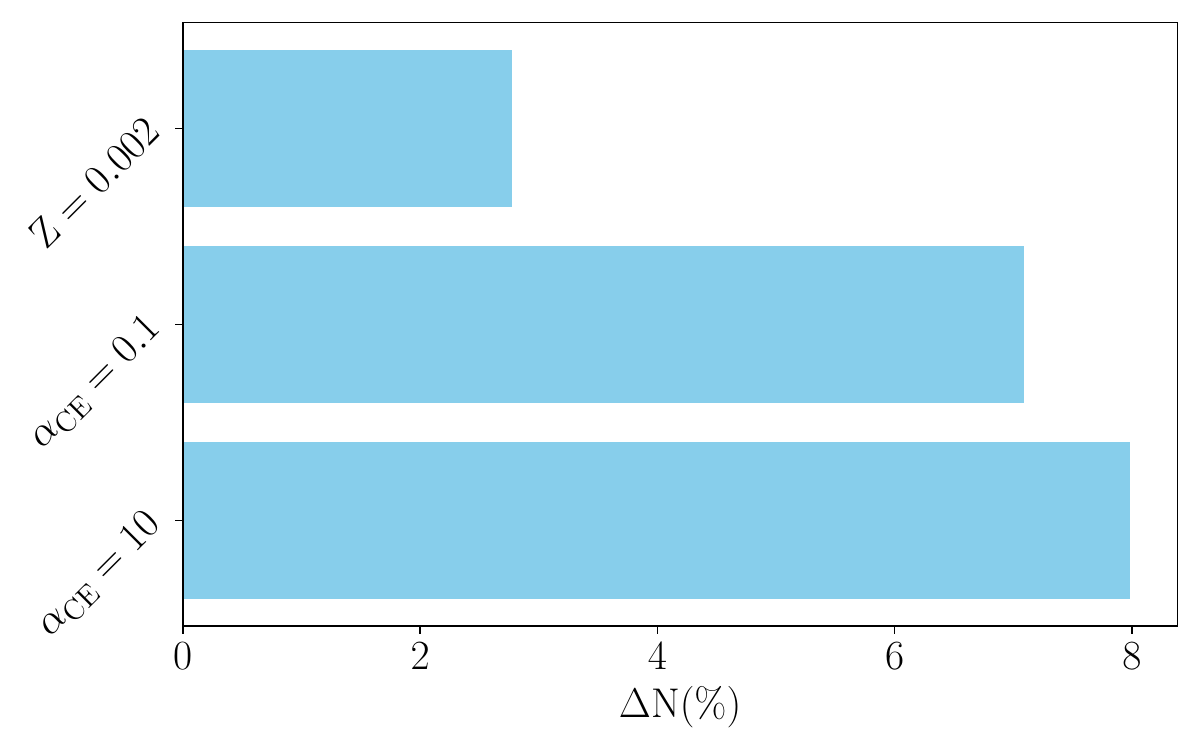}
    \caption{Fractional difference of LISA-detectable double white dwarfs from models with varying common envelope efficiency, $\alpha_{\mathrm{CE}}$, and subsolar metallicity relative to the default values ($\alpha_{\mathrm{CE}} = 1$, $Z = 0.02$). The number of double white dwarfs varies by up to $8\,\%$ when the common envelope parameters are modified.}
    \label{fig:example_figure}
\end{figure}

In our normalization calculations, we assume constant multiplicity fractions $0.2$, $0.3$, and $0.5$ for stars in triple, binary, and single systems respectively. The multiplicity fractions play a crucial role in the mass normalization and, hence, the total number of LISA double dwarfs. All previous works assume a zero triple fraction. To compare our results with previous works, we have also calculated the number of LISA double dwarfs with a 0.5 binary fraction and zero triple fraction. We found $\sim 9 \times 10^{6}$ systems, showing consistency with previous studies in scale. For the same isolated binary population, assuming a 0.2 triple fraction, we only obtain $\sim 3.8 \times 10^{6}$ systems, half as many. This highlights the sensitivity of the results to the assumed multiplicity fraction.

We adopt the default critical mass-ratio criteria in \texttt{MSE} to model the stability of mass transfer. Studies investigating alternative stability criteria find that mass transfer tends to be more stable than we have assumed, suggesting that fewer systems undergo a common envelope phase \citep{Tauris:2000toa, Podsiadlowski_2002, Ge_2010, Woods_2012,Passy_2012,Ge_2015,Ge_2020,Temmink_2023}. This affects the formation of double white dwarfs \citep{2012ApJ...744...12W}, particularly in the LISA band \citep{2023A&A...669A..82L,van_Zeist_2024}. Additionally, we assume that the orbit circularizes following the common envelope phase. However, there is ongoing debate about whether residual eccentricity may persist at the end of the common envelope evolution. The modeling of unstable mass transfer leading to a common envelope phase follows the approximate $\alpha$-$\lambda$ formalism. We investigated the effect of the common-envelope efficiency parameter, $\alpha_{\text{CE}}$, on our results. As discussed in \cite{2013A&ARv..21...59I}, the commonly used parameter $\alpha_{\text{CE}}$ only describes an energetic ``efficiency'' under restrictive assumptions which are not known to be true.  Contributions from sources of energy not included in the standard $\alpha$-$\lambda$ formalism may perhaps allow for an effective $\alpha_{\text{CE}} > 1$. We would nonetheless be surprised if our model with effective $\alpha_{\text{CE}} = 10 $ is a good description of reality, but include that model to explore how sensitive our predictions are to extremely different outcomes of common-envelope phases.

For lower ($\alpha_{\text{CE}} = 0.1$) and higher efficiency parameters ($\alpha_{\text{CE}} = 10$), our simulations yield approximately $7.5 \times 10^{6}$ and $7.6 \times 10^{6}$ LISA double white dwarfs, respectively. For a lower value of the common envelope efficiency, the common envelope phase is more effective at shrinking the orbit. Consequently, these systems enter the LISA frequency bandwidth within a Hubble time. In contrast, for the default value ($\alpha_{\text{CE}} = 1$), the common envelope phase results in a wider orbital period. However, for a high common envelope efficiency parameter, there is an increase in the number of LISA-detectable double white dwarf binaries. This is because the more efficient common envelope phase is less likely to merge binaries that would otherwise have merged for $\alpha_{\text{CE}} = 1$.

We assume that all progenitors of the Galactic LISA double white dwarfs are formed with solar metallicity. About $\sim 2\%$ of the star particles in the selected Milky Way model have subsolar metallicity. However, our Galaxy encompasses a larger range of metallicities. We also explore the impact of subsolar metallicity by constructing the galaxy using the procedure described in Sect.~\ref{sec:gal_pop} but with an initial metallicity of Z = 0.002. It results in approximately $\sim 7.5 \times 10^{6}$ LISA double white dwarfs, an increase of $\sim 0.3 \times 10^{6}$ compared to those with solar metallicity. Stars with lower metallicity evolve faster. For example, a $0.91\, \mathrm{M_{\odot}}$ star evolves into a white dwarf within a Hubble time at subsolar metallicity, whereas it does not at solar metallicity. This increase in the number of low-mass stars that evolve into white dwarfs significantly contributes to the population of LISA-detectable double white dwarfs.

The ZLK effect induces eccentric oscillations in the inner binary, increasing the possibility of mass transfer at periapsis or eccentric mass transfer. The {\tt MSE} model treats eccentric mass transfer using an approximate prescription. Furthermore, {\tt MSE} also uses approximate prescriptions for mass transfer from the third star onto the inner binary. In our simulation, about $5.5\,\%$ of Galactic LISA double white dwarfs undergo a TCE phase before reaching the LISA frequency bandwidth. Inputs from hydrodynamical \citep{10.1093/mnras/staa3242} simulations of eccentric mass transfer and triple mass transfer are needed to improve the eccentric mass transfer and TCE prescriptions.

We highlight that simulations of triple systems are computationally expensive. We simulated $10^5$ triple systems, producing $\sim 3 \times 10^3$ LISA double white dwarfs. Deriving from the mass function in Fig.~\ref{fig:initial_dist}, there is a very low probability of producing high-mass white dwarfs. Hence, uncertainties in the statistics on the number of white dwarfs increase with mass. There is also a possibility of producing ONe white dwarfs from 8-10 M$_\odot$ stars. Our models are created from progenitors in the mass range 1-8 M$_\odot$. The sampling uncertainties in this mass range are too high to draw meaningful conclusions. In addition, we use random sampling to construct the initial triple population. However, more targeted sampling algorithms, such as \texttt{STROOPWAFEL} \citep{Broekgaarden_2019}, could be employed to address the impact of sampling uncertainties in the initial population on rare events.

Finally, we note that {\tt MSE} is a population synthesis code designed for statistical studies, and individual system modeling is not recommended. Due to variations in floating-point representation and rounding across different machines, the code may yield slightly different results when executed on different machines. However, these numerical difficulties average out on a large sample of the population. Given this inherent complexity, we limited the computational time to 10 hours per system. The majority of systems completed their evolution in under 2 hours. Within the ten-hour limit, $4\,\%$ of systems did not complete their evolution. Of these, around $0.4\,\%$ were dynamically unstable, while the remaining systems were undergoing stable secular evolution.  Only $0.1\,\%$ of the systems did not complete their evolution within ten hours and had white dwarfs or both components massive enough to evolve into white dwarfs within a Hubble time. This $0.1\,\%$ of the systems contribute to some uncertainty in our predictions for LISA double white dwarfs, while the remaining incomplete systems were excluded without affecting the overall results. 

\subsection{Possibility of electromagnetic constraints to LISA observations}

While millions of double white dwarfs will emit gravitational waves in the LISA frequency bandwidth, only a few hundred double white dwarfs are well-characterized in electromagnetic observations \citep[e.g.,][]{2023MNRAS.518.5123M}. Of these, 48 of the binaries will have a high $\rho$ ratio and will serve as verification binaries for the LISA mission \citep[e.g.,][]{2006CQGra..23S.809S,2018MNRAS.480..302K, 2024ApJ...963..100K, 2023MNRAS.522.5358F}. To date, it is unknown is any of these known LISA binaries are part of triple systems with wide tertiary companion.

LISA is set to be launched in 2035 \citep{LISARedBook}. However, some electromagnetic observations/surveys planned in the near future will already increase the sample of short-period double white dwarfs and offer better constraints for the modeling, data analysis, and detection techniques of LISA double white dwarfs. This includes new binary insights from Gaia's next data release (DR4), SDSS-V \citep{2017A&A...598L...7K}, LAMOST \citep{2012arXiv1206.3569Z}, ZTF \citep{2019PASP..131a8002B} , 4MOST \citep{2019Msngr.175....3D}, WEAVE \citep{2014MNRAS.438.1909D}, the Asteroid Terrestrial-impact Last Alert System \citep{2018AJ....156..241H,2018PASP..130f4505T}, the Gravitational-wave Optical Transient Observer \citep{2022MNRAS.511.2405S}, Euclid \citep{Euclid_2011}, the Nancy Roman Space Telescope \citep{Roman_2019}, and Vera Rubin-LSST \citep{LSST_2009}.
Similarly to the approach taken by \cite{2024ApJ...969...68H}, who used the age discrepancy from observational data to quantify the contribution of triple systems, applying comparable modeling techniques to forthcoming double white dwarf observations could provide further constraints on the contribution of triples to double white dwarfs in the LISA frequency bandwidth.

\section{Conclusion}
\label{sec:Conclusion}

We estimated the Galactic LISA double white dwarfs from triple systems. We combined the triple population synthesis code \texttt{MSE} and the \texttt{TNG50} cosmological simulations to seed a Milky Way-like galaxy and obtain a population of LISA-detectable double white dwarfs. To compare our results with LISA double white dwarfs from isolated binaries, we also estimated the LISA double white dwarf population from isolated binaries using the same \texttt{MSE} code. Our main conclusions are listed below:
\begin{enumerate}
    \item Galactic LISA double white dwarfs have comparable contributions from both binary and triple channels. We estimate $\sim 7.2 \times 10^{6}$ and $\sim 3.8 \times 10^{6}$ LISA double white dwarfs from triple and binary star systems, respectively. Of these systems, $\sim 1.09 \times 10^{4}$ and $\sim 6.5 \times 10^{3}$ are individually resolvable double white dwarfs from triple and binary star systems, respectively. In addition, we find that the confusion foreground produced from our population (triples and isolated binaries) is below the instrument noise level.
    \item We identify five different key processes that shape the evolutionary pathways of triple systems and lead to the formation of double white dwarfs that emit gravitational waves in the LISA frequency bandwidth.
        \begin{enumerate}
            \item Induced mass transfer: The gravitational perturbations from the third star triggers a mass transfer episode in the inner binary that leads to a shorter orbital period.
            \item Outer binary channel: The inner binary merges to form a rejuvenated star that combines with the third star to enter the LISA frequency bandwidth.
            \item Ejected tertiary: Perturbations from the third star alter the inner binary configurations and are responsible for the short period of the inner binary, but the third star is later ejected before the inner binary enters the LISA frequency bandwidth.
            \item Triple common envelope: The tertiary star overflows its Roche lobe and transfers mass onto the inner binary, triggering a triple common envelope phase that tightens the inner orbit to short periods.
            \item Inner binary channel: A binary with a third star that is too distant to have any effect enters the LISA frequency bandwidth due to effectively isolated binary interactions.
        \end{enumerate}
    \item Our models show no major distinguishable differences in the population properties of systems originating from triples compared to those from isolated binaries.
    \item Of the LISA double white dwarfs from triple systems, about $ 50\,\%$ of the systems have a bound third star. However, the tertiary is typically too distant to have an observable imprint in the gravitational wave signal of the inner binary.
    \item Of the predicted LISA double white dwarfs from triple systems, we estimate that the majority of the systems that enter the LISA frequency bandwidth have circular orbits, and only $3 \times 10^{-6}$ (i.e., 31 systems in total in the Milky Way) of these systems have eccentric orbits. Meanwhile, in the estimated LISA double white dwarf population from isolated binary star systems, all systems have circular orbits. All the eccentric systems are found to have highly eccentric orbits ($e > 0.9$) and will emit gravitational bursts, with a typical periapsis period of a few hours. However, they are unlikely to be observable due to their small gravitational wave strain amplitude.
    \item In our Galactic LISA double white dwarf population, we observe all types of double white dwarfs. Nonetheless, containing a He white dwarf and a CO white dwarf is the most common configuration.
\end{enumerate}

Our study is the first to investigate the role of the triple evolution channel in the context of future LISA observations. We quantitatively assessed the impact of including this channel on the number of observable LISA sources. While our results indicate no major distinguishable differences in the population properties of systems originating from triples compared to those formed through isolated binary evolution, they are particularly relevant for the future interpretation of LISA data on the Galactic population. Additionally, we show that the triple channel produces highly eccentric sources; although rare, these systems generate burst-like signals, in contrast to the predominantly monochromatic continuous signals emitted by the majority of the Galactic population. Thus, it is also important to consider these systems in the context of LISA data analysis.

\section*{Data availability} The catalog used to generate the astrophysical noise background from double white dwarfs (both triple and isolated binary origin) is available at \href{https://zenodo.org/records/16743708?token=eyJhbGciOiJIUzUxMiJ9.eyJpZCI6IjVlNTBlMDcwLWY0ZTctNGM4OC1iODIxLWU3MjI2M2QzZGY5MSIsImRhdGEiOnt9LCJyYW5kb20iOiJkYTE0OTBjMDMwYjkwNTdiZTNlNmEzMmIxNzVmYzE5NCJ9.YGwOg3WJxUZ_k_a4TcVsuTT_i4DzN8-LE9pN98U49-QKpwzN5Ek6KePVTtsDizGn3MW8jkvUrxRVLNtkuNBckQ}{Zenodo}.

\begin{acknowledgements}
ASR would like to thank Adrian Hamers for his guidance on the fundamentals of triple dynamics and the development of the \texttt{MSE} code, which was instrumental in this work. ASR also thanks Nikolaos Karnesis for constructing the confusion background using our population.  We thank Onno Pols for clarifying to us the assumptions adopted in the binding energy parameter ($\lambda$) fits provided by \citet{2014A&A...563A..83C}. JS, SJ, and SDM acknowledge funding from the Netherlands Organisation for Scientific Research (NWO), as part of the Vidi research program BinWaves (project number 639.042.728, PI: de Mink). ST also acknowledges support from the Netherlands Research Council NWO (VENI 639.041.645 and VIDI 203.061 grants).
Software: \texttt{MSE} \citep{2021MNRAS.502.4479H}, \texttt{Matplotlib} \citep{2007CSE.....9...90H}, \texttt{NumPy} \citep{2011CSE....13b..22V}, \texttt{pandas} \citep{mckinney2010}, \texttt{gwg} \citep{kar21}.

\end{acknowledgements}

\bibliographystyle{aa} 
\bibliography{paper} 

@ARTICLE{2023A&A...669A..82L,
       author = {{Li}, Zhenwei and {Chen}, Xuefei and {Ge}, Hongwei and {Chen}, Hai-Liang and {Han}, Zhanwen},
        title = "{Influence of a mass transfer stability criterion on double white dwarf populations}",
      journal = {\aap},
         year = 2023,
        month = jan,
       volume = {669},
          eid = {A82},
        pages = {A82},
          doi = {10.1051/0004-6361/202243893},
archivePrefix = {arXiv},
       eprint = {2211.01861},
 primaryClass = {astro-ph.SR},
       adsurl = {https://ui.adsabs.harvard.edu/abs/2023A&A...669A..82L}
}

@ARTICLE{2024arXiv240520484T,
       author = {{Tang}, Petra and {Eldridge}, Jan and {Meyer}, Renate and {Lamberts}, Astrid and {Boileau}, Guillaume and {van Zeist}, Wouter},
        title = "{Predicting gravitational wave signals from BPASS White Dwarf Binary and Black Hole Binary populations of a Milky Way-like galaxy model for LISA}",
      journal = {arXiv e-prints},
         year = 2024,
        month = may,
          eid = {arXiv:2405.20484},
        pages = {arXiv:2405.20484},
          doi = {10.48550/arXiv.2405.20484},
archivePrefix = {arXiv},
       eprint = {2405.20484},
 primaryClass = {astro-ph.GA},
       adsurl = {https://ui.adsabs.harvard.edu/abs/2024arXiv240520484T}
}

@ARTICLE{2023ApJ...945..162T,
       author = {{Thiele}, Sarah and {Breivik}, Katelyn and {Sanderson}, Robyn E. and {Luger}, Rodrigo},
        title = "{Applying the Metallicity-dependent Binary Fraction to Double White Dwarf Formation: Implications for LISA}",
      journal = {\apj},
         year = 2023,
        month = mar,
       volume = {945},
       number = {2},
          eid = {162},
        pages = {162},
          doi = {10.3847/1538-4357/aca7be},
archivePrefix = {arXiv},
       eprint = {2111.13700},
 primaryClass = {astro-ph.HE},
       adsurl = {https://ui.adsabs.harvard.edu/abs/2023ApJ...945..162T}
}

@ARTICLE{2021MNRAS.500.4958W,
       author = {{Wilhelm}, Martijn J.~C. and {Korol}, Valeriya and {Rossi}, Elena M. and {D'Onghia}, Elena},
        title = "{The Milky Way's bar structural properties from gravitational waves}",
      journal = {\mnras},
         year = 2021,
        month = jan,
       volume = {500},
       number = {4},
        pages = {4958-4971},
          doi = {10.1093/mnras/staa3457},
archivePrefix = {arXiv},
       eprint = {2003.11074},
 primaryClass = {astro-ph.GA},
       adsurl = {https://ui.adsabs.harvard.edu/abs/2021MNRAS.500.4958W}
}

@article{Stegmann:2024rnk,
    author = "Stegmann, Jakob and Vigna-G\'omez, Alejandro and Rantala, Antti and Wagg, Tom and Zwick, Lorenz and Renzo, Mathieu and van Son, Lieke A. C. and de Mink, Selma E. and White, Simon D. M.",
    title = "{Close Encounters of Wide Binaries Induced by the Galactic Tide: Implications for Stellar Mergers and Gravitational-wave Sources}",
    eprint = "2405.02912",
    archivePrefix = "arXiv",
    primaryClass = "astro-ph.GA",
    doi = "10.3847/2041-8213/ad70bb",
    journal = "Astrophys. J. Lett.",
    volume = "972",
    number = "2",
    pages = "L19",
    year = "2024"
}

@ARTICLE{1999MNRAS.307..122M,
       author = {{Maxted}, P.~F.~L. and {Marsh}, T.~R.},
        title = "{The fraction of double degenerates among DA white dwarfs}",
      journal = {\mnras},
         year = 1999,
        month = jul,
       volume = {307},
       number = {1},
        pages = {122-132},
          doi = {10.1046/j.1365-8711.1999.02635.x},
archivePrefix = {arXiv},
       eprint = {astro-ph/9901273},
 primaryClass = {astro-ph},
       adsurl = {https://ui.adsabs.harvard.edu/abs/1999MNRAS.307..122M}
}

@ARTICLE{2024ApJ...969...68H,
       author = {{Heintz}, Tyler M. and {Hermes}, J.~J. and {Tremblay}, P. -E. and {Ould Rouis}, Lou Baya and {Reding}, Joshua S. and {Kaiser}, B.~C. and {van Saders}, Jennifer L.},
        title = "{A Test of Spectroscopic Age Estimates of White Dwarfs Using Wide WD+WD Binaries}",
      journal = {\apj},
         year = 2024,
        month = jul,
       volume = {969},
       number = {1},
          eid = {68},
        pages = {68},
          doi = {10.3847/1538-4357/ad479b},
archivePrefix = {arXiv},
       eprint = {2405.02423},
 primaryClass = {astro-ph.SR},
       adsurl = {https://ui.adsabs.harvard.edu/abs/2024ApJ...969...68H}
}

@ARTICLE{GrishinPerets,
       author = {{Grishin}, Evgeni and {Perets}, Hagai B.},
        title = "{Chaotic dynamics of wide triples induced by galactic tides: a novel channel for producing compact binaries, mergers, and collisions}",
      journal = {\mnras},
         year = 2022,
        month = jun,
       volume = {512},
       number = {4},
        pages = {4993-5009},
          doi = {10.1093/mnras/stac706},
archivePrefix = {arXiv},
       eprint = {2112.11475},
 primaryClass = {astro-ph.SR},
       adsurl = {https://ui.adsabs.harvard.edu/abs/2022MNRAS.512.4993G}
}

@article{Zeipel1910,
author = {von Zeipel, H. V.},
title = {Sur l'application des séries de M. Lindstedt à l'étude du mouvement des comètes périodiques},
journal = {Astronomische Nachrichten},
volume = {183},
number = {22-24},
pages = {345-418},
doi = {https://doi.org/10.1002/asna.19091832202},
url = {https://onlinelibrary.wiley.com/doi/abs/10.1002/asna.19091832202},
eprint = {https://onlinelibrary.wiley.com/doi/pdf/10.1002/asna.19091832202},
year = {1909}
}

@ARTICLE{2022MNRAS.516.1406S,
       author = {{Stegmann}, Jakob and {Antonini}, Fabio and {Moe}, Maxwell},
        title = "{Evolution of massive stellar triples and implications for compact object binary formation}",
      journal = {\mnras},
         year = 2022,
        month = oct,
       volume = {516},
       number = {1},
        pages = {1406-1427},
          doi = {10.1093/mnras/stac2192},
archivePrefix = {arXiv},
       eprint = {2112.10786},
 primaryClass = {astro-ph.SR},
       adsurl = {https://ui.adsabs.harvard.edu/abs/2022MNRAS.516.1406S}
}

@ARTICLE{2022PhRvD.106b3014S,
       author = {{Stegmann}, Jakob and {Antonini}, Fabio and {Schneider}, Fabian R.~N. and {Tiwari}, Vaibhav and {Chattopadhyay}, Debatri},
        title = "{Binary black hole mergers from merged stars in the Galactic field}",
      journal = {\prd},
         year = 2022,
        month = jul,
       volume = {106},
       number = {2},
          eid = {023014},
        pages = {023014},
          doi = {10.1103/PhysRevD.106.023014},
archivePrefix = {arXiv},
       eprint = {2203.16544},
 primaryClass = {astro-ph.GA},
       adsurl = {https://ui.adsabs.harvard.edu/abs/2022PhRvD.106b3014S}
}

@ARTICLE{2023ApJ...955L..14S,
       author = {{Shariat}, Cheyanne and {Naoz}, Smadar and {Hansen}, Bradley M.~S. and {Angelo}, Isabel and {Michaely}, Erez and {Stephan}, Alexander P.},
        title = "{Dynamical Evolution of White Dwarfs in Triples in the Era of Gaia}",
      journal = {\apjl},
         year = 2023,
        month = sep,
       volume = {955},
       number = {1},
          eid = {L14},
        pages = {L14},
          doi = {10.3847/2041-8213/acf76b},
archivePrefix = {arXiv},
       eprint = {2306.13130},
 primaryClass = {astro-ph.SR},
       adsurl = {https://ui.adsabs.harvard.edu/abs/2023ApJ...955L..14S}
}

@ARTICLE{2024arXiv240706257S,
       author = {{Shariat}, Cheyanne and {Naoz}, Smadar and {El-Badry}, Kareem and {Rodriguez}, Antonio C. and {Hansen}, Bradley M.~S. and {Angelo}, Isabel and {Stephan}, Alexander P.},
        title = "{Once a Triple, Not Always a Triple: The Evolution of Hierarchical Triples that Yield Merged Inner Binaries}",
      journal = {arXiv e-prints},
         year = 2024,
        month = jul,
          eid = {arXiv:2407.06257},
        pages = {arXiv:2407.06257},
          doi = {10.48550/arXiv.2407.06257},
archivePrefix = {arXiv},
       eprint = {2407.06257},
 primaryClass = {astro-ph.SR},
       adsurl = {https://ui.adsabs.harvard.edu/abs/2024arXiv240706257S}
}

@ARTICLE{2019NatAs...3..858T,
       author = {{Tamanini}, Nicola and {Danielski}, Camilla},
        title = "{The gravitational-wave detection of exoplanets orbiting white dwarf binaries using LISA}",
      journal = {Nature Astronomy},
         year = 2019,
        month = jul,
       volume = {3},
        pages = {858-866},
          doi = {10.1038/s41550-019-0807-y},
archivePrefix = {arXiv},
       eprint = {1812.04330},
 primaryClass = {astro-ph.EP},
       adsurl = {https://ui.adsabs.harvard.edu/abs/2019NatAs...3..858T}
}

@ARTICLE{2021AJ....162..247K,
       author = {{Kang}, Yacheng and {Liu}, Chang and {Shao}, Lijing},
        title = "{Prospects for Detecting Exoplanets around Double White Dwarfs with LISA and Taiji}",
      journal = {\aj},
         year = 2021,
        month = dec,
       volume = {162},
       number = {6},
          eid = {247},
        pages = {247},
          doi = {10.3847/1538-3881/ac23d8},
archivePrefix = {arXiv},
       eprint = {2108.01357},
 primaryClass = {astro-ph.EP},
       adsurl = {https://ui.adsabs.harvard.edu/abs/2021AJ....162..247K}
}

@ARTICLE{2019A&A...632A.113D,
       author = {{Danielski}, C. and {Korol}, V. and {Tamanini}, N. and {Rossi}, E.~M.},
        title = "{Circumbinary exoplanets and brown dwarfs with the Laser Interferometer Space Antenna}",
      journal = {\aap},
         year = 2019,
        month = dec,
       volume = {632},
          eid = {A113},
        pages = {A113},
          doi = {10.1051/0004-6361/201936729},
archivePrefix = {arXiv},
       eprint = {1910.05414},
 primaryClass = {astro-ph.EP},
       adsurl = {https://ui.adsabs.harvard.edu/abs/2019A&A...632A.113D}
}

@ARTICLE{2022MNRAS.517..697K,
       author = {{Katz}, Michael L. and {Danielski}, Camilla and {Karnesis}, Nikolaos and {Korol}, Valeriya and {Tamanini}, Nicola and {Cornish}, Neil J. and {Littenberg}, Tyson B.},
        title = "{Bayesian characterization of circumbinary sub-stellar objects with LISA}",
      journal = {\mnras},
         year = 2022,
        month = nov,
       volume = {517},
       number = {1},
        pages = {697-711},
          doi = {10.1093/mnras/stac2555},
archivePrefix = {arXiv},
       eprint = {2205.03461},
 primaryClass = {astro-ph.EP},
       adsurl = {https://ui.adsabs.harvard.edu/abs/2022MNRAS.517..697K}
}

@ARTICLE{2018PhRvD..98f4012R,
       author = {{Robson}, Travis and {Cornish}, Neil J. and {Tamanini}, Nicola and {Toonen}, Silvia},
        title = "{Detecting hierarchical stellar systems with LISA}",
      journal = {\prd},
         year = 2018,
        month = sep,
       volume = {98},
       number = {6},
          eid = {064012},
        pages = {064012},
          doi = {10.1103/PhysRevD.98.064012},
archivePrefix = {arXiv},
       eprint = {1806.00500},
 primaryClass = {gr-qc},
       adsurl = {https://ui.adsabs.harvard.edu/abs/2018PhRvD..98f4012R}
}

@ARTICLE{2017ApJ...841...77A,
       author = {{Antonini}, Fabio and {Toonen}, Silvia and {Hamers}, Adrian S.},
        title = "{Binary Black Hole Mergers from Field Triples: Properties, Rates, and the Impact of Stellar Evolution}",
      journal = {\apj},
         year = 2017,
        month = jun,
       volume = {841},
       number = {2},
          eid = {77},
        pages = {77},
          doi = {10.3847/1538-4357/aa6f5e},
archivePrefix = {arXiv},
       eprint = {1703.06614},
 primaryClass = {astro-ph.GA},
       adsurl = {https://ui.adsabs.harvard.edu/abs/2017ApJ...841...77A}
}

@ARTICLE{2014A&A...563A..83C,
       author = {{Claeys}, J.~S.~W. and {Pols}, O.~R. and {Izzard}, R.~G. and {Vink}, J. and {Verbunt}, F.~W.~M.},
        title = "{Theoretical uncertainties of the Type Ia supernova rate}",
      journal = {\aap},
         year = 2014,
        month = mar,
       volume = {563},
          eid = {A83},
        pages = {A83},
          doi = {10.1051/0004-6361/201322714},
archivePrefix = {arXiv},
       eprint = {1401.2895},
 primaryClass = {astro-ph.SR},
       adsurl = {https://ui.adsabs.harvard.edu/abs/2014A&A...563A..83C}
}

@ARTICLE{1983ApJ...268..368E,
       author = {{Eggleton}, P.~P.},
        title = "{Aproximations to the radii of Roche lobes.}",
      journal = {\apj},
         year = 1983,
        month = may,
       volume = {268},
        pages = {368-369},
          doi = {10.1086/160960},
       adsurl = {https://ui.adsabs.harvard.edu/abs/1983ApJ...268..368E}
}

@ARTICLE{2008MNRAS.389..869E,
       author = {{Eggleton}, P.~P. and {Tokovinin}, A.~A.},
        title = "{A catalogue of multiplicity among bright stellar systems}",
      journal = {\mnras},
         year = 2008,
        month = sep,
       volume = {389},
       number = {2},
        pages = {869-879},
          doi = {10.1111/j.1365-2966.2008.13596.x},
archivePrefix = {arXiv},
       eprint = {0806.2878},
 primaryClass = {astro-ph},
       adsurl = {https://ui.adsabs.harvard.edu/abs/2008MNRAS.389..869E}
}

@ARTICLE{2007ApJ...669.1298F,
       author = {{Fabrycky}, Daniel and {Tremaine}, Scott},
        title = "{Shrinking Binary and Planetary Orbits by Kozai Cycles with Tidal Friction}",
      journal = {\apj},
         year = 2007,
        month = nov,
       volume = {669},
       number = {2},
        pages = {1298-1315},
          doi = {10.1086/521702},
archivePrefix = {arXiv},
       eprint = {0705.4285},
 primaryClass = {astro-ph},
       adsurl = {https://ui.adsabs.harvard.edu/abs/2007ApJ...669.1298F}
}

@ARTICLE{2013MNRAS.430.2262H,
       author = {{Hamers}, A.~S. and {Pols}, O.~R. and {Claeys}, J.~S.~W. and {Nelemans}, G.},
        title = "{Population synthesis of triple systems in the context of mergers of carbon-oxygen white dwarfs}",
      journal = {\mnras},
         year = 2013,
        month = apr,
       volume = {430},
       number = {3},
        pages = {2262-2280},
          doi = {10.1093/mnras/stt046},
archivePrefix = {arXiv},
       eprint = {1301.1469},
 primaryClass = {astro-ph.SR},
       adsurl = {https://ui.adsabs.harvard.edu/abs/2013MNRAS.430.2262H}
}

@ARTICLE{2019ApJ...882...24H,
       author = {{Hamers}, Adrian S. and {Thompson}, Todd A.},
        title = "{The Impact of White Dwarf Natal Kicks and Stellar Flybys on the Rates of Type Ia Supernovae in Triple-star Systems}",
      journal = {\apj},
         year = 2019,
        month = sep,
       volume = {882},
       number = {1},
          eid = {24},
        pages = {24},
          doi = {10.3847/1538-4357/ab321f},
archivePrefix = {arXiv},
       eprint = {1904.12881},
 primaryClass = {astro-ph.HE},
       adsurl = {https://ui.adsabs.harvard.edu/abs/2019ApJ...882...24H}
}

@ARTICLE{2000MNRAS.315..543H,
       author = {{Hurley}, Jarrod R. and {Pols}, Onno R. and {Tout}, Christopher A.},
        title = "{Comprehensive analytic formulae for stellar evolution as a function of mass and metallicity}",
      journal = {\mnras},
         year = 2000,
        month = jul,
       volume = {315},
       number = {3},
        pages = {543-569},
          doi = {10.1046/j.1365-8711.2000.03426.x},
archivePrefix = {arXiv},
       eprint = {astro-ph/0001295},
 primaryClass = {astro-ph},
       adsurl = {https://ui.adsabs.harvard.edu/abs/2000MNRAS.315..543H}
}

@ARTICLE{2002MNRAS.329..897H,
       author = {{Hurley}, Jarrod R. and {Tout}, Christopher A. and {Pols}, Onno R.},
        title = "{Evolution of binary stars and the effect of tides on binary populations}",
      journal = {\mnras},
         year = 2002,
        month = feb,
       volume = {329},
       number = {4},
        pages = {897-928},
          doi = {10.1046/j.1365-8711.2002.05038.x},
archivePrefix = {arXiv},
       eprint = {astro-ph/0201220},
 primaryClass = {astro-ph},
       adsurl = {https://ui.adsabs.harvard.edu/abs/2002MNRAS.329..897H}
}

@ARTICLE{2011PhRvL.107r1101K,
       author = {{Katz}, Boaz and {Dong}, Subo and {Malhotra}, Renu},
        title = "{Long-Term Cycling of Kozai-Lidov Cycles: Extreme Eccentricities and Inclinations Excited by a Distant Eccentric Perturber}",
      journal = {\prl},
         year = 2011,
        month = oct,
       volume = {107},
       number = {18},
          eid = {181101},
        pages = {181101},
          doi = {10.1103/PhysRevLett.107.181101},
archivePrefix = {arXiv},
       eprint = {1106.3340},
 primaryClass = {astro-ph.EP},
       adsurl = {https://ui.adsabs.harvard.edu/abs/2011PhRvL.107r1101K}
}

@ARTICLE{1962AJ.....67..591K,
       author = {{Kozai}, Yoshihide},
        title = "{Secular perturbations of asteroids with high inclination and eccentricity}",
      journal = {\aj},
         year = 1962,
        month = nov,
       volume = {67},
        pages = {591-598},
          doi = {10.1086/108790},
       adsurl = {https://ui.adsabs.harvard.edu/abs/1962AJ.....67..591K}
}

@ARTICLE{1962P&SS....9..719L,
       author = {{Lidov}, M.~L.},
        title = "{The evolution of orbits of artificial satellites of planets under the action of gravitational perturbations of external bodies}",
      journal = {\planss},
         year = 1962,
        month = oct,
       volume = {9},
       number = {10},
        pages = {719-759},
          doi = {10.1016/0032-0633(62)90129-0},
       adsurl = {https://ui.adsabs.harvard.edu/abs/1962P&SS....9..719L}
}

@ARTICLE{2001MNRAS.321..398M,
       author = {{Mardling}, Rosemary A. and {Aarseth}, Sverre J.},
        title = "{Tidal interactions in star cluster simulations}",
      journal = {\mnras},
         year = 2001,
        month = mar,
       volume = {321},
       number = {3},
        pages = {398-420},
          doi = {10.1046/j.1365-8711.2001.03974.x},
       adsurl = {https://ui.adsabs.harvard.edu/abs/2001MNRAS.321..398M}
}

@ARTICLE{2014ApJ...794..122M,
       author = {{Michaely}, Erez and {Perets}, Hagai B.},
        title = "{Secular Dynamics in Hierarchical Three-body Systems with Mass Loss and Mass Transfer}",
      journal = {\apj},
         year = 2014,
        month = oct,
       volume = {794},
       number = {2},
          eid = {122},
        pages = {122},
          doi = {10.1088/0004-637X/794/2/122},
archivePrefix = {arXiv},
       eprint = {1406.3035},
 primaryClass = {astro-ph.SR},
       adsurl = {https://ui.adsabs.harvard.edu/abs/2014ApJ...794..122M}
}

@ARTICLE{2010MNRAS.401..977J,
       author = {{Jiang}, Yan-Fei and {Tremaine}, Scott},
        title = "{The evolution of wide binary stars}",
      journal = {\mnras},
         year = 2010,
        month = jan,
       volume = {401},
       number = {2},
        pages = {977-994},
          doi = {10.1111/j.1365-2966.2009.15744.x},
archivePrefix = {arXiv},
       eprint = {0907.2952},
 primaryClass = {astro-ph.GA},
       adsurl = {https://ui.adsabs.harvard.edu/abs/2010MNRAS.401..977J}
}

@ARTICLE{2017ApJS..230...15M,
       author = {{Moe}, Maxwell and {Di Stefano}, Rosanne},
        title = "{Mind Your Ps and Qs: The Interrelation between Period (P) and Mass-ratio (Q) Distributions of Binary Stars}",
      journal = {\apjs},
         year = 2017,
        month = jun,
       volume = {230},
       number = {2},
          eid = {15},
        pages = {15},
          doi = {10.3847/1538-4365/aa6fb6},
archivePrefix = {arXiv},
       eprint = {1606.05347},
 primaryClass = {astro-ph.SR},
       adsurl = {https://ui.adsabs.harvard.edu/abs/2017ApJS..230...15M}
}

@ARTICLE{2012ApJ...760...99P,
       author = {{Perets}, Hagai B. and {Kratter}, Kaitlin M.},
        title = "{The Triple Evolution Dynamical Instability: Stellar Collisions in the Field and the Formation of Exotic Binaries}",
      journal = {\apj},
         year = 2012,
        month = dec,
       volume = {760},
       number = {2},
          eid = {99},
        pages = {99},
          doi = {10.1088/0004-637X/760/2/99},
archivePrefix = {arXiv},
       eprint = {1203.2914},
 primaryClass = {astro-ph.SR},
       adsurl = {https://ui.adsabs.harvard.edu/abs/2012ApJ...760...99P}
}

@ARTICLE{2019MNRAS.483..901P,
       author = {{Perpiny{\`a}-Vall{\`e}s}, M. and {Rebassa-Mansergas}, A. and {G{\"a}nsicke}, B.~T. and {Toonen}, S. and {Hermes}, J.~J. and {Gentile Fusillo}, N.~P. and {Tremblay}, P. -E.},
        title = "{Discovery of the first resolved triple white dwarf}",
      journal = {\mnras},
         year = 2019,
        month = feb,
       volume = {483},
       number = {1},
        pages = {901-907},
          doi = {10.1093/mnras/sty3149},
archivePrefix = {arXiv},
       eprint = {1811.07752},
 primaryClass = {astro-ph.SR},
       adsurl = {https://ui.adsabs.harvard.edu/abs/2019MNRAS.483..901P}
}

@ARTICLE{1964PhRv..136.1224P,
       author = {{Peters}, P.~C.},
        title = "{Gravitational Radiation and the Motion of Two Point Masses}",
      journal = {Physical Review},
         year = 1964,
        month = nov,
       volume = {136},
       number = {4B},
        pages = {1224-1232},
          doi = {10.1103/PhysRev.136.B1224},
       adsurl = {https://ui.adsabs.harvard.edu/abs/1964PhRv..136.1224P}
}

@ARTICLE{1996A&A...309..179P,
       author = {{Portegies Zwart}, S.~F. and {Verbunt}, F.},
        title = "{Population synthesis of high-mass binaries.}",
      journal = {\aap},
         year = 1996,
        month = may,
       volume = {309},
        pages = {179-196},
       adsurl = {https://ui.adsabs.harvard.edu/abs/1996A&A...309..179P}
}

@ARTICLE{2010ApJS..190....1R,
       author = {{Raghavan}, Deepak and {McAlister}, Harold A. and {Henry}, Todd J. and {Latham}, David W. and {Marcy}, Geoffrey W. and {Mason}, Brian D. and {Gies}, Douglas R. and {White}, Russel J. and {ten Brummelaar}, Theo A.},
        title = "{A Survey of Stellar Families: Multiplicity of Solar-type Stars}",
      journal = {\apjs},
         year = 2010,
        month = sep,
       volume = {190},
       number = {1},
        pages = {1-42},
          doi = {10.1088/0067-0049/190/1/1},
archivePrefix = {arXiv},
       eprint = {1007.0414},
 primaryClass = {astro-ph.SR},
       adsurl = {https://ui.adsabs.harvard.edu/abs/2010ApJS..190....1R}
}

@ARTICLE{2013ApJ...766...64S,
       author = {{Shappee}, Benjamin J. and {Thompson}, Todd A.},
        title = "{The Mass-loss-induced Eccentric Kozai Mechanism: A New Channel for the Production of Close Compact Object-Stellar Binaries}",
      journal = {\apj},
         year = 2013,
        month = mar,
       volume = {766},
       number = {1},
          eid = {64},
        pages = {64},
          doi = {10.1088/0004-637X/766/1/64},
archivePrefix = {arXiv},
       eprint = {1204.1053},
 primaryClass = {astro-ph.SR},
       adsurl = {https://ui.adsabs.harvard.edu/abs/2013ApJ...766...64S}
}

@ARTICLE{2014AJ....147...87T,
       author = {{Tokovinin}, Andrei},
        title = "{From Binaries to Multiples. II. Hierarchical Multiplicity of F and G Dwarfs}",
      journal = {\aj},
         year = 2014,
        month = apr,
       volume = {147},
       number = {4},
          eid = {87},
        pages = {87},
          doi = {10.1088/0004-6256/147/4/87},
archivePrefix = {arXiv},
       eprint = {1401.6827},
 primaryClass = {astro-ph.SR},
       adsurl = {https://ui.adsabs.harvard.edu/abs/2014AJ....147...87T}
}

@ARTICLE{2014A&A...562A..14T,
       author = {{Toonen}, S. and {Claeys}, J.~S.~W. and {Mennekens}, N. and {Ruiter}, A.~J.},
        title = "{PopCORN: Hunting down the differences between binary population synthesis codes}",
      journal = {\aap},
         year = 2014,
        month = feb,
       volume = {562},
          eid = {A14},
        pages = {A14},
          doi = {10.1051/0004-6361/201321576},
archivePrefix = {arXiv},
       eprint = {1311.6503},
 primaryClass = {astro-ph.SR},
       adsurl = {https://ui.adsabs.harvard.edu/abs/2014A&A...562A..14T}
}

@ARTICLE{2022ApJ...927L..31R,
       author = {{Rebassa-Mansergas}, Alberto and {Xu}, Siyi and {Raddi}, Roberto and {Pala}, Anna F. and {Solano}, Enrique and {Torres}, Santiago and {Jim{\'e}nez-Esteban}, Francisco and {Cruz}, Patricia},
        title = "{Gaia 0007-1605: An Old Triple System with an Inner Brown Dwarf-White Dwarf Binary and an Outer White Dwarf Companion}",
      journal = {\apjl},
         year = 2022,
        month = mar,
       volume = {927},
       number = {2},
          eid = {L31},
        pages = {L31},
          doi = {10.3847/2041-8213/ac5a55},
archivePrefix = {arXiv},
       eprint = {2203.05901},
 primaryClass = {astro-ph.SR},
       adsurl = {https://ui.adsabs.harvard.edu/abs/2022ApJ...927L..31R}
}

@ARTICLE{2017A&A...598L...7K,
       author = {{Kervella}, P. and {Th{\'e}venin}, F. and {Lovis}, C.},
        title = "{Proxima's orbit around {\ensuremath{\alpha}} Centauri}",
      journal = {\aap},
         year = 2017,
        month = feb,
       volume = {598},
          eid = {L7},
        pages = {L7},
          doi = {10.1051/0004-6361/201629930},
archivePrefix = {arXiv},
       eprint = {1611.03495},
 primaryClass = {astro-ph.SR},
       adsurl = {https://ui.adsabs.harvard.edu/abs/2017A&A...598L...7K}
}

@ARTICLE{2021A&A...653A..40L,
       author = {{Lillo-Box}, J. and {Ribas}, {\'A}. and {Montesinos}, B. and {Santos}, N.~C. and {Campante}, T. and {Cunha}, M. and {Barrado}, D. and {Villaver}, E. and {Sousa}, S. and {Bouy}, H. and {Aller}, A. and {Corsaro}, E. and {Li}, T. and {Ong}, J.~M.~J. and {Rebollido}, I. and {Audenaert}, J. and {Pereira}, F.},
        title = "{Uncovering the ultimate planet impostor. An eclipsing brown dwarf in a hierarchical triple with two evolved stars}",
      journal = {\aap},
         year = 2021,
        month = sep,
       volume = {653},
          eid = {A40},
        pages = {A40},
          doi = {10.1051/0004-6361/202141158},
archivePrefix = {arXiv},
       eprint = {2106.05011},
 primaryClass = {astro-ph.EP},
       adsurl = {https://ui.adsabs.harvard.edu/abs/2021A&A...653A..40L}
}

@ARTICLE{2020A&A...638A.131N,
       author = {{Napiwotzki}, R. and {Karl}, C.~A. and {Lisker}, T. and {Catal{\'a}n}, S. and {Drechsel}, H. and {Heber}, U. and {Homeier}, D. and {Koester}, D. and {Leibundgut}, B. and {Marsh}, T.~R. and {Moehler}, S. and {Nelemans}, G. and {Reimers}, D. and {Renzini}, A. and {Str{\"o}er}, A. and {Yungelson}, L.},
        title = "{The ESO supernovae type Ia progenitor survey (SPY). The radial velocities of 643 DA white dwarfs}",
      journal = {\aap},
         year = 2020,
        month = jun,
       volume = {638},
          eid = {A131},
        pages = {A131},
          doi = {10.1051/0004-6361/201629648},
archivePrefix = {arXiv},
       eprint = {1906.10977},
 primaryClass = {astro-ph.SR},
       adsurl = {https://ui.adsabs.harvard.edu/abs/2020A&A...638A.131N}
}

@ARTICLE{2014Natur.505..520R,
       author = {{Ransom}, S.~M. and {Stairs}, I.~H. and {Archibald}, A.~M. and {Hessels}, J.~W.~T. and {Kaplan}, D.~L. and {van Kerkwijk}, M.~H. and {Boyles}, J. and {Deller}, A.~T. and {Chatterjee}, S. and {Schechtman-Rook}, A. and {Berndsen}, A. and {Lynch}, R.~S. and {Lorimer}, D.~R. and {Karako-Argaman}, C. and {Kaspi}, V.~M. and {Kondratiev}, V.~I. and {McLaughlin}, M.~A. and {van Leeuwen}, J. and {Rosen}, R. and {Roberts}, M.~S.~E. and {Stovall}, K.},
        title = "{A millisecond pulsar in a stellar triple system}",
      journal = {\nat},
         year = 2014,
        month = jan,
       volume = {505},
       number = {7484},
        pages = {520-524},
          doi = {10.1038/nature12917},
archivePrefix = {arXiv},
       eprint = {1401.0535},
 primaryClass = {astro-ph.SR},
       adsurl = {https://ui.adsabs.harvard.edu/abs/2014Natur.505..520R}
}

@ARTICLE{2020NatAs...4..650T,
       author = {{Triaud}, Amaury H.~M.~J. and {Burgasser}, Adam J. and {Burdanov}, Artem and {Kunovac Hod{\v{z}}i{\'c}}, Vedad and {Alonso}, Roi and {Bardalez Gagliuffi}, Daniella and {Delrez}, Laetitia and {Demory}, Brice-Olivier and {de Wit}, Julien and {Ducrot}, Elsa and {Hessman}, Frederic V. and {Husser}, Tim-Oliver and {Jehin}, Emmanu{\"e}l and {Pedersen}, Peter P. and {Queloz}, Didier and {McCormac}, James and {Murray}, Catriona and {Sebastian}, Daniel and {Thompson}, Samantha and {Van Grootel}, Val{\'e}rie and {Gillon}, Micha{\"e}l},
        title = "{An eclipsing substellar binary in a young triple system discovered by SPECULOOS}",
      journal = {Nature Astronomy},
         year = 2020,
        month = mar,
       volume = {4},
        pages = {650-657},
          doi = {10.1038/s41550-020-1018-2},
archivePrefix = {arXiv},
       eprint = {2001.07175},
 primaryClass = {astro-ph.SR},
       adsurl = {https://ui.adsabs.harvard.edu/abs/2020NatAs...4..650T}
}

@ARTICLE{2018MNRAS.476.2584M,
       author = {{Maoz}, Dan and {Hallakoun}, Na'ama and {Badenes}, Carles},
        title = "{The separation distribution and merger rate of double white dwarfs: improved constraints}",
      journal = {\mnras},
         year = 2018,
        month = may,
       volume = {476},
       number = {2},
        pages = {2584-2590},
          doi = {10.1093/mnras/sty339},
archivePrefix = {arXiv},
       eprint = {1801.04275},
 primaryClass = {astro-ph.SR},
       adsurl = {https://ui.adsabs.harvard.edu/abs/2018MNRAS.476.2584M}
}

@ARTICLE{2022MNRAS.511.5936K,
       author = {{Korol}, Valeriya and {Hallakoun}, Na'ama and {Toonen}, Silvia and {Karnesis}, Nikolaos},
        title = "{Observationally driven Galactic double white dwarf population for LISA}",
      journal = {\mnras},
         year = 2022,
        month = apr,
       volume = {511},
       number = {4},
        pages = {5936-5947},
          doi = {10.1093/mnras/stac415},
archivePrefix = {arXiv},
       eprint = {2109.10972},
 primaryClass = {astro-ph.HE},
       adsurl = {https://ui.adsabs.harvard.edu/abs/2022MNRAS.511.5936K}
}

@ARTICLE{2023LRR....26....2A,
       author = {{Amaro-Seoane}, Pau and {Andrews}, Jeff and {Arca Sedda}, Manuel and {Askar}, Abbas and {Baghi}, Quentin and {Balasov}, Razvan and {Bartos}, Imre and {Bavera}, Simone S. and {Bellovary}, Jillian and {Berry}, Christopher P.~L. and {Berti}, Emanuele and {Bianchi}, Stefano and {Blecha}, Laura and {Blondin}, St{\'e}phane and {Bogdanovi{\'c}}, Tamara and {Boissier}, Samuel and {Bonetti}, Matteo and {Bonoli}, Silvia and {Bortolas}, Elisa and {Breivik}, Katelyn and {Capelo}, Pedro R. and {Caramete}, Laurentiu and {Cattorini}, Federico and {Charisi}, Maria and {Chaty}, Sylvain and {Chen}, Xian and {Chru{\'s}li{\'n}ska}, Martyna and {Chua}, Alvin J.~K. and {Church}, Ross and {Colpi}, Monica and {D'Orazio}, Daniel and {Danielski}, Camilla and {Davies}, Melvyn B. and {Dayal}, Pratika and {De Rosa}, Alessandra and {Derdzinski}, Andrea and {Destounis}, Kyriakos and {Dotti}, Massimo and {Dutan}, Ioana and {Dvorkin}, Irina and {Fabj}, Gaia and {Foglizzo}, Thierry and {Ford}, Saavik and {Fouvry}, Jean-Baptiste and {Franchini}, Alessia and {Fragos}, Tassos and {Fryer}, Chris and {Gaspari}, Massimo and {Gerosa}, Davide and {Graziani}, Luca and {Groot}, Paul and {Habouzit}, Melanie and {Haggard}, Daryl and {Haiman}, Zoltan and {Han}, Wen-Biao and {Istrate}, Alina and {Johansson}, Peter H. and {Khan}, Fazeel Mahmood and {Kimpson}, Tomas and {Kokkotas}, Kostas and {Kong}, Albert and {Korol}, Valeriya and {Kremer}, Kyle and {Kupfer}, Thomas and {Lamberts}, Astrid and {Larson}, Shane and {Lau}, Mike and {Liu}, Dongliang and {Lloyd-Ronning}, Nicole and {Lodato}, Giuseppe and {Lupi}, Alessandro and {Ma}, Chung-Pei and {Maccarone}, Tomas and {Mandel}, Ilya and {Mangiagli}, Alberto and {Mapelli}, Michela and {Mathis}, St{\'e}phane and {Mayer}, Lucio and {McGee}, Sean and {McKernan}, Barry and {Miller}, M. Coleman and {Mota}, David F. and {Mumpower}, Matthew and {Nasim}, Syeda S. and {Nelemans}, Gijs and {Noble}, Scott and {Pacucci}, Fabio and {Panessa}, Francesca and {Paschalidis}, Vasileios and {Pfister}, Hugo and {Porquet}, Delphine and {Quenby}, John and {Ricarte}, Angelo and {R{\"o}pke}, Friedrich K. and {Regan}, John and {Rosswog}, Stephan and {Ruiter}, Ashley and {Ruiz}, Milton and {Runnoe}, Jessie and {Schneider}, Raffaella and {Schnittman}, Jeremy and {Secunda}, Amy and {Sesana}, Alberto and {Seto}, Naoki and {Shao}, Lijing and {Shapiro}, Stuart and {Sopuerta}, Carlos and {Stone}, Nicholas C. and {Suvorov}, Arthur and {Tamanini}, Nicola and {Tamfal}, Tomas and {Tauris}, Thomas and {Temmink}, Karel and {Tomsick}, John and {Toonen}, Silvia and {Torres-Orjuela}, Alejandro and {Toscani}, Martina and {Tsokaros}, Antonios and {Unal}, Caner and {V{\'a}zquez-Aceves}, Ver{\'o}nica and {Valiante}, Rosa and {van Putten}, Maurice and {van Roestel}, Jan and {Vignali}, Christian and {Volonteri}, Marta and {Wu}, Kinwah and {Younsi}, Ziri and {Yu}, Shenghua and {Zane}, Silvia and {Zwick}, Lorenz and {Antonini}, Fabio and {Baibhav}, Vishal and {Barausse}, Enrico and {Bonilla Rivera}, Alexander and {Branchesi}, Marica and {Branduardi-Raymont}, Graziella and {Burdge}, Kevin and {Chakraborty}, Srija and {Cuadra}, Jorge and {Dage}, Kristen and {Davis}, Benjamin and {de Mink}, Selma E. and {Decarli}, Roberto and {Doneva}, Daniela and {Escoffier}, Stephanie and {Gandhi}, Poshak and {Haardt}, Francesco and {Lousto}, Carlos O. and {Nissanke}, Samaya and {Nordhaus}, Jason and {O'Shaughnessy}, Richard and {Portegies Zwart}, Simon and {Pound}, Adam and {Schussler}, Fabian and {Sergijenko}, Olga and {Spallicci}, Alessandro and {Vernieri}, Daniele and {Vigna-G{\'o}mez}, Alejandro},
        title = "{Astrophysics with the Laser Interferometer Space Antenna}",
      journal = {Living Reviews in Relativity},
         year = 2023,
        month = dec,
       volume = {26},
       number = {1},
          eid = {2},
        pages = {2},
          doi = {10.1007/s41114-022-00041-y},
archivePrefix = {arXiv},
       eprint = {2203.06016},
 primaryClass = {gr-qc},
       adsurl = {https://ui.adsabs.harvard.edu/abs/2023LRR....26....2A}
}

@ARTICLE{2017MNRAS.470.1894K,
       author = {{Korol}, Valeriya and {Rossi}, Elena M. and {Groot}, Paul J. and {Nelemans}, Gijs and {Toonen}, Silvia and {Brown}, Anthony G.~A.},
        title = "{Prospects for detection of detached double white dwarf binaries with Gaia, LSST and LISA}",
      journal = {\mnras},
         year = 2017,
        month = sep,
       volume = {470},
       number = {2},
        pages = {1894-1910},
          doi = {10.1093/mnras/stx1285},
archivePrefix = {arXiv},
       eprint = {1703.02555},
 primaryClass = {astro-ph.HE},
       adsurl = {https://ui.adsabs.harvard.edu/abs/2017MNRAS.470.1894K}
}

@ARTICLE{2019MNRAS.490.5888L,
       author = {{Lamberts}, Astrid and {Blunt}, Sarah and {Littenberg}, Tyson B. and {Garrison-Kimmel}, Shea and {Kupfer}, Thomas and {Sanderson}, Robyn E.},
        title = "{Predicting the LISA white dwarf binary population in the Milky Way with cosmological simulations}",
      journal = {\mnras},
         year = 2019,
        month = dec,
       volume = {490},
       number = {4},
        pages = {5888-5903},
          doi = {10.1093/mnras/stz2834},
archivePrefix = {arXiv},
       eprint = {1907.00014},
 primaryClass = {astro-ph.HE},
       adsurl = {https://ui.adsabs.harvard.edu/abs/2019MNRAS.490.5888L}
}

@ARTICLE{2023ApJ...950....9R,
       author = {{Rajamuthukumar}, Abinaya Swaruba and {Hamers}, Adrian S. and {Neunteufel}, Patrick and {Pakmor}, R{\"u}diger and {de Mink}, Selma E.},
        title = "{Triple Evolution: An Important Channel in the Formation of Type Ia Supernovae}",
      journal = {\apj},
         year = 2023,
        month = jun,
       volume = {950},
       number = {1},
          eid = {9},
        pages = {9},
          doi = {10.3847/1538-4357/acc86c},
archivePrefix = {arXiv},
       eprint = {2211.04463},
 primaryClass = {astro-ph.SR},
       adsurl = {https://ui.adsabs.harvard.edu/abs/2023ApJ...950....9R}
}

@ARTICLE{2020A&A...640A..16T,
       author = {{Toonen}, S. and {Portegies Zwart}, S. and {Hamers}, A.~S. and {Bandopadhyay}, D.},
        title = "{The evolution of stellar triples. The most common evolutionary pathways}",
      journal = {\aap},
         year = 2020,
        month = aug,
       volume = {640},
          eid = {A16},
        pages = {A16},
          doi = {10.1051/0004-6361/201936835},
archivePrefix = {arXiv},
       eprint = {2004.07848},
 primaryClass = {astro-ph.SR},
       adsurl = {https://ui.adsabs.harvard.edu/abs/2020A&A...640A..16T}
}

@INPROCEEDINGS{2023ASPC..534..275O,
       author = {{Offner}, S.~S.~R. and {Moe}, M. and {Kratter}, K.~M. and {Sadavoy}, S.~I. and {Jensen}, E.~L.~N. and {Tobin}, J.~J.},
        title = "{The Origin and Evolution of Multiple Star Systems}",
    booktitle = {Protostars and Planets VII},
         year = 2023,
       editor = {{Inutsuka}, S. and {Aikawa}, Y. and {Muto}, T. and {Tomida}, K. and {Tamura}, M.},
       series = {Astronomical Society of the Pacific Conference Series},
       volume = {534},
        month = jul,
        pages = {275},
          doi = {10.48550/arXiv.2203.10066},
archivePrefix = {arXiv},
       eprint = {2203.10066},
 primaryClass = {astro-ph.SR},
       adsurl = {https://ui.adsabs.harvard.edu/abs/2023ASPC..534..275O}
}

@ARTICLE{2019ApJ...872..119H,
       author = {{Hamers}, Adrian S. and {Dosopoulou}, Fani},
        title = "{An Analytic Model for Mass Transfer in Binaries with Arbitrary Eccentricity, with Applications to Triple-star Systems}",
      journal = {\apj},
         year = 2019,
        month = feb,
       volume = {872},
       number = {2},
          eid = {119},
        pages = {119},
          doi = {10.3847/1538-4357/ab001d},
archivePrefix = {arXiv},
       eprint = {1812.05624},
 primaryClass = {astro-ph.SR},
       adsurl = {https://ui.adsabs.harvard.edu/abs/2019ApJ...872..119H}
}

@ARTICLE{2022ApJS..259...25H,
       author = {{Hamers}, Adrian S. and {Glanz}, Hila and {Neunteufel}, Patrick},
        title = "{A Statistical View of the Stable and Unstable Roche Lobe Overflow of a Tertiary Star onto the Inner Binary in Triple Systems}",
      journal = {\apjs},
         year = 2022,
        month = mar,
       volume = {259},
       number = {1},
          eid = {25},
        pages = {25},
          doi = {10.3847/1538-4365/ac49e7},
archivePrefix = {arXiv},
       eprint = {2110.00024},
 primaryClass = {astro-ph.SR},
       adsurl = {https://ui.adsabs.harvard.edu/abs/2022ApJS..259...25H}
}

@ARTICLE{2001MNRAS.322..231K,
       author = {{Kroupa}, Pavel},
        title = "{On the variation of the initial mass function}",
      journal = {\mnras},
         year = 2001,
        month = apr,
       volume = {322},
       number = {2},
        pages = {231-246},
          doi = {10.1046/j.1365-8711.2001.04022.x},
archivePrefix = {arXiv},
       eprint = {astro-ph/0009005},
 primaryClass = {astro-ph},
       adsurl = {https://ui.adsabs.harvard.edu/abs/2001MNRAS.322..231K}
}

@ARTICLE{2021MNRAS.502.4479H,
       author = {{Hamers}, Adrian S. and {Rantala}, Antti and {Neunteufel}, Patrick and {Preece}, Holly and {Vynatheya}, Pavan},
        title = "{Multiple Stellar Evolution: a population synthesis algorithm to model the stellar, binary, and dynamical evolution of multiple-star systems}",
      journal = {\mnras},
         year = 2021,
        month = apr,
       volume = {502},
       number = {3},
        pages = {4479-4512},
          doi = {10.1093/mnras/stab287},
archivePrefix = {arXiv},
       eprint = {2011.04513},
 primaryClass = {astro-ph.SR},
       adsurl = {https://ui.adsabs.harvard.edu/abs/2021MNRAS.502.4479H}
}

@ARTICLE{2020MNRAS.492.4131R,
       author = {{Rantala}, Antti and {Pihajoki}, Pauli and {Mannerkoski}, Matias and {Johansson}, Peter H. and {Naab}, Thorsten},
        title = "{MSTAR - a fast parallelized algorithmically regularized integrator with minimum spanning tree coordinates}",
      journal = {\mnras},
         year = 2020,
        month = mar,
       volume = {492},
       number = {3},
        pages = {4131-4148},
          doi = {10.1093/mnras/staa084},
archivePrefix = {arXiv},
       eprint = {2001.03180},
 primaryClass = {astro-ph.IM},
       adsurl = {https://ui.adsabs.harvard.edu/abs/2020MNRAS.492.4131R}
}

@ARTICLE{2022MNRAS.516.4146V,
       author = {{Vynatheya}, Pavan and {Hamers}, Adrian S. and {Mardling}, Rosemary A. and {Bellinger}, Earl P.},
        title = "{Algebraic and machine learning approach to hierarchical triple-star stability}",
      journal = {\mnras},
         year = 2022,
        month = nov,
       volume = {516},
       number = {3},
        pages = {4146-4155},
          doi = {10.1093/mnras/stac2540},
archivePrefix = {arXiv},
       eprint = {2207.03151},
 primaryClass = {astro-ph.SR},
       adsurl = {https://ui.adsabs.harvard.edu/abs/2022MNRAS.516.4146V}
}

@ARTICLE{2019MNRAS.490.3234N,
       author = {{Nelson}, Dylan and {Pillepich}, Annalisa and {Springel}, Volker and {Pakmor}, R{\"u}diger and {Weinberger}, Rainer and {Genel}, Shy and {Torrey}, Paul and {Vogelsberger}, Mark and {Marinacci}, Federico and {Hernquist}, Lars},
        title = "{First results from the TNG50 simulation: galactic outflows driven by supernovae and black hole feedback}",
      journal = {\mnras},
         year = 2019,
        month = dec,
       volume = {490},
       number = {3},
        pages = {3234-3261},
          doi = {10.1093/mnras/stz2306},
archivePrefix = {arXiv},
       eprint = {1902.05554},
 primaryClass = {astro-ph.GA},
       adsurl = {https://ui.adsabs.harvard.edu/abs/2019MNRAS.490.3234N}
}

@ARTICLE{2023MNRAS.518.5123M,
       author = {{Munday}, James and {Marsh}, T.~R. and {Hollands}, Mark and {Pelisoli}, Ingrid and {Steeghs}, Danny and {Hakala}, Pasi and {Breedt}, Elm{\'e} and {Brown}, Alex and {Dhillon}, V.~S. and {Dyer}, Martin J. and {Green}, Matthew and {Kerry}, Paul and {Littlefair}, S.~P. and {Parsons}, Steven G. and {Sahman}, Dave and {Somjit}, Sorawit and {Sukaum}, Boonchoo and {Wild}, James},
        title = "{Two decades of optical timing of the shortest-period binary star system HM Cancri}",
      journal = {\mnras},
         year = 2023,
        month = feb,
       volume = {518},
       number = {4},
        pages = {5123-5139},
          doi = {10.1093/mnras/stac3385},
archivePrefix = {arXiv},
       eprint = {2211.09834},
 primaryClass = {astro-ph.SR},
       adsurl = {https://ui.adsabs.harvard.edu/abs/2023MNRAS.518.5123M}
}

@ARTICLE{1984ApJS...54..335I,
       author = {{Iben}, I., Jr. and {Tutukov}, A.~V.},
        title = "{Supernovae of type I as end products of the evolution of binaries with components of moderate initial mass.}",
      journal = {\apjs},
         year = 1984,
        month = feb,
       volume = {54},
        pages = {335-372},
          doi = {10.1086/190932},
       adsurl = {https://ui.adsabs.harvard.edu/abs/1984ApJS...54..335I}
}

@ARTICLE{2016ARA&A..54..529B,
       author = {{Bland-Hawthorn}, Joss and {Gerhard}, Ortwin},
        title = "{The Galaxy in Context: Structural, Kinematic, and Integrated Properties}",
      journal = {\araa},
         year = 2016,
        month = sep,
       volume = {54},
        pages = {529-596},
          doi = {10.1146/annurev-astro-081915-023441},
archivePrefix = {arXiv},
       eprint = {1602.07702},
 primaryClass = {astro-ph.GA},
       adsurl = {https://ui.adsabs.harvard.edu/abs/2016ARA&A..54..529B}
}

@ARTICLE{2014MNRAS.438.1909D,
       author = {{de Vries}, N. and {Portegies Zwart}, S. and {Figueira}, J.},
        title = "{The evolution of triples with a Roche lobe filling outer star}",
      journal = {\mnras},
         year = 2014,
        month = mar,
       volume = {438},
       number = {3},
        pages = {1909-1921},
          doi = {10.1093/mnras/stt1688},
archivePrefix = {arXiv},
       eprint = {1309.1475},
 primaryClass = {astro-ph.SR},
       adsurl = {https://ui.adsabs.harvard.edu/abs/2014MNRAS.438.1909D}
}

@ARTICLE{2010ApJ...717.1006R,
       author = {{Ruiter}, Ashley J. and {Belczynski}, Krzysztof and {Benacquista}, Matthew and {Larson}, Shane L. and {Williams}, Gabriel},
        title = "{The LISA Gravitational Wave Foreground: A Study of Double White Dwarfs}",
      journal = {\apj},
         year = 2010,
        month = jul,
       volume = {717},
       number = {2},
        pages = {1006-1021},
          doi = {10.1088/0004-637X/717/2/1006},
archivePrefix = {arXiv},
       eprint = {0705.3272},
 primaryClass = {astro-ph},
       adsurl = {https://ui.adsabs.harvard.edu/abs/2010ApJ...717.1006R}
}

@ARTICLE{2010A&A...521A..85Y,
       author = {{Yu}, S. and {Jeffery}, C.~S.},
        title = "{The gravitational wave signal from diverse populations of double white dwarf binaries in the Galaxy}",
      journal = {\aap},
         year = 2010,
        month = oct,
       volume = {521},
          eid = {A85},
        pages = {A85},
          doi = {10.1051/0004-6361/201014827},
archivePrefix = {arXiv},
       eprint = {1007.4267},
 primaryClass = {astro-ph.SR},
       adsurl = {https://ui.adsabs.harvard.edu/abs/2010A&A...521A..85Y}
}

@ARTICLE{2019Msngr.175....3D,
       author = {{de Jong}, R.~S. and {Agertz}, O. and {Berbel}, A.~A. and {Aird}, J. and {Alexander}, D.~A. and {Amarsi}, A. and {Anders}, F. and {Andrae}, R. and {Ansarinejad}, B. and {Ansorge}, W. and {Antilogus}, P. and {Anwand-Heerwart}, H. and {Arentsen}, A. and {Arnadottir}, A. and {Asplund}, M. and {Auger}, M. and {Azais}, N. and {Baade}, D. and {Baker}, G. and {Baker}, S. and {Balbinot}, E. and {Baldry}, I.~K. and {Banerji}, M. and {Barden}, S. and {Barklem}, P. and {Barth{\'e}l{\'e}my-Mazot}, E. and {Battistini}, C. and {Bauer}, S. and {Bell}, C.~P.~M. and {Bellido-Tirado}, O. and {Bellstedt}, S. and {Belokurov}, V. and {Bensby}, T. and {Bergemann}, M. and {Bestenlehner}, J.~M. and {Bielby}, R. and {Bilicki}, M. and {Blake}, C. and {Bland-Hawthorn}, J. and {Boeche}, C. and {Boland}, W. and {Boller}, T. and {Bongard}, S. and {Bongiorno}, A. and {Bonifacio}, P. and {Boudon}, D. and {Brooks}, D. and {Brown}, M.~J.~I. and {Brown}, R. and {Br{\"u}ggen}, M. and {Brynnel}, J. and {Brzeski}, J. and {Buchert}, T. and {Buschkamp}, P. and {Caffau}, E. and {Caillier}, P. and {Carrick}, J. and {Casagrande}, L. and {Case}, S. and {Casey}, A. and {Cesarini}, I. and {Cescutti}, G. and {Chapuis}, D. and {Chiappini}, C. and {Childress}, M. and {Christlieb}, N. and {Church}, R. and {Cioni}, M. -R.~L. and {Cluver}, M. and {Colless}, M. and {Collett}, T. and {Comparat}, J. and {Cooper}, A. and {Couch}, W. and {Courbin}, F. and {Croom}, S. and {Croton}, D. and {Daguis{\'e}}, E. and {Dalton}, G. and {Davies}, L.~J.~M. and {Davis}, T. and {de Laverny}, P. and {Deason}, A. and {Dionies}, F. and {Disseau}, K. and {Doel}, P. and {D{\"o}scher}, D. and {Driver}, S.~P. and {Dwelly}, T. and {Eckert}, D. and {Edge}, A. and {Edvardsson}, B. and {Youssoufi}, D.~E. and {Elhaddad}, A. and {Enke}, H. and {Erfanianfar}, G. and {Farrell}, T. and {Fechner}, T. and {Feiz}, C. and {Feltzing}, S. and {Ferreras}, I. and {Feuerstein}, D. and {Feuillet}, D. and {Finoguenov}, A. and {Ford}, D. and {Fotopoulou}, S. and {Fouesneau}, M. and {Frenk}, C. and {Frey}, S. and {Gaessler}, W. and {Geier}, S. and {Gentile Fusillo}, N. and {Gerhard}, O. and {Giannantonio}, T. and {Giannone}, D. and {Gibson}, B. and {Gillingham}, P. and {Gonz{\'a}lez-Fern{\'a}ndez}, C. and {Gonzalez-Solares}, E. and {Gottloeber}, S. and {Gould}, A. and {Grebel}, E.~K. and {Gueguen}, A. and {Guiglion}, G. and {Haehnelt}, M. and {Hahn}, T. and {Hansen}, C.~J. and {Hartman}, H. and {Hauptner}, K. and {Hawkins}, K. and {Haynes}, D. and {Haynes}, R. and {Heiter}, U. and {Helmi}, A. and {Aguayo}, C.~H. and {Hewett}, P. and {Hinton}, S. and {Hobbs}, D. and {Hoenig}, S. and {Hofman}, D. and {Hook}, I. and {Hopgood}, J. and {Hopkins}, A. and {Hourihane}, A. and {Howes}, L. and {Howlett}, C. and {Huet}, T. and {Irwin}, M. and {Iwert}, O. and {Jablonka}, P. and {Jahn}, T. and {Jahnke}, K. and {Jarno}, A. and {Jin}, S. and {Jofre}, P. and {Johl}, D. and {Jones}, D. and {J{\"o}nsson}, H. and {Jordan}, C. and {Karovicova}, I. and {Khalatyan}, A. and {Kelz}, A. and {Kennicutt}, R. and {King}, D. and {Kitaura}, F. and {Klar}, J. and {Klauser}, U. and {Kneib}, J. -P. and {Koch}, A. and {Koposov}, S. and {Kordopatis}, G. and {Korn}, A. and {Kosmalski}, J. and {Kotak}, R. and {Kovalev}, M. and {Kreckel}, K. and {Kripak}, Y. and {Krumpe}, M. and {Kuijken}, K. and {Kunder}, A. and {Kushniruk}, I. and {Lam}, M.~I. and {Lamer}, G. and {Laurent}, F. and {Lawrence}, J. and {Lehmitz}, M. and {Lemasle}, B. and {Lewis}, J. and {Li}, B. and {Lidman}, C. and {Lind}, K. and {Liske}, J. and {Lizon}, J. -L. and {Loveday}, J. and {Ludwig}, H. -G. and {McDermid}, R.~M. and {Maguire}, K. and {Mainieri}, V. and {Mali}, S. and {Mandel}, H. and {Mandel}, K. and {Mannering}, L. and {Martell}, S. and {Martinez Delgado}, D. and {Matijevic}, G. and {McGregor}, H. and {McMahon}, R. and {McMillan}, P. and {Mena}, O. and {Merloni}, A. and {Meyer}, M.~J. and {Michel}, C. and {Micheva}, G. and {Migniau}, J. -E. and {Minchev}, I. and {Monari}, G. and {Muller}, R. and {Murphy}, D. and {Muthukrishna}, D. and {Nandra}, K. and {Navarro}, R. and {Ness}, M. and {Nichani}, V. and {Nichol}, R. and {Nicklas}, H. and {Niederhofer}, F. and {Norberg}, P. and {Obreschkow}, D. and {Oliver}, S. and {Owers}, M. and {Pai}, N. and {Pankratow}, S. and {Parkinson}, D. and {Paschke}, J. and {Paterson}, R. and {Pecontal}, A. and {Parry}, I. and {Phillips}, D. and {Pillepich}, A. and {Pinard}, L. and {Pirard}, J. and {Piskunov}, N. and {Plank}, V. and {Pl{\"u}schke}, D. and {Pons}, E. and {Popesso}, P. and {Power}, C. and {Pragt}, J. and {Pramskiy}, A. and {Pryer}, D. and {Quattri}, M. and {Queiroz}, A.~B. d. A. and {Quirrenbach}, A. and {Rahurkar}, S. and {Raichoor}, A. and {Ramstedt}, S. and {Rau}, A. and {Recio-Blanco}, A. and {Reiss}, R. and {Renaud}, F. and {Revaz}, Y. and {Rhode}, P. and {Richard}, J. and {Richter}, A.~D. and {Rix}, H. -W. and {Robotham}, A.~S.~G. and {Roelfsema}, R. and {Romaniello}, M. and {Rosario}, D. and {Rothmaier}, F. and {Roukema}, B. and {Ruchti}, G. and {Rupprecht}, G. and {Rybizki}, J. and {Ryde}, N. and {Saar}, A. and {Sadler}, E. and {Sahl{\'e}n}, M. and {Salvato}, M. and {Sassolas}, B. and {Saunders}, W. and {Saviauk}, A. and {Sbordone}, L. and {Schmidt}, T. and {Schnurr}, O. and {Scholz}, R. -D. and {Schwope}, A. and {Seifert}, W. and {Shanks}, T. and {Sheinis}, A. and {Sivov}, T. and {Sk{\'u}lad{\'o}ttir}, {\'A}. and {Smartt}, S. and {Smedley}, S. and {Smith}, G. and {Smith}, R. and {Sorce}, J. and {Spitler}, L. and {Starkenburg}, E. and {Steinmetz}, M. and {Stilz}, I. and {Storm}, J. and {Sullivan}, M. and {Sutherland}, W. and {Swann}, E. and {Tamone}, A. and {Taylor}, E.~N. and {Teillon}, J. and {Tempel}, E. and {ter Horst}, R. and {Thi}, W. -F. and {Tolstoy}, E. and {Trager}, S. and {Traven}, G. and {Tremblay}, P. -E. and {Tresse}, L. and {Valentini}, M. and {van de Weygaert}, R. and {van den Ancker}, M. and {Veljanoski}, J. and {Venkatesan}, S. and {Wagner}, L. and {Wagner}, K. and {Walcher}, C.~J. and {Waller}, L. and {Walton}, N. and {Wang}, L. and {Winkler}, R. and {Wisotzki}, L. and {Worley}, C.~C. and {Worseck}, G. and {Xiang}, M. and {Xu}, W. and {Yong}, D. and {Zhao}, C. and {Zheng}, J. and {Zscheyge}, F. and {Zucker}, D.},
        title = "{4MOST: Project overview and information for the First Call for Proposals}",
      journal = {The Messenger},
         year = 2019,
        month = mar,
       volume = {175},
        pages = {3-11},
          doi = {10.18727/0722-6691/5117},
archivePrefix = {arXiv},
       eprint = {1903.02464},
 primaryClass = {astro-ph.IM},
       adsurl = {https://ui.adsabs.harvard.edu/abs/2019Msngr.175....3D}
}

@ARTICLE{2012arXiv1206.3569Z,
       author = {{Zhao}, Gang and {Zhao}, Yongheng and {Chu}, Yaoquan and {Jing}, Yipeng and {Deng}, Licai},
        title = "{LAMOST Spectral Survey}",
      journal = {arXiv e-prints},
         year = 2012,
        month = jun,
          eid = {arXiv:1206.3569},
        pages = {arXiv:1206.3569},
          doi = {10.48550/arXiv.1206.3569},
archivePrefix = {arXiv},
       eprint = {1206.3569},
 primaryClass = {astro-ph.IM},
       adsurl = {https://ui.adsabs.harvard.edu/abs/2012arXiv1206.3569Z}
}

@ARTICLE{2018AJ....156..241H,
       author = {{Heinze}, A.~N. and {Tonry}, J.~L. and {Denneau}, L. and {Flewelling}, H. and {Stalder}, B. and {Rest}, A. and {Smith}, K.~W. and {Smartt}, S.~J. and {Weiland}, H.},
        title = "{A First Catalog of Variable Stars Measured by the Asteroid Terrestrial-impact Last Alert System (ATLAS)}",
      journal = {\aj},
         year = 2018,
        month = nov,
       volume = {156},
       number = {5},
          eid = {241},
        pages = {241},
          doi = {10.3847/1538-3881/aae47f},
archivePrefix = {arXiv},
       eprint = {1804.02132},
 primaryClass = {astro-ph.SR},
       adsurl = {https://ui.adsabs.harvard.edu/abs/2018AJ....156..241H}
}

@ARTICLE{2018PASP..130f4505T,
       author = {{Tonry}, J.~L. and {Denneau}, L. and {Heinze}, A.~N. and {Stalder}, B. and {Smith}, K.~W. and {Smartt}, S.~J. and {Stubbs}, C.~W. and {Weiland}, H.~J. and {Rest}, A.},
        title = "{ATLAS: A High-cadence All-sky Survey System}",
      journal = {\pasp},
         year = 2018,
        month = jun,
       volume = {130},
       number = {988},
        pages = {064505},
          doi = {10.1088/1538-3873/aabadf},
archivePrefix = {arXiv},
       eprint = {1802.00879},
 primaryClass = {astro-ph.IM},
       adsurl = {https://ui.adsabs.harvard.edu/abs/2018PASP..130f4505T}
}

@ARTICLE{2022MNRAS.511.2405S,
       author = {{Steeghs}, D. and {Galloway}, D.~K. and {Ackley}, K. and {Dyer}, M.~J. and {Lyman}, J. and {Ulaczyk}, K. and {Cutter}, R. and {Mong}, Y. -L. and {Dhillon}, V. and {O'Brien}, P. and {Ramsay}, G. and {Poshyachinda}, S. and {Kotak}, R. and {Nuttall}, L.~K. and {Pall{\'e}}, E. and {Breton}, R.~P. and {Pollacco}, D. and {Thrane}, E. and {Aukkaravittayapun}, S. and {Awiphan}, S. and {Burhanudin}, U. and {Chote}, P. and {Chrimes}, A. and {Daw}, E. and {Duffy}, C. and {Eyles-Ferris}, R. and {Gompertz}, B. and {Heikkil{\"a}}, T. and {Irawati}, P. and {Kennedy}, M.~R. and {Killestein}, T. and {Kuncarayakti}, H. and {Levan}, A.~J. and {Littlefair}, S. and {Makrygianni}, L. and {Marsh}, T. and {Mata-Sanchez}, D. and {Mattila}, S. and {Maund}, J. and {McCormac}, J. and {Mkrtichian}, D. and {Mullaney}, J. and {Noysena}, K. and {Patel}, M. and {Rol}, E. and {Sawangwit}, U. and {Stanway}, E.~R. and {Starling}, R. and {Str{\o}m}, P. and {Tooke}, S. and {West}, R. and {White}, D.~J. and {Wiersema}, K.},
        title = "{The Gravitational-wave Optical Transient Observer (GOTO): prototype performance and prospects for transient science}",
      journal = {\mnras},
         year = 2022,
        month = apr,
       volume = {511},
       number = {2},
        pages = {2405-2422},
          doi = {10.1093/mnras/stac013},
archivePrefix = {arXiv},
       eprint = {2110.05539},
 primaryClass = {astro-ph.IM},
       adsurl = {https://ui.adsabs.harvard.edu/abs/2022MNRAS.511.2405S}
}

@ARTICLE{2024ApJ...965..148X,
       author = {{Xuan}, Zeyuan and {Naoz}, Smadar and {Kocsis}, Bence and {Michaely}, Erez},
        title = "{Detecting Gravitational Wave Bursts from Stellar-mass Binaries in the mHz Band}",
      journal = {\apj},
         year = 2024,
        month = apr,
       volume = {965},
       number = {2},
          eid = {148},
        pages = {148},
          doi = {10.3847/1538-4357/ad2c94},
archivePrefix = {arXiv},
       eprint = {2310.00042},
 primaryClass = {astro-ph.HE},
       adsurl = {https://ui.adsabs.harvard.edu/abs/2024ApJ...965..148X}
}

@ARTICLE{kar21,
       author = {{Karnesis}, Nikolaos and {Babak}, Stanislav and {Pieroni}, Mauro and {Cornish}, Neil and {Littenberg}, Tyson},
        title = "{Characterization of the stochastic signal originating from compact binary populations as measured by LISA}",
      journal = {\prd},
         year = 2021,
        month = aug,
       volume = {104},
       number = {4},
          eid = {043019},
        pages = {043019},
          doi = {10.1103/PhysRevD.104.043019},
archivePrefix = {arXiv},
       eprint = {2103.14598},
 primaryClass = {astro-ph.IM},
       adsurl = {https://ui.adsabs.harvard.edu/abs/2021PhRvD.104d3019K}
}

@ARTICLE{cro07,
       author = {{Crowder}, Jeff and {Cornish}, Neil J.},
        title = "{Solution to the galactic foreground problem for LISA}",
      journal = {\prd},
         year = 2007,
        month = feb,
       volume = {75},
       number = {4},
          eid = {043008},
        pages = {043008},
          doi = {10.1103/PhysRevD.75.043008},
archivePrefix = {arXiv},
       eprint = {astro-ph/0611546},
 primaryClass = {astro-ph},
       adsurl = {https://ui.adsabs.harvard.edu/abs/2007PhRvD..75d3008C}
}

@ARTICLE{nis12,
       author = {{Nissanke}, Samaya and {Vallisneri}, Michele and {Nelemans}, Gijs and {Prince}, Thomas A.},
        title = "{Gravitational-wave Emission from Compact Galactic Binaries}",
      journal = {\apj},
         year = 2012,
        month = oct,
       volume = {758},
       number = {2},
          eid = {131},
        pages = {131},
          doi = {10.1088/0004-637X/758/2/131},
archivePrefix = {arXiv},
       eprint = {1201.4613},
 primaryClass = {astro-ph.GA},
       adsurl = {https://ui.adsabs.harvard.edu/abs/2012ApJ...758..131N}
}

@ARTICLE{tim06,
       author = {{Timpano}, Seth E. and {Rubbo}, Louis J. and {Cornish}, Neil J.},
        title = "{Characterizing the galactic gravitational wave background with LISA}",
      journal = {\prd},
         year = 2006,
        month = jun,
       volume = {73},
       number = {12},
          eid = {122001},
        pages = {122001},
          doi = {10.1103/PhysRevD.73.122001},
archivePrefix = {arXiv},
       eprint = {gr-qc/0504071},
 primaryClass = {gr-qc},
       adsurl = {https://ui.adsabs.harvard.edu/abs/2006PhRvD..73l2001T}
}

@techreport{LISAdoc,
	Author = {{LISA Science Study Team}},
	Month = May,
	Note = {\url{www.cosmos.esa.int/web/lisa/lisa-documents/}},
	Number = {ESA-L3-EST-SCI-RS-001},
	Title = {{LISA} Science Requirements Document},
	Url = {www.cosmos.esa.int/web/lisa/lisa-documents/},
	Institution = {ESA},
	Year = 2018
	}

@ARTICLE{LISAWaveform,
       author = {{LISA Consortium Waveform Working Group} and {Afshordi}, Niayesh and {Ak{\c{c}}ay}, Sarp and {Amaro Seoane}, Pau and {Antonelli}, Andrea and {Aurrekoetxea}, Josu C. and {Barack}, Leor and {Barausse}, Enrico and {Benkel}, Robert and {Bernard}, Laura and {Bernuzzi}, Sebastiano and {Berti}, Emanuele and {Bonetti}, Matteo and {Bonga}, B{\'e}atrice and {Bozzola}, Gabriele and {Brito}, Richard and {Buonanno}, Alessandra and {C{\'a}rdenas-Avenda{\~n}o}, Alejandro and {Casals}, Marc and {Chernoff}, David F. and {Chua}, Alvin J.~K. and {Clough}, Katy and {Colleoni}, Marta and {Dhesi}, Mekhi and {Druart}, Adrien and {Durkan}, Leanne and {Faye}, Guillaume and {Ferguson}, Deborah and {Field}, Scott E. and {Gabella}, William E. and {Garc{\'\i}a-Bellido}, Juan and {Gracia-Linares}, Miguel and {Gerosa}, Davide and {Green}, Stephen R. and {Haney}, Maria and {Hannam}, Mark and {Heffernan}, Anna and {Hinderer}, Tanja and {Helfer}, Thomas and {Hughes}, Scott A. and {Husa}, Sascha and {Isoyama}, Soichiro and {Katz}, Michael L. and {Kavanagh}, Chris and {Khanna}, Gaurav and {Kidder}, Larry E. and {Korol}, Valeriya and {K{\"u}chler}, Lorenzo and {Laguna}, Pablo and {Larrouturou}, Fran{\c{c}}ois and {Le Tiec}, Alexandre and {Leather}, Benjamin and {Lim}, Eugene A. and {Lim}, Hyun and {Littenberg}, Tyson B. and {Long}, Oliver and {Lousto}, Carlos O. and {Lovelace}, Geoffrey and {Lukes-Gerakopoulos}, Georgios and {Lynch}, Philip and {Macedo}, Rodrigo P. and {Markakis}, Charalampos and {Maggio}, Elisa and {Mandel}, Ilya and {Maselli}, Andrea and {Mathews}, Josh and {Mourier}, Pierre and {Neilsen}, David and {Nagar}, Alessandro and {Nichols}, David A. and {Nov{\'a}k}, Jan and {Okounkova}, Maria and {O'Shaughnessy}, Richard and {Oshita}, Naritaka and {O'Toole}, Conor and {Pan}, Zhen and {Pani}, Paolo and {Pappas}, George and {Paschalidis}, Vasileios and {Pfeiffer}, Harald P. and {Pompili}, Lorenzo and {Pound}, Adam and {Pratten}, Geraint and {R{\"u}ter}, Hannes R. and {Ruiz}, Milton and {Sam}, Zeyd and {Sberna}, Laura and {Shapiro}, Stuart L. and {Shoemaker}, Deirdre M. and {Sopuerta}, Carlos F. and {Spiers}, Andrew and {Sundar}, Hari and {Tamanini}, Nicola and {Thompson}, Jonathan E. and {Toubiana}, Alexandre and {Tsokaros}, Antonios and {Upton}, Samuel D. and {van de Meent}, Maarten and {Vernieri}, Daniele and {Wachter}, Jeremy M. and {Warburton}, Niels and {Wardell}, Barry and {Witek}, Helvi and {Witzany}, Vojt{\v{e}}ch and {Yang}, Huan and {Zilh{\~a}o}, Miguel and {Albertini}, Angelica and {Arun}, K.~G. and {Bezares}, Miguel and {Bonilla}, Alexander and {Chapman-Bird}, Christian and {Cownden}, Bradley and {Cunningham}, Kevin and {Devitt}, Chris and {Dolan}, Sam and {Duque}, Francisco and {Dyson}, Conor and {Fryer}, Chris L. and {Gair}, Jonathan R. and {Giacomazzo}, Bruno and {Gupta}, Priti and {Han}, Wen-Biao and {Haas}, Roland and {Hirschmann}, Eric W. and {Huerta}, E.~A. and {Jetzer}, Philippe and {Kelly}, Bernard and {Khalil}, Mohammed and {Lewis}, Jack and {Lloyd-Ronning}, Nicole and {Marsat}, Sylvain and {Nardini}, Germano and {Neef}, Jakob and {Ottewill}, Adrian and {Pantelidou}, Christiana and {Piovano}, Gabriel Andres and {Redondo-Yuste}, Jaime and {Sagunski}, Laura and {Stein}, Leo C. and {Skoup{\'y}}, Viktor and {Sperhake}, Ulrich and {Speri}, Lorenzo and {Spieksma}, Thomas F.~M. and {Stevens}, Chris and {Trestini}, David and {Va{\~n}{\'o}-Vi{\~n}uales}, Alex},
        title = "{Waveform Modelling for the Laser Interferometer Space Antenna}",
      journal = {arXiv e-prints},
         year = 2023,
        month = nov,
          eid = {arXiv:2311.01300},
        pages = {arXiv:2311.01300},
          doi = {10.48550/arXiv.2311.01300},
archivePrefix = {arXiv},
       eprint = {2311.01300},
 primaryClass = {gr-qc},
       adsurl = {https://ui.adsabs.harvard.edu/abs/2023arXiv231101300L}
}

@ARTICLE{LISAproposal,
       author = {{Amaro-Seoane}, Pau and {Audley}, Heather and {Babak}, Stanislav and {Baker}, John and {Barausse}, Enrico and {Bender}, Peter and {Berti}, Emanuele and {Binetruy}, Pierre and {Born}, Michael and {Bortoluzzi}, Daniele and {Camp}, Jordan and {Caprini}, Chiara and {Cardoso}, Vitor and {Colpi}, Monica and {Conklin}, John and {Cornish}, Neil and {Cutler}, Curt and {Danzmann}, Karsten and {Dolesi}, Rita and {Ferraioli}, Luigi and {Ferroni}, Valerio and {Fitzsimons}, Ewan and {Gair}, Jonathan and {Gesa Bote}, Lluis and {Giardini}, Domenico and {Gibert}, Ferran and {Grimani}, Catia and {Halloin}, Hubert and {Heinzel}, Gerhard and {Hertog}, Thomas and {Hewitson}, Martin and {Holley-Bockelmann}, Kelly and {Hollington}, Daniel and {Hueller}, Mauro and {Inchauspe}, Henri and {Jetzer}, Philippe and {Karnesis}, Nikos and {Killow}, Christian and {Klein}, Antoine and {Klipstein}, Bill and {Korsakova}, Natalia and {Larson}, Shane L and {Livas}, Jeffrey and {Lloro}, Ivan and {Man}, Nary and {Mance}, Davor and {Martino}, Joseph and {Mateos}, Ignacio and {McKenzie}, Kirk and {McWilliams}, Sean T and {Miller}, Cole and {Mueller}, Guido and {Nardini}, Germano and {Nelemans}, Gijs and {Nofrarias}, Miquel and {Petiteau}, Antoine and {Pivato}, Paolo and {Plagnol}, Eric and {Porter}, Ed and {Reiche}, Jens and {Robertson}, David and {Robertson}, Norna and {Rossi}, Elena and {Russano}, Giuliana and {Schutz}, Bernard and {Sesana}, Alberto and {Shoemaker}, David and {Slutsky}, Jacob and {Sopuerta}, Carlos F. and {Sumner}, Tim and {Tamanini}, Nicola and {Thorpe}, Ira and {Troebs}, Michael and {Vallisneri}, Michele and {Vecchio}, Alberto and {Vetrugno}, Daniele and {Vitale}, Stefano and {Volonteri}, Marta and {Wanner}, Gudrun and {Ward}, Harry and {Wass}, Peter and {Weber}, William and {Ziemer}, John and {Zweifel}, Peter},
        title = "{Laser Interferometer Space Antenna}",
      journal = {arXiv e-prints},
         year = 2017,
        month = feb,
          eid = {arXiv:1702.00786},
        pages = {arXiv:1702.00786},
          doi = {10.48550/arXiv.1702.00786},
archivePrefix = {arXiv},
       eprint = {1702.00786},
 primaryClass = {astro-ph.IM},
       adsurl = {https://ui.adsabs.harvard.edu/abs/2017arXiv170200786A}
}

@ARTICLE{LISARedBook,
       author = {{Colpi}, Monica and {Danzmann}, Karsten and {Hewitson}, Martin and {Holley-Bockelmann}, Kelly and {Jetzer}, Philippe and {Nelemans}, Gijs and {Petiteau}, Antoine and {Shoemaker}, David and {Sopuerta}, Carlos and {Stebbins}, Robin and {Tanvir}, Nial and {Ward}, Henry and {Weber}, William Joseph and {Thorpe}, Ira and {Daurskikh}, Anna and {Deep}, Atul and {Fern{\'a}ndez N{\'u}{\~n}ez}, Ignacio and {Garc{\'\i}a Marirrodriga}, C{\'e}sar and {Gehler}, Martin and {Halain}, Jean-Philippe and {Jennrich}, Oliver and {Lammers}, Uwe and {Larra{\~n}aga}, Jonan and {Lieser}, Maike and {L{\"u}tzgendorf}, Nora and {Martens}, Waldemar and {Mondin}, Linda and {Piris Ni{\~n}o}, Ana and {Amaro-Seoane}, Pau and {Arca Sedda}, Manuel and {Auclair}, Pierre and {Babak}, Stanislav and {Baghi}, Quentin and {Baibhav}, Vishal and {Baker}, Tessa and {Bayle}, Jean-Baptiste and {Berry}, Christopher and {Berti}, Emanuele and {Boileau}, Guillaume and {Bonetti}, Matteo and {Brito}, Richard and {Buscicchio}, Riccardo and {Calcagni}, Gianluca and {Capelo}, Pedro R. and {Caprini}, Chiara and {Caputo}, Andrea and {Castelli}, Eleonora and {Chen}, Hsin-Yu and {Chen}, Xian and {Chua}, Alvin and {Davies}, Gareth and {Derdzinski}, Andrea and {Domcke}, Valerie Fiona and {Doneva}, Daniela and {Dvorkin}, Irna and {Mar{\'\i}a Ezquiaga}, Jose and {Gair}, Jonathan and {Haiman}, Zoltan and {Harry}, Ian and {Hartwig}, Olaf and {Hees}, Aurelien and {Heffernan}, Anna and {Husa}, Sascha and {Izquierdo-Villalba}, David and {Karnesis}, Nikolaos and {Klein}, Antoine and {Korol}, Valeriya and {Korsakova}, Natalia and {Kupfer}, Thomas and {Laghi}, Danny and {Lamberts}, Astrid and {Larson}, Shane and {Le Jeune}, Maude and {Lewicki}, Marek and {Littenberg}, Tyson and {Madge}, Eric and {Mangiagli}, Alberto and {Marsat}, Sylvain and {Vilchez}, Ivan Martin and {Maselli}, Andrea and {Mathews}, Josh and {van de Meent}, Maarten and {Muratore}, Martina and {Nardini}, Germano and {Pani}, Paolo and {Peloso}, Marco and {Pieroni}, Mauro and {Pound}, Adam and {Quelquejay-Leclere}, Hippolyte and {Ricciardone}, Angelo and {Rossi}, Elena Maria and {Sartirana}, Andrea and {Savalle}, Etienne and {Sberna}, Laura and {Sesana}, Alberto and {Shoemaker}, Deirdre and {Slutsky}, Jacob and {Sotiriou}, Thomas and {Speri}, Lorenzo and {Staab}, Martin and {Steer}, Dani{\`e}le and {Tamanini}, Nicola and {Tasinato}, Gianmassimo and {Torrado}, Jesus and {Torres-Orjuela}, Alejandro and {Toubiana}, Alexandre and {Vallisneri}, Michele and {Vecchio}, Alberto and {Volonteri}, Marta and {Yagi}, Kent and {Zwick}, Lorenz},
        title = "{LISA Definition Study Report}",
      journal = {arXiv e-prints},
         year = 2024,
        month = feb,
          eid = {arXiv:2402.07571},
        pages = {arXiv:2402.07571},
          doi = {10.48550/arXiv.2402.07571},
archivePrefix = {arXiv},
       eprint = {2402.07571},
 primaryClass = {astro-ph.CO},
       adsurl = {https://ui.adsabs.harvard.edu/abs/2024arXiv240207571C}
}

@ARTICLE{2004MNRAS.349..181N,
       author = {{Nelemans}, G. and {Yungelson}, L.~R. and {Portegies Zwart}, S.~F.},
        title = "{Short-period AM CVn systems as optical, X-ray and gravitational-wave sources}",
      journal = {\mnras},
         year = 2004,
        month = mar,
       volume = {349},
       number = {1},
        pages = {181-192},
          doi = {10.1111/j.1365-2966.2004.07479.x},
archivePrefix = {arXiv},
       eprint = {astro-ph/0312193},
 primaryClass = {astro-ph},
       adsurl = {https://ui.adsabs.harvard.edu/abs/2004MNRAS.349..181N}
}

@ARTICLE{2023arXiv231103431V,
       author = {{Valli}, Ruggero and {Graziani}, Luca and {the LISA Synthetic UCB Catalogue Group}},
        title = "{BinCodex: a common output format for binary population synthesis}",
      journal = {arXiv e-prints},
         year = 2023,
        month = nov,
          eid = {arXiv:2311.03431},
        pages = {arXiv:2311.03431},
          doi = {10.48550/arXiv.2311.03431},
archivePrefix = {arXiv},
       eprint = {2311.03431},
 primaryClass = {astro-ph.IM},
       adsurl = {https://ui.adsabs.harvard.edu/abs/2023arXiv231103431V}
}

@ARTICLE{2008ApJ...677L..55S,
       author = {{Seto}, Naoki},
        title = "{Detecting Planets around Compact Binaries with Gravitational Wave Detectors in Space}",
      journal = {\apjl},
         year = 2008,
        month = apr,
       volume = {677},
       number = {1},
        pages = {L55},
          doi = {10.1086/587785},
archivePrefix = {arXiv},
       eprint = {0802.3411},
 primaryClass = {astro-ph},
       adsurl = {https://ui.adsabs.harvard.edu/abs/2008ApJ...677L..55S}
}

@ARTICLE{2023MNRAS.522.5358F,
       author = {{Finch}, Eliot and {Bartolucci}, Giorgia and {Chucherko}, Daniel and {Patterson}, Ben G. and {Korol}, Valeriya and {Klein}, Antoine and {Bandopadhyay}, Diganta and {Middleton}, Hannah and {Moore}, Christopher J. and {Vecchio}, Alberto},
        title = "{Identifying LISA verification binaries among the Galactic population of double white dwarfs}",
      journal = {\mnras},
         year = 2023,
        month = jul,
       volume = {522},
       number = {4},
        pages = {5358-5373},
          doi = {10.1093/mnras/stad1288},
archivePrefix = {arXiv},
       eprint = {2210.10812},
 primaryClass = {astro-ph.SR},
       adsurl = {https://ui.adsabs.harvard.edu/abs/2023MNRAS.522.5358F}
}

@BOOK{1997nceg.book.....P,
       author = {{Pagel}, Bernard E.~J.},
        title = "{Nucleosynthesis and Chemical Evolution of Galaxies}",
         year = 1997,
       adsurl = {https://ui.adsabs.harvard.edu/abs/1997nceg.book.....P}
}

@article{van_Zeist_2024,
   title={Evaluating the gravitational wave detectability of globular clusters and the Magellanic Clouds for LISA},
   volume={691},
   ISSN={1432-0746},
   url={http://dx.doi.org/10.1051/0004-6361/202451026},
   DOI={10.1051/0004-6361/202451026},
   journal={Astronomy and Astrophysics},
   publisher={EDP Sciences},
   author={van Zeist, Wouter G. J. and Nelemans, Gijs and Zwart, Simon F. Portegies and Eldridge, Jan J.},
   year={2024},
   month=nov, pages={A316} }

@ARTICLE{2007CSE.....9...90H,
       author = {{Hunter}, John D.},
        title = "{Matplotlib: A 2D Graphics Environment}",
      journal = {Computing in Science and Engineering},
         year = 2007,
        month = may,
       volume = {9},
       number = {3},
        pages = {90-95},
          doi = {10.1109/MCSE.2007.55},
       adsurl = {https://ui.adsabs.harvard.edu/abs/2007CSE.....9...90H}
}

@ARTICLE{2011CSE....13b..22V,
       author = {{van der Walt}, St{\'e}fan and {Colbert}, S. Chris and {Varoquaux}, Ga{\"e}l},
        title = "{The NumPy Array: A Structure for Efficient Numerical Computation}",
      journal = {Computing in Science and Engineering},
         year = 2011,
        month = mar,
       volume = {13},
       number = {2},
        pages = {22-30},
          doi = {10.1109/MCSE.2011.37},
archivePrefix = {arXiv},
       eprint = {1102.1523},
 primaryClass = {cs.MS},
       adsurl = {https://ui.adsabs.harvard.edu/abs/2011CSE....13b..22V}
}

@inproceedings{mckinney2010,
  author    = {Wes McKinney},
  title     = {Data Structures for Statistical Computing in Python},
  booktitle = {Proceedings of the 9th Python in Science Conference (SciPy 2010)},
  pages     = {51--56},
  year      = {2010},
}

@ARTICLE{1996MNRAS.281..257T,
       author = {{Tout}, Christopher A. and {Pols}, Onno R. and {Eggleton}, Peter P. and {Han}, Zhanwen},
        title = "{Zero-age main-seqence radii and luminosities as analytic functions of mass and metallicity}",
      journal = {\mnras},
         year = 1996,
        month = jul,
       volume = {281},
       number = {1},
        pages = {257-262},
          doi = {10.1093/mnras/281.1.257},
       adsurl = {https://ui.adsabs.harvard.edu/abs/1996MNRAS.281..257T}
}

@INPROCEEDINGS{1976IAUS...73...75P,
       author = {{Paczynski}, B.},
        title = "{Common Envelope Binaries}",
    booktitle = {Structure and Evolution of Close Binary Systems},
         year = 1976,
       editor = {{Eggleton}, Peter and {Mitton}, Simon and {Whelan}, John},
       series = {IAU Symposium},
       volume = {73},
        month = jan,
        pages = {75},
       adsurl = {https://ui.adsabs.harvard.edu/abs/1976IAUS...73...75P}
}

@INPROCEEDINGS{1976IAUS...73...35V,
       author = {{van den Heuvel}, E.~P.~J.},
        title = "{Late Stages of Close Binary Systems}",
    booktitle = {Structure and Evolution of Close Binary Systems},
         year = 1976,
       editor = {{Eggleton}, Peter and {Mitton}, Simon and {Whelan}, John},
       series = {IAU Symposium},
       volume = {73},
        month = jan,
        pages = {35},
       adsurl = {https://ui.adsabs.harvard.edu/abs/1976IAUS...73...35V}
}

@ARTICLE{1988ApJ...329..764L,
       author = {{Livio}, Mario and {Soker}, Noam},
        title = "{The Common Envelope Phase in the Evolution of Binary Stars}",
      journal = {\apj},
         year = 1988,
        month = jun,
       volume = {329},
        pages = {764},
          doi = {10.1086/166419},
       adsurl = {https://ui.adsabs.harvard.edu/abs/1988ApJ...329..764L}
}

@ARTICLE{1993PASP..105.1373I,
       author = {{Iben}, Jr., Icko and {Livio}, Mario},
        title = "{Common Envelopes in Binary Star Evolution}",
      journal = {\pasp},
         year = 1993,
        month = dec,
       volume = {105},
        pages = {1373},
          doi = {10.1086/133321},
       adsurl = {https://ui.adsabs.harvard.edu/abs/1993PASP..105.1373I}
}

@ARTICLE{2019MNRAS.490.3196P,
       author = {{Pillepich}, Annalisa and {Nelson}, Dylan and {Springel}, Volker and {Pakmor}, R{\"u}diger and {Torrey}, Paul and {Weinberger}, Rainer and {Vogelsberger}, Mark and {Marinacci}, Federico and {Genel}, Shy and {van der Wel}, Arjen and {Hernquist}, Lars},
        title = "{First results from the TNG50 simulation: the evolution of stellar and gaseous discs across cosmic time}",
      journal = {\mnras},
         year = 2019,
        month = dec,
       volume = {490},
       number = {3},
        pages = {3196-3233},
          doi = {10.1093/mnras/stz2338},
archivePrefix = {arXiv},
       eprint = {1902.05553},
 primaryClass = {astro-ph.GA},
       adsurl = {https://ui.adsabs.harvard.edu/abs/2019MNRAS.490.3196P}
}

@ARTICLE{2020MNRAS.498.2957C,
       author = {{Comerford}, T.~A.~F. and {Izzard}, R.~G.},
        title = "{Estimating the outcomes of common envelope evolution in triple stellar systems}",
      journal = {\mnras},
         year = 2020,
        month = oct,
       volume = {498},
       number = {2},
        pages = {2957-2967},
          doi = {10.1093/mnras/staa2539},
archivePrefix = {arXiv},
       eprint = {2008.09671},
 primaryClass = {astro-ph.SR},
       adsurl = {https://ui.adsabs.harvard.edu/abs/2020MNRAS.498.2957C}
}

@article{10.1093/mnras/staa3242,
    author = {Glanz, Hila and Perets, Hagai B},
    title = {Simulations of common envelope evolution in triple systems: circumstellar case},
    journal = {Monthly Notices of the Royal Astronomical Society},
    volume = {500},
    number = {2},
    pages = {1921-1932},
    year = {2020},
    month = {10},
    issn = {0035-8711},
    doi = {10.1093/mnras/staa3242},
    url = {https://doi.org/10.1093/mnras/staa3242},
    eprint = {https://academic.oup.com/mnras/article-pdf/500/2/1921/34463113/staa3242.pdf},
}

@ARTICLE{2013A&ARv..21...59I,
       author = {{Ivanova}, N. and {Justham}, S. and {Chen}, X. and {De Marco}, O. and {Fryer}, C.~L. and {Gaburov}, E. and {Ge}, H. and {Glebbeek}, E. and {Han}, Z. and {Li}, X. -D. and {Lu}, G. and {Marsh}, T. and {Podsiadlowski}, P. and {Potter}, A. and {Soker}, N. and {Taam}, R. and {Tauris}, T.~M. and {van den Heuvel}, E.~P.~J. and {Webbink}, R.~F.},
        title = "{Common envelope evolution: where we stand and how we can move forward}",
      journal = {\aapr},
         year = 2013,
        month = feb,
       volume = {21},
          eid = {59},
        pages = {59},
          doi = {10.1007/s00159-013-0059-2},
archivePrefix = {arXiv},
       eprint = {1209.4302},
 primaryClass = {astro-ph.HE},
       adsurl = {https://ui.adsabs.harvard.edu/abs/2013A&ARv..21...59I}
}

@article{10.1046/j.1365-8711.2000.03343.x,
    author = {Maxted, P. F. L. and Marsh, T. R. and Moran, C. K. J. and Han, Z.},
    title = {The triple degenerate star WD 1704+481},
    journal = {Monthly Notices of the Royal Astronomical Society},
    volume = {314},
    number = {2},
    pages = {334-337},
    year = {2000},
    month = {05},
    issn = {0035-8711},
    doi = {10.1046/j.1365-8711.2000.03343.x},
    url = {https://doi.org/10.1046/j.1365-8711.2000.03343.x},
    eprint = {https://academic.oup.com/mnras/article-pdf/314/2/334/3994908/314-2-334.pdf},
}

@article{Eggleton_2001,
doi = {10.1086/323843},
url = {https://dx.doi.org/10.1086/323843},
year = {2001},
month = {dec},
publisher = {},
volume = {562},
number = {2},
pages = {1012},
author = {Eggleton, Peter P. and Kiseleva-Eggleton, Ludmila},
title = {Orbital Evolution in Binary and Triple Stars, with an Application to SS Lacertae},
journal = {The Astrophysical Journal}
}

@article{10.1046/j.1365-8711.1998.01903.x,
    author = {Kiseleva, L. G. and Eggleton, P. P. and Mikkola, S.},
    title = {Tidal friction in triple stars},
    journal = {Monthly Notices of the Royal Astronomical Society},
    volume = {300},
    number = {1},
    pages = {292-302},
    year = {1998},
    month = {10},
    issn = {0035-8711},
    doi = {10.1046/j.1365-8711.1998.01903.x},
    url = {https://doi.org/10.1046/j.1365-8711.1998.01903.x},
    eprint = {https://academic.oup.com/mnras/article-pdf/300/1/292/3920716/300-1-292.pdf},
}

@article{Temmink_2023,
   title={Coping with loss: Stability of mass transfer from post-main-sequence donor stars},
   volume={669},
   ISSN={1432-0746},
   url={http://dx.doi.org/10.1051/0004-6361/202244137},
   DOI={10.1051/0004-6361/202244137},
   journal={Astronomy and; Astrophysics},
   publisher={EDP Sciences},
   author={Temmink, K. D. and Pols, O. R. and Justham, S. and Istrate, A. G. and Toonen, S.},
   year={2023},
   month=jan, pages={A45} }

@article{Woods_2012,
doi = {10.1088/0004-637X/744/1/12},
url = {https://dx.doi.org/10.1088/0004-637X/744/1/12},
year = {2011},
month = {dec},
publisher = {The American Astronomical Society},
volume = {744},
number = {1},
pages = {12},
author = {Woods, T. E. and Ivanova, N. and van der Sluys, M. V. and Chaichenets, S.},
title = {ON THE FORMATION OF DOUBLE WHITE DWARFS THROUGH STABLE MASS TRANSFER AND A COMMON ENVELOPE},
journal = {The Astrophysical Journal}
}

@article{Passy_2012,
   title={THE RESPONSE OF GIANT STARS TO DYNAMICAL-TIMESCALE MASS LOSS},
   volume={760},
   ISSN={1538-4357},
   url={http://dx.doi.org/10.1088/0004-637X/760/1/90},
   DOI={10.1088/0004-637x/760/1/90},
   number={1},
   journal={The Astrophysical Journal},
   publisher={American Astronomical Society},
   author={Passy, Jean-Claude and Herwig, Falk and Paxton, Bill},
   year={2012},
   month=nov, pages={90} }

@article{Ge_2010,
doi = {10.1088/0004-637X/717/2/724},
url = {https://dx.doi.org/10.1088/0004-637X/717/2/724},
year = {2010},
month = {jun},
publisher = {The American Astronomical Society},
volume = {717},
number = {2},
pages = {724},
author = {Ge, Hongwei and Hjellming, Michael S. and Webbink, Ronald F. and Chen, Xuefei and Han, Zhanwen},
title = {ADIABATIC MASS LOSS IN BINARY STARS. I. COMPUTATIONAL METHOD},
journal = {The Astrophysical Journal}
}

@article{Ge_2015,
doi = {10.1088/0004-637X/812/1/40},
url = {https://dx.doi.org/10.1088/0004-637X/812/1/40},
year = {2015},
month = {oct},
publisher = {The American Astronomical Society},
volume = {812},
number = {1},
pages = {40},
author = {Ge, Hongwei and Webbink, Ronald F. and Chen, Xuefei and Han, Zhanwen},
title = {ADIABATIC MASS LOSS IN BINARY STARS. II. FROM ZERO-AGE MAIN SEQUENCE TO THE BASE OF THE GIANT BRANCH},
journal = {The Astrophysical Journal}
}

@article{Ge_2020,
doi = {10.3847/1538-4357/aba7b7},
url = {https://dx.doi.org/10.3847/1538-4357/aba7b7},
year = {2020},
month = {aug},
publisher = {The American Astronomical Society},
volume = {899},
number = {2},
pages = {132},
author = {Ge, Hongwei and Webbink, Ronald F and Chen, Xuefei and Han, Zhanwen},
title = {Adiabatic Mass Loss in Binary Stars. III. From the Base of the Red Giant Branch to the Tip of the Asymptotic Giant Branch},
journal = {The Astrophysical Journal}
}

@article{Cornish_2007,
   title={Tests of Bayesian model selection techniques for gravitational wave astronomy},
   volume={76},
   ISSN={1550-2368},
   url={http://dx.doi.org/10.1103/PhysRevD.76.083006},
   DOI={10.1103/physrevd.76.083006},
   number={8},
   journal={Physical Review D},
   publisher={American Physical Society (APS)},
   author={Cornish, Neil J. and Littenberg, Tyson B.},
   year={2007},
   month=oct }

@article{Littenberg_2023,
   title={Prototype global analysis of LISA data with multiple source types},
   volume={107},
   ISSN={2470-0029},
   url={http://dx.doi.org/10.1103/PhysRevD.107.063004},
   DOI={10.1103/physrevd.107.063004},
   number={6},
   journal={Physical Review D},
   publisher={American Physical Society (APS)},
   author={Littenberg, Tyson B. and Cornish, Neil J.},
   year={2023},
   month=mar }

@misc{katz2024efficientgpuacceleratedmultisourceglobal,
      title={An efficient GPU-accelerated multi-source global fit pipeline for LISA data analysis}, 
      author={Michael L. Katz and Nikolaos Karnesis and Natalia Korsakova and Jonathan R. Gair and Nikolaos Stergioulas},
      year={2024},
      eprint={2405.04690},
      archivePrefix={arXiv},
      primaryClass={gr-qc},
      url={https://arxiv.org/abs/2405.04690}, 
}

@misc{deng2025modularglobalfitpipelinelisa,
      title={Modular global-fit pipeline for LISA data analysis}, 
      author={Senwen Deng and Stanislav Babak and Maude Le Jeune and Sylvain Marsat and Éric Plagnol and Andrea Sartirana},
      year={2025},
      eprint={2501.10277},
      archivePrefix={arXiv},
      primaryClass={gr-qc},
      url={https://arxiv.org/abs/2501.10277}, 
}

@article{Broekgaarden_2019,
   title={<scp>stroopwafel</scp>: simulating rare outcomes from astrophysical populations, with application to gravitational-wave sources},
   volume={490},
   ISSN={1365-2966},
   url={http://dx.doi.org/10.1093/mnras/stz2558},
   DOI={10.1093/mnras/stz2558},
   number={4},
   journal={Monthly Notices of the Royal Astronomical Society},
   publisher={Oxford University Press (OUP)},
   author={Broekgaarden, Floor S and Justham, Stephen and de Mink, Selma E and Gair, Jonathan and Mandel, Ilya and Stevenson, Simon and Barrett, Jim W and Vigna-Gómez, Alejandro and Neijssel, Coenraad J},
   year={2019},
   month=sep, pages={5228–5248} }

@article{Tauris:2000toa,
    author = "Tauris, T. M. and den Heuvel, E. P. J. van and Savonije, G. J.",
    title = "{Formation of millisecond pulsars with heavy white dwarf companions - extreme mass transfer on sub-thermal timescales}",
    eprint = "astro-ph/0001013",
    archivePrefix = "arXiv",
    doi = "10.1086/312496",
    journal = "Astrophys. J. Lett.",
    volume = "530",
    pages = "L93",
    year = "2000"
}

@article{Podsiadlowski_2002,
doi = {10.1086/324686},
url = {https://dx.doi.org/10.1086/324686},
year = {2002},
month = {feb},
publisher = {},
volume = {565},
number = {2},
pages = {1107},
author = {Podsiadlowski, Ph. and Rappaport, S. and Pfahl, E. D.},
title = {Evolutionary Sequences for Low- and Intermediate-Mass X-Ray Binaries},
journal = {The Astrophysical Journal}
}

@ARTICLE{too12,
       author = {{Toonen}, S. and {Nelemans}, G. and {Portegies Zwart}, S.},
        title = "{Supernova Type Ia progenitors from merging double white dwarfs. Using a new population synthesis model}",
      journal = {\aap},
         year = 2012,
        month = oct,
       volume = {546},
          eid = {A70},
        pages = {A70},
          doi = {10.1051/0004-6361/201218966},
archivePrefix = {arXiv},
       eprint = {1208.6446},
 primaryClass = {astro-ph.HE},
       adsurl = {https://ui.adsabs.harvard.edu/abs/2012A&A...546A..70T}
}

@ARTICLE{2023MNRAS.521.1088K,
       author = {{Keim}, Michael A. and {Korol}, Valeriya and {Rossi}, Elena M.},
        title = "{The large magellanic cloud revealed in gravitational waves with LISA}",
      journal = {\mnras},
         year = 2023,
        month = may,
       volume = {521},
       number = {1},
        pages = {1088-1098},
          doi = {10.1093/mnras/stad554},
archivePrefix = {arXiv},
       eprint = {2207.14277},
 primaryClass = {astro-ph.GA},
       adsurl = {https://ui.adsabs.harvard.edu/abs/2023MNRAS.521.1088K}
}

@ARTICLE{2017PASA...34...58E,
       author = {{Eldridge}, J.~J. and {Stanway}, E.~R. and {Xiao}, L. and {McClelland}, L.~A.~S. and {Taylor}, G. and {Ng}, M. and {Greis}, S.~M.~L. and {Bray}, J.~C.},
        title = "{Binary Population and Spectral Synthesis Version 2.1: Construction, Observational Verification, and New Results}",
      journal = {\pasa},
         year = 2017,
        month = nov,
       volume = {34},
          eid = {e058},
        pages = {e058},
          doi = {10.1017/pasa.2017.51},
archivePrefix = {arXiv},
       eprint = {1710.02154},
 primaryClass = {astro-ph.SR},
       adsurl = {https://ui.adsabs.harvard.edu/abs/2017PASA...34...58E}
}

@ARTICLE{2018MNRAS.480..302K,
       author = {{Kupfer}, T. and {Korol}, V. and {Shah}, S. and {Nelemans}, G. and {Marsh}, T.~R. and {Ramsay}, G. and {Groot}, P.~J. and {Steeghs}, D.~T.~H. and {Rossi}, E.~M.},
        title = "{LISA verification binaries with updated distances from Gaia Data Release 2}",
      journal = {\mnras},
         year = 2018,
        month = oct,
       volume = {480},
       number = {1},
        pages = {302-309},
          doi = {10.1093/mnras/sty1545},
archivePrefix = {arXiv},
       eprint = {1805.00482},
 primaryClass = {astro-ph.SR},
       adsurl = {https://ui.adsabs.harvard.edu/abs/2018MNRAS.480..302K}
}

@ARTICLE{2024ApJ...963..100K,
       author = {{Kupfer}, Thomas and {Korol}, Valeriya and {Littenberg}, Tyson B. and {Shah}, Sweta and {Savalle}, Etienne and {Groot}, Paul J. and {Marsh}, Thomas R. and {Le Jeune}, Maude and {Nelemans}, Gijs and {Pala}, Anna F. and {Petiteau}, Antoine and {Ramsay}, Gavin and {Steeghs}, Danny and {Babak}, Stanislav},
        title = "{LISA Galactic Binaries with Astrometry from Gaia DR3}",
      journal = {\apj},
         year = 2024,
        month = mar,
       volume = {963},
       number = {2},
          eid = {100},
        pages = {100},
          doi = {10.3847/1538-4357/ad2068},
archivePrefix = {arXiv},
       eprint = {2302.12719},
 primaryClass = {astro-ph.SR},
       adsurl = {https://ui.adsabs.harvard.edu/abs/2024ApJ...963..100K}
}

@article{Euclid_2011,
  author = {R. Laureijs and J. Amiaux and S. Arduini and others},
  title = "{Euclid Definition Study Report}",
  year = {2011},
  journal = {arXiv e-prints},
  eprint = {1110.3193},
  primaryClass = {astro-ph.CO}
}

@article{Roman_2019,
  author = {Akeson, R. and others},
  title = "{The Wide-Field InfraRed Survey Telescope (WFIRST) Science Investigations}",
  year = {2019},
  journal = {arXiv e-prints},
  eprint = {1902.05569},
  primaryClass = {astro-ph.IM}
}

@article{LSST_2009,
  author = {LSST Science Collaboration and others},
  title = "{LSST Science Book, Version 2.0}",
  year = {2009},
  journal = {arXiv e-prints},
  eprint = {0912.0201},
  primaryClass = {astro-ph.IM}
}

@ARTICLE{2006CQGra..23S.809S,
       author = {{Stroeer}, A. and {Vecchio}, A.},
        title = "{The LISA verification binaries}",
      journal = {Classical and Quantum Gravity},
         year = 2006,
        month = oct,
       volume = {23},
       number = {19},
        pages = {S809-S817},
          doi = {10.1088/0264-9381/23/19/S19},
archivePrefix = {arXiv},
       eprint = {astro-ph/0605227},
 primaryClass = {astro-ph},
       adsurl = {https://ui.adsabs.harvard.edu/abs/2006CQGra..23S.809S}
}

@ARTICLE{2012ApJ...744...12W,
       author = {{Woods}, T.~E. and {Ivanova}, N. and {van der Sluys}, M.~V. and {Chaichenets}, S.},
        title = "{On the Formation of Double White Dwarfs through Stable Mass Transfer and a Common Envelope}",
      journal = {\apj},
         year = 2012,
        month = jan,
       volume = {744},
       number = {1},
          eid = {12},
        pages = {12},
          doi = {10.1088/0004-637X/744/1/12},
archivePrefix = {arXiv},
       eprint = {1102.1039},
 primaryClass = {astro-ph.SR},
       adsurl = {https://ui.adsabs.harvard.edu/abs/2012ApJ...744...12W}
}

@ARTICLE{2019PASP..131a8002B,
       author = {{Bellm}, Eric C. and {Kulkarni}, Shrinivas R. and {Graham}, Matthew J. and {Dekany}, Richard and {Smith}, Roger M. and {Riddle}, Reed and {Masci}, Frank J. and {Helou}, George and {Prince}, Thomas A. and {Adams}, Scott M. and {Barbarino}, C. and {Barlow}, Tom and {Bauer}, James and {Beck}, Ron and {Belicki}, Justin and {Biswas}, Rahul and {Blagorodnova}, Nadejda and {Bodewits}, Dennis and {Bolin}, Bryce and {Brinnel}, Valery and {Brooke}, Tim and {Bue}, Brian and {Bulla}, Mattia and {Burruss}, Rick and {Cenko}, S. Bradley and {Chang}, Chan-Kao and {Connolly}, Andrew and {Coughlin}, Michael and {Cromer}, John and {Cunningham}, Virginia and {De}, Kishalay and {Delacroix}, Alex and {Desai}, Vandana and {Duev}, Dmitry A. and {Eadie}, Gwendolyn and {Farnham}, Tony L. and {Feeney}, Michael and {Feindt}, Ulrich and {Flynn}, David and {Franckowiak}, Anna and {Frederick}, S. and {Fremling}, C. and {Gal-Yam}, Avishay and {Gezari}, Suvi and {Giomi}, Matteo and {Goldstein}, Daniel A. and {Golkhou}, V. Zach and {Goobar}, Ariel and {Groom}, Steven and {Hacopians}, Eugean and {Hale}, David and {Henning}, John and {Ho}, Anna Y.~Q. and {Hover}, David and {Howell}, Justin and {Hung}, Tiara and {Huppenkothen}, Daniela and {Imel}, David and {Ip}, Wing-Huen and {Ivezi{\'c}}, {\v{Z}}eljko and {Jackson}, Edward and {Jones}, Lynne and {Juric}, Mario and {Kasliwal}, Mansi M. and {Kaspi}, S. and {Kaye}, Stephen and {Kelley}, Michael S.~P. and {Kowalski}, Marek and {Kramer}, Emily and {Kupfer}, Thomas and {Landry}, Walter and {Laher}, Russ R. and {Lee}, Chien-De and {Lin}, Hsing Wen and {Lin}, Zhong-Yi and {Lunnan}, Ragnhild and {Giomi}, Matteo and {Mahabal}, Ashish and {Mao}, Peter and {Miller}, Adam A. and {Monkewitz}, Serge and {Murphy}, Patrick and {Ngeow}, Chow-Choong and {Nordin}, Jakob and {Nugent}, Peter and {Ofek}, Eran and {Patterson}, Maria T. and {Penprase}, Bryan and {Porter}, Michael and {Rauch}, Ludwig and {Rebbapragada}, Umaa and {Reiley}, Dan and {Rigault}, Mickael and {Rodriguez}, Hector and {van Roestel}, Jan and {Rusholme}, Ben and {van Santen}, Jakob and {Schulze}, S. and {Shupe}, David L. and {Singer}, Leo P. and {Soumagnac}, Maayane T. and {Stein}, Robert and {Surace}, Jason and {Sollerman}, Jesper and {Szkody}, Paula and {Taddia}, F. and {Terek}, Scott and {Van Sistine}, Angela and {van Velzen}, Sjoert and {Vestrand}, W. Thomas and {Walters}, Richard and {Ward}, Charlotte and {Ye}, Quan-Zhi and {Yu}, Po-Chieh and {Yan}, Lin and {Zolkower}, Jeffry},
        title = "{The Zwicky Transient Facility: System Overview, Performance, and First Results}",
      journal = {\pasp},
         year = 2019,
        month = jan,
       volume = {131},
       number = {995},
        pages = {018002},
          doi = {10.1088/1538-3873/aaecbe},
archivePrefix = {arXiv},
       eprint = {1902.01932},
 primaryClass = {astro-ph.IM},
       adsurl = {https://ui.adsabs.harvard.edu/abs/2019PASP..131a8002B}
}

@ARTICLE{2024A&A...691A..44K,
       author = {{Korol}, Valeriya and {Buscicchio}, Riccardo and {Pakmor}, Ruediger and {Mor{\'a}n-Fraile}, Javier and {Moore}, Christopher J. and {de Mink}, Selma E.},
        title = "{Expected insights into Type Ia supernovae from LISA's gravitational wave observations}",
      journal = {\aap},
         year = 2024,
        month = nov,
       volume = {691},
          eid = {A44},
        pages = {A44},
          doi = {10.1051/0004-6361/202451380},
archivePrefix = {arXiv},
       eprint = {2407.03935},
 primaryClass = {astro-ph.HE},
       adsurl = {https://ui.adsabs.harvard.edu/abs/2024A&A...691A..44K}
}

@ARTICLE{2024Natur.635..316B,
       author = {{Burdge}, Kevin B. and {El-Badry}, Kareem and {Kara}, Erin and {Canizares}, Claude and {Chakrabarty}, Deepto and {Frebel}, Anna and {Millholland}, Sarah C. and {Rappaport}, Saul and {Simcoe}, Rob and {Vanderburg}, Andrew},
        title = "{The black hole low-mass X-ray binary V404 Cygni is part of a wide triple}",
      journal = {\nat},
         year = 2024,
        month = nov,
       volume = {635},
       number = {8038},
        pages = {316-320},
          doi = {10.1038/s41586-024-08120-6},
archivePrefix = {arXiv},
       eprint = {2404.03719},
 primaryClass = {astro-ph.HE},
       adsurl = {https://ui.adsabs.harvard.edu/abs/2024Natur.635..316B}
}
\clearpage
\appendix

\section{LISA double white dwarfs from isolated binary population} \label{appendix:A}

This section explains how we seed the double white dwarfs from isolated binary channels to a Milky Way-like galaxy. We do it in two steps. 

In the first step, we construct a synthetic population of isolated zero-age main-sequence binaries. The primary mass ($m_1$) is sampled from Kroupa's IMF \citep{2001MNRAS.322..231K}. Following this, we sample the orbital period of these binaries following empirically derived functions from \cite{2017ApJS..230...15M}. Using $m_1$ and $T_1$, we sample the initial mass ratio and eccentricities using functions from \cite{2017ApJS..230...15M}. This mass ratio is then used to calculate $m_2$. For a fair comparison to the triple systems, all masses are restricted to be between $1 - 8 \, \mathrm{M_{\odot}}$. Furthermore, we reject any systems that are Roche lobe filling \citep{1983ApJ...268..368E}. We note that this population inherently differs from the inner binaries of triple systems. The crucial, though not the only,  distinction arises from the dynamical stability requirement in triples, which forces the inner binary to be more compact. As a result, the semi-major axis distribution of inner binaries in triples is significantly more compact compared to that of isolated binaries \citep{2023ApJ...950....9R}. 
  We repeat the process described in Sect.~\ref{sec:gal_pop} to create a population of $10^5$ isolated binaries. Based on  Fig.~\ref{fig:initial_dist} we note that this isolated binary population looks distinctly different from the inner binary population of the triple systems. We evolve this synthetic population until Hubble time using {\tt MSE} and select the double white dwarf that enters LISA frequency bandwidth.

In the second step, We use the same galaxy and a similar method as mentioned in Sect.~\ref{sec:gal_pop} to seed the LISA double white dwarfs in the Galaxy. However, the total stellar mass in the simulated population is calculated after using an isolated binary population and  is given by
\begin{equation}
M_{\text{tot, MSE}} = \frac{N_{b,\,\mathrm{in\,range}}}{f_{b,\,\mathrm{in\,range}} \cdot f_b} \cdot \left[ f_t \cdot m_t + f_b \cdot m_b + (1 - f_t - f_b) \cdot m_s \right],
\end{equation}
where $N_{b,\,\mathrm{in\,range}} = 10^5$ is the number of simulated isolated binary-star systems with \texttt{MSE}, $f_{b,\,\mathrm{in\,range}}$ is the fraction of binaries in this range relative to the full mass range, and $f_t = 0.2$, $f_b =0.3$, and $1 - f_t - f_b = 0.5$ represent the fractions of triples, binaries, and singles in the full stellar population; $m_t =3.5\, \mathrm{M_\odot}$, $m_b = 0.9 \, \mathrm{M_\odot}$, and $m_s = 0.5 \, \mathrm{M_\odot}$ denote the average masses of triple, binary, and single systems, respectively. This gives an estimate of about $\sim 3.8 \times 10^{6}$ LISA double white dwarfs.

To assess the impact of the triple fraction on normalization, we repeat the same procedure with zero triple fraction ($f_t = 0$). Under this assumption, the estimated number of Galactic LISA double white dwarfs originating from isolated binaries is approximately $\sim 9 \times 10^{6}$. All these results are presented in Table~\ref{tab:counts}.

\section{Estimation of sampling uncertainty}\label{appendix:B}

We estimate the sampling uncertainty introduced by the stochastic seeding of the galaxy in our population using the bootstrapping method. Specifically, we repeat the procedure described in Sect. \ref{sec:gal_pop} fifty times. In each iteration, we sample from the intrinsic MSE double white dwarf population, separately for systems with triple and isolated binary origins, with replacement (\texttt{replacement=True}). This approach allows for random selection with repeated entries from each population. As a result, we generate fifty distinct mock populations for Galactic double white dwarfs from systems with triple origins and fifty more from systems with isolated binary origins.

To quantify the uncertainty, we calculate error bars based on the $ \approx 15.9$th and $ \approx 84.1$th percentiles of the mock populations, corresponding to the $1\sigma$ ($\approx 68.3\%$) confidence interval.

\end{document}